\documentclass[useAMS]{mn2e} 
\usepackage{epsfig,lscape} 
\usepackage{graphicx}

\usepackage{amsmath}
\usepackage{amssymb}
\usepackage{multirow}

\DeclareGraphicsExtensions{.eps,.eps.gz,.epsi}

\newcommand{\procspie}{Proc. SPIE}%
\newcommand{\aj}{AJ}%
\newcommand{\apj}{ApJ}%
\newcommand{\apjl}{ApJL}%
\newcommand{\aap}{A\&A}%
\newcommand{\pasp}{PASP}%
\newcommand{\mnras}{MNRAS}%
\newcommand{\nat}{Nature}%
\newcommand{\apjs}{ApJS}%

\newcommand{\teff}{\mbox{$T_{\rm eff}$}} 
 
\newcommand{\logg}{{\rm{log}~$g$}}
\newcommand{\feh}{{\rm [Fe/H]}} 
\newcommand{\ebv}{$E(B-V)$}

\newcommand{\kms}{km s$^{-1}$}

\begin{document}

\title[LSS-GAC: target selection and value-added catalogues]{
LAMOST Spectroscopic Survey of the Galactic Anti-centre (LSS-GAC): target selection 
and the first release of value-added catalogues
}

\author[Yuan et al.]{
H. B. Yuan$^{1}$\thanks{LAMOST Fellow}, X. W. Liu$^{1,2}$\thanks{E-mail: x.liu@pku.edu.cn}, Z. Y. Huo$^{3}$, 
M. S. Xiang$^{2}$, Y. Huang$^{2}$, B. Q. Chen$^{2}$, 
\newauthor H.-H. Zhang$^{2}$, N. C. Sun$^{2}$, C. Wang$^{2}$, H. W. Zhang$^{2}$, Y.-H. Zhao$^{3}$, A.-L. Luo$^{3}$, 
\newauthor J.-R. Shi$^{3}$, G.-P. Li$^{4}$, H.-L. Yuan$^{3}$, Y.-Q. Dong$^{3}$, G.-W. Li$^{3}$, Y.-H. Hou$^{4}$, and 
\newauthor Y. Zhang$^{4}$ \\\\
$1$ Kavli Institute for Astronomy and Astrophysics, Peking University, Beijing 100871, P. R. China \\ 
$2$ Department of Astronomy, Peking University, Beijing 100871, P. R. China\\
$3$ Key Laboratory of Optical Astronomy, National Astronomical Observatories, Chinese Academy of Sciences, Beijing 100012, P. R. China\\
$4$ Nanjing Institute of Astronomical Optics \& Technology, National Astronomical Observatories, Chinese Academy of Sciences, \\ Nanjing 210042, P. R. China
}

\date{Received:}
\pagerange{\pageref{firstpage}--\pageref{lastpage}} \pubyear{2014}

\maketitle
\label{firstpage}
\begin{abstract}{
As a major component of the LAMOST Galactic surveys, the LAMOST Spectroscopic
Survey of the Galactic Anti-centre (LSS-GAC) aims to survey a significant
volume of the Galactic thin/thick discs and halo for a contiguous sky area of
over 3,400 deg$^2$ centred on the Galactic anti-centre ($|b| \leq 30^{\circ}$, $150 \leq l \leq 210^{\circ}$),
and obtain $\lambda\lambda$3700 -- 9000 low resolution ($R \sim 1,800$) spectra for a statistically
complete sample of $\sim 3$\,M stars of all colours down to a limiting
magnitued of $r$ $\sim$ 17.8\,mag (to 18.5\,mag for limited fields).
Together with Gaia, the LSS-GAC will yield a unique dataset to advance our
understanding of the structure and assemblage history of
the Galaxy, in particular its disk(s). In addition to the main survey, the LSS-GAC will also target hundreds of
thousands objects in the vicinity fields of M\,31 and M\,33 and survey a significant fraction
(over a million) of randomly selected very bright stars (VB; $r \le 14$ mag) in the northern hemisphere.
During the Pilot and the first year Regular Surveys of LAMOST,
a total of 1,042,586 [750,867] spectra of a signal to noise ratio  S/N(7450\AA) $\ge$ 10 [S/N(4650\AA) $\ge$ 10] have been collected.
In this paper, we present a detailed description of the target selection algorithm, survey design, observations
and the first data release of value-added catalogues (including radial velocities, effective temperatures, surface gravities,
metallicities, values of interstellar extinction, distances, proper motions and orbital parameters) of the LSS-GAC.
}
\end{abstract}
\section{Introduction} 
As a Chinese national research facility, the Guoshoujing Telescope (also named the LAMOST and Wang-Su Reflecting 
Schmidt Telescope, Cui et al. 2012) 
is an innovative quasi-meridian reflecting Schmidt Telescope capable of simultaneously recording spectra 
of up to 4,000 objects in a large field of view (FoV) of 5\degr~in diameter. 
The telescope is equipped with 16 low-resolution spectrographs, 32 CCDs and 4,000 fibres. 
Each spectrograph is fed with the light from 250 fibres and covering 
a wavelength range 3700 -- 9100\,\AA~at a resolving power $R\sim 1800$ 
(with slit masks of width of two thirds the fibre diameter, i.e. 2.2\,arcsec; Zhu et al. 2006; Zhu et al. 2010).
With an effective aperture ranging from 3.6 -- 4.9 m in diameter, depending on the pointing direction, the LAMOST can reach a faint 
limiting magnitude of $r = 18.5$\,mag in a total integration time of 90 minutes under favourable observing conditions.  
The telescope is located in Xinglong Station and operated by the National Astronomical Observatories, Chinese Academy of Sciences (NAOC). 
Following  a two-year period of commissioning initiated in September 2009 that solved the problem of auto-positioning of the 
4,000 fibres and a year long Pilot Surveys (Luo et al. 2012)  
from October 2011 -- June 2012 to test and verify the designs and scientific goals of the various 
components of the Surveys, the LAMOST Regular Surveys began in October 2012.
    
The LAMOST Regular Surveys consist of two parts (Zhao et al. 2012): the LAMOST ExtraGAlactic Survey (LEGAS) and
the LAMOST Experiment for Galactic Understanding and Exploration (LEGUE; Deng et al. 2012).
The LEGUE has three components: 
the spheroid, the anti-centre and the disc surveys,
each focusing on a distinct component of the Galaxy and having different survey footprints and target selection algorithms.
The spheroid survey plans to observe at least 2.5 million stars in the Galactic caps 
($|b| \ge 20\degr$), down to a limiting magnitude of $r \sim 17.8$ mag (to 18.5 mag for limited fields), 
with targets selected from the fourth United States Naval Observatory 
CCD Astrograph Catalog (UCAC4; Zacharias et al. 2013) and the PanSTARRS 1 (PS1; Kaiser 2004) catalogues  
using a preferential target selection algorithm (Carlin et al. 2012; Zhang et al. 2012; Yang et al. 2012). 
The disc survey (Chen et al. 2012) plans to cover as much area of low Galactic latitudes 
($-20\degr \le b \le +20 \degr$) as visible from Xinglong Station and allowed by the available observing time,
focusing on open clusters and selected star-forming regions in the Galactic (thin) disk. The targets 
are selected from a combination of the UCAC4, PS1 and the Two Micro All Sky Survey (2MASS; Skrutskie et al. 2006) catalogues,
down to a limiting magnitude of $r \sim 17.8$ mag (to 18.5 mag for limited fields).

As a major component of the LEGUE, the LAMOST Spectroscopic Survey of the Galactic Anti-centre (LSS-GAC; Liu et al. 2014) 
aims to collect spectra for a statistically complete sample of $\sim 3$ million stars of all colours  
down to a limiting magnitude of $r \sim 17.8$ mag (to 18.5 mag for limited fields),
distributed in a large and continuous sky area of over 3,400 deg$^2$ centred on the Galactic
anti-centre ($|b| \leq 30^{\circ}$, $150^{\circ} \leq l \leq 210^{\circ}$)
and covering a significant volume of the Galactic thin/thick discs and halo.
Of the LEGUE survey footprints, the area of Galactic anti-centre is particularly interesting for several reasons: 
1) Disk is the defining component of a giant spiral galaxy like our own Milky Way 
in terms of stellar mass, angular momentum and star formation activities. It is  
also the most challenging component of a galaxy to study. 
The Galactic disc has already been known to exhibit rich yet poorly-understood (sub-)structures 
(e.g., Minniti et al. 2011; Bovy et al. 2012; G{\'o}mez et al. 2012; Li et al. 2012; Liu et al. 2012; 
Widrow et al. 2012; Carlin et al. 2013; Williams et al. 2013). 

Restricted by the observing capability, previous spectroscopic surveys including  
the Sloan Extension for Galactic Understanding and Exploration (SEGUE; Yanny et al. 2009),
the Apache Point Observatory Galactic Evolution Experiment (APOGEE; Eisenstein et al. 2011) and 
the Radial Velocity Experiment (RAVE; Steinmetz et al. 2006) suffer from various limitations, including complicated 
and hard to characterize target selection criteria, small or sporadic sky coverage, low sampling density or shallow survey depth,  
making it difficult to use them to establish a global census of the structure, stellar populations, kinematics and chemistry 
of the Galactic disks. With an unprecedented number of 4,000 fibres afforded by the LAMOST, 
building a deep, systematic spectroscopic survey targeting millions of stars drawn from a significant and contiguous  
volume of the Galactic discs with simple yet nontrivial algorithms has become, for the first time, feasible;
2) The LSS-GAC is ideally matched to the seasonal variation of weather conditions
of the site, which has most observable nights spreading from September to March the following year  (Yao et al. 2012); 
3) The stellar number density in the GAC direction is high enough to ensure full and efficient use of 
the unique multiplexing capability of LAMOST, i.e. 4,000 fibres; 
4) The interstellar extinction in the direction of GAC (Schlegel et al. 1998; hereafter SFD) is moderate even in the disk.
Within the footprint of LSS-GAC, about 70 per cent of the region has a colour excess \ebv~less than 0.5 mag,
and only about 10 per cent has \ebv~larger than 1.0\,mag.

Being the largest and most luminous galaxy of the Local Group,
M\,31 hosts a variety of interesting targets including planetary nebulae (PNe), H\,{\sc ii} regions, supergiants and globular clusters
that are easily detectable with the LAMOST (Yuan et al. 2010; Huo et al. 2010, 2013).
Kinematic and elemental abundance studies of PNe in M\,31 provide important information of the
chemical composition, kinematics and structure of M\,31 as well as the surrounding, extremely extended 
and complex stellar streams revealed by the recent deep imaging surveys
(e.g., Ibata et al. 2001, 2007; McConnachie et al. 2009), and 
are thus of great interest for the understanding of the assemblage history of the Local Group.
Background quasars in the vicinity fields of M\,31 and M\,33 can potentially provide an
excellent astrometric reference frame to measure the minute proper motions of M\,31 and M\,33 and their associated substructures.
Bright quasars are also excellent tracers to probe via absorption line spectroscopy the kinematics and 
chemistry of the interstellar/intergalactic medium of the Local Group.
The direction towards M\,31 and M\,33 also harbours rich foreground Galactic substructures 
(Bonaca et al. 2012; Chou et al. 2011; Majewski et al. 2004; Martin et al. 2007, 2014; Rocha-Pinto et al. 2004). 
For the above reasons, M\,31, M\,33 and the vicinity fields have  been 
observed with the LAMOST since the commissioning phase as parts of the LSS-GAC program, 
targeting foreground Galactic stars, accessible objects of special interests in M\,31 and M\,33 as well as background quasars.

To make full use of the telescope, randomly selected very bright (VB) stars of 9 $\le r < 14$\,mag 
in sky areas accessible to the LAMOST ($-10\degr \le$ Dec $\le 60\degr$) are targeted 
using observing time of non-ideal conditions, including bright/grey lunar nights. 
The VB survey is supplementary to the LSS-GAC main survey in the sense that it has a much larger
sky coverage but brighter magnitude limit and uses a simpler target selection algorithm.
The VB survey thus yields an excellent sample of the local stars to probe the solar neighbourhood.
A concise description of the scientific motivations, survey design and target selection, 
observations and data reduction and preliminary scientific results of LSS-GAC has been previously presented in Liu et al. (2014).

The LSS-GAC will deliver classifications, radial velocities and stellar
atmospheric  parameters (\teff, \logg, \feh, $\alpha$-element to iron abundance ratio {\rm [$\alpha$/Fe]}
and carbon to iron abundance ratio {\rm [C/Fe]}) for a magnitude-limited, spatially continuous and statistically complete sample of about 3 million stars 
in the Galactic thin/thick discs and halo.
The recently successfully launched astrometric satellite Gaia (Perryman et al.
2001; Katz et al. 2004) will yield accurate proper motions and parallaxes
for one billion Galactic stars to $V \sim 20$\,mag, radial velocities of 60 -- 100 million stars to $V \sim$ 15 -- 16\,mag,
and atmospheric parameters of $\sim$ 5 million stars to $V \sim$ 13\,mag\footnote{
See http://www.cosmos.esa.int/web/gaia/science-performance.}.
The LSS-GAC forms excellent synergy with Gaia.
Together with measurements of distances and tangential velocities provided by Gaia, the LSS-GAC will yield a unique dataset to:

a) characterize the stellar populations, chemical composition,
kinematics and structure of the thin/thick discs and their interface with the
halo; 

b) identify tidal streams and debris of disrupted dwarf galaxies and star clusters;

c) understand how resilient galaxy discs are to gravitational interactions;

d) study the temporal and secular evolution of the disks;

e) probe the gravitational potential and dark matter distribution;

f) map the interstellar extinction as a function of distance;

g) search for rare objects (e.g. stars of peculiar chemical composition or hyper-velocities);

h) study variable stars and binaries with multi-epoch spectroscopy;

i) and ultimately advance our understanding of the assemblage of galaxies and the origin of their regularity and diversity.

Following year-long Pilot Surveys, the LAMOST Regular Surveys were initiated 
in the fall of 2012. The Surveys are expected to complete in June 2017.  
By June 2013, about 1 million spectra of ${\rm S/N}(7450\AA) \ge
10$ per pixel\footnote{In this paper, the S/N refers to that per pixel. One pixel corresponds to 1.07\,\AA~ 
at 4650\,\AA~and 1.70\,\AA~at 7450\,\AA, respectively.} 
have been collected, including 0.44, 0.12 and 0.48 million 
from the main, M\,31/M\,33 and VB surveys, respectively. 
The LAMOST early data release (Luo et al. 2012), containing 319,000 spectra taken in the Pilot surveys 
and a catalogue of those objects, was publicly released to the international community in August 2012.
The first LAMOST regular data release (DR1; Bai et al. 2014), 
available to the Chinese astronomical community and the international partners from September 2013
and scheduled to be released to the whole international community in December 2014 
according to the LAMOST data policy, 
contains 1.8 million calibrated stellar spectra 
collected during the Pilot and the first year Regular Surveys and radial velocities derived from them.  
About 21.0, 3.7 and 28.4 per cent of the released spectra of LAMOST DR1 are 
from the LSS-GAC main, M\,31/M\,33 and VB surveys, respectively.
The LAMOST DR1 also includes a main catalogue providing stellar atmospheric parameters   
\teff, \logg~and \feh~of 1,085,404 A-, F-, G- and K-type stars, 
a supplementary catalogue giving spectral subtypes, 
luminosity classes and magnetic activities of 122,678 M-type stars (Yi et al. 2014) and another one listing
spectral subtypes and luminosity classes of 101,513 A-type stars. In the main catalog, about 13.9, 2.3 and 29.5 per cent stars are 
from the LSS-GAC main, M\,31/M\,33 and VB surveys, respectively.

In this paper, we present a detailed description of the LSS-GAC, including the target selection, survey design, 
observations and data reduction. The first release of LSS-GAC value-added catalogues, complementary to the LAMOST DR1 
and publicly available to the world-wide community, is described. 
Stellar parameters in the main catalogue of LAMOST DR1 are determined with the LAMOST
stellar parameter pipeline (LASP; Wu et al. 2014) by template matching with the ELODIE spectral library (Prugniel \& Soubiran 2001; Prugniel et al. 2007).
The value-added catalogues presented in the current work contain values of $V_{\rm r}$, \teff, \logg~and \feh~derived 
from LAMOST spectra with the LAMOST Stellar Parameter Pipeline at Peking University (LSP3 hereafter; Xiang et al. 2014a, b), 
values of the interstellar extinction and stellar distance determined by a variety of methods,
multi-band photometry collected from the Galaxy Evolution Explorer (GALEX; Martin et al. 2005) in the ultraviolet,
the Xuyi Schmidt Telescope Photometric Survey of the Galactic Anti-center (XSTPS-GAC; Liu et al. 2014; Zhang et al. 2014) in the optical
and the 2MASS and the Wide-field Infrared Survey Explorer (WISE; Wright et al. 2010) in the near infrared,
proper motions compiled from various sources and corrected for systematics, as well as estimated orbital parameters for about 0.67 million 
stars surveyed under the umbrella of LSS-GAC for the survey period covered by the LAMOST official DR1 release. 
The properties of LSS-GAC samples in this data release are discussed.  
The paper is organized as follows. 
Section 2 presents a detailed description of the target selection algorithm and survey design of LSS-GAC.
Survey progress and data reduction are described in Section 3. The LSP3 stellar parameters are discussed in Section 4.
The methods and results of extinction and distance determinations are given in Sections 5 and 6, respectively. 
Compilations and comparisons of proper motions and estimates of stellar orbital parameters are described in Section 7. 
The properties of LSS-GAC sample stars accumulated hitherto are discussed and the value-added catalogues described in Section 8.
We close with a summary of the main results in Section 9.

\section{Target selection and survey design}

\subsection{The LSS-GAC main survey}
\begin{figure*}
\includegraphics[width=140mm]{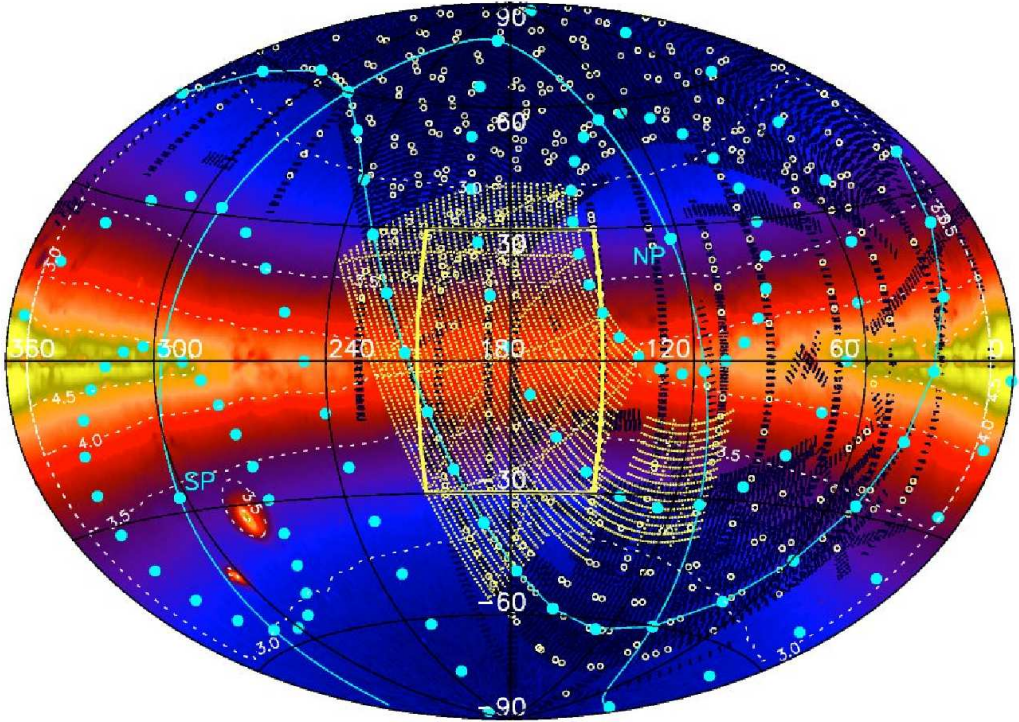}
\caption{
Footprints of the LSS-GAC (the central yellow bucket box) and 
XSTPS-GAC (the central yellow shaded area) surveys in a Galactic coordinate system centred on
the Galactic anti-centre ($l = 180^{\circ}$, $b = 0^{\circ}$). The background
pseudo colour image shows stellar number densities per square degree from
the 2MASS survey, overlaid with contours of constant
logarithmic number densities (white dotted lines). The black shaded areas
delineate footprints of the SDSS and SDSS-II imaging surveys. White open
circles denotes the SDSS/SEGUE spectroscopic plates. The cyan lines and dots
delineate the equatorial coordinate system, with the north and south celestial
poles marked.}
\label{footprint}
\end{figure*}

\begin{figure}
\includegraphics[width=90mm]{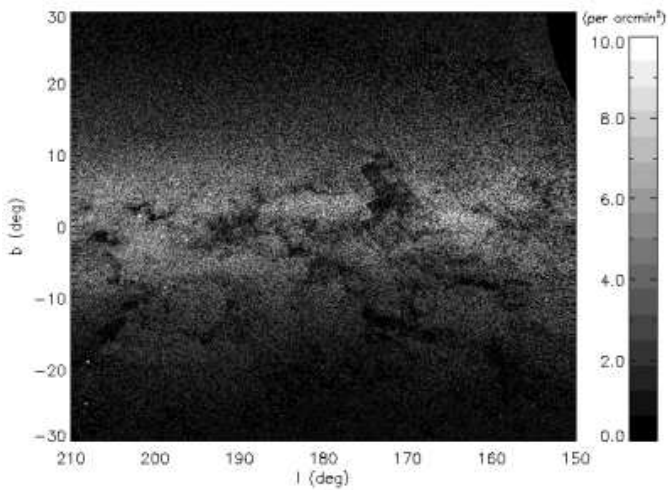}
\caption{
Distribution of stellar number densities of the clean photometric sample constructed 
from the XSTPS-GAC catalogues within the footprint of the LSS-GAC main survey.}
\label{r18.5}
\end{figure}

\begin{figure}
\includegraphics[width=90mm]{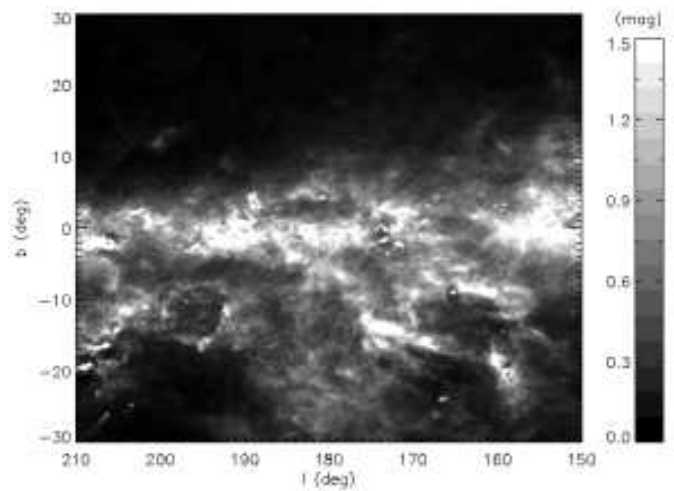}
\caption{
Same as Fig.\,\ref{r18.5} but for values of \ebv~of the interstellar extinction.
The map is constructed from the data of Schlegel et al. (1998).
}
\label{SFD}
\end{figure}

The LSS-GAC aims to deliver spectral classifications, radial velocities and atmospheric parameters (\teff, \logg, \feh, [$\alpha$/Fe]
and [C/Fe]) for a statistically complete sample of about 3 million stars of all colours down to 
a limiting magnitude of $r \sim$ 17.8\,mag (18.5\,mag for selected fields),
distributed in a contiguous area of 3,438\,deg$^2$ ($-30\degr \le b \le +30 \degr, 150 \le l \le 210 \degr$),
sampling a significant volume of the Galactic thin/thick discs and halo.
There are several constraints needed to be considered when designing the LSS-GAC survey: 
1) For the desired limiting magnitude, there are over 10,000 stars per deg$^2$ on the disc and 
about 3,000 stars per deg$^2$ at 30$^{\circ}$ above/below the disc within the LSS-GAC footprint (Figs.\,\ref{footprint} and \ref{r18.5}),
too many to survey for a reasonable survey lifetime, say 5 years, even with 4,000 fibres offered by the LAMOST, six times more than available to the 
Sloan Digital Sky Survey (SDSS; York et al. 2000); 
2) The spatial distribution of stars is highly non-uniform, and this is further compounded by the relatively high interstellar extinction in this region;
3) As a fully developed grand-design spiral, the Galaxy, in particular its disk, consists of extremely complex and varying 
stellar populations; 
4) The FoV of LAMOST plates must be centred on stars brighter than $\sim 8$\,mag for the active optics to work 
so as to bring the individual segments of the primary and corrector mirrors into focus.  
Stars of such brightness are relatively rare and not uniformly distributed on the sky;
5) The LAMOST has a circular FoV, 5\,deg in diameter, so field overlapping can not be avoided in order to make a contiguous sky coverage;
and 6) Except for a few holes occupied by the four guiding cameras and the central Shack-Hartmann sensor, the 
4,000 fibres of LAMOST, each controlled by two motors and patrolling a sky area of $\sim 6.5$\,arcmin in diameter 
around its parking position, are more or less uniformly distributed on the focal plane.
Given these constraints, a simple yet non-trivial target selection algorithm and sophisticated survey strategy must be developed in order  
to achieve the defining scientific goals of LSS-GAC.

LSS-GAC targets are selected from the XSTPS-GAC photometric catalogues (Liu et al. 2014; Zhang et al. 2013, 2014; Yuan et al. in prep.). 
The basic idea of LSS-GAC target selection is to uniformly and randomly select stars 
from the colour-magnitude diagrams using a Monte Carlo method.
It has the advantages that: 1) It has a well defined selection function, such that whatever class of objects 
(e.g. white dwarfs, white-dwarf-main-sequence binaries, extremely metal-poor or hyper-velocity stars) 
are revealed by spectroscopic observations, they can be studied in a statistically meaningful way in 
terms of the underlying stellar populations for the survey volume, 
after various selection effects have been properly taken into account; 
2) Stars of all colours (spectral types) and magnitudes (distances) are selected in large numbers as much and as evenly as possible;
3) Rare objects of extreme colours are first selected and targeted with high priorities,  thus vastly increasing the discovery space of the survey.
Based on the actual measured performance of LAMOST, the scientific goals and data quality requirements of LSS-GAC,
the bright and faint limiting magnitudes of the main survey of LSS-GAC have been set at 14.0 and $\sim$ 17.8\,mag in $r$-band, respectively.
For a small number of selected fields, the limiting magnitude are set at 18.5\,mag. 
To make efficient use of observing time of different qualities and avoid fibre
cross-talking, three categories of spectroscopic plates are designated, bright (B), 
medium-bright (M) and faint (F), targeting sources of brightness
14.0\,mag$ < r \lesssim m_1$, $m_1 \lesssim r \lesssim m_2$ and $m_2 \lesssim r \leq
18.5$\,mag, respectively.  Here $m_1$ and $m_2$ are the border magnitudes separating B, M and F plates (discussed later).
Above the Galactic plane ($|b|$ $\ge$ 3.5$\degr$), $\sim 1,000$ stars will be surveyed per deg$^2$, i.e. about five
plates on average (2\,B, 2\,M and 1\,F) will be allocated for a given patch of sky.   
On the Galactic plane ($|b|$ $\le$ 3.5$\degr$), the sampling is doubled and 
about ten plates (4\,B, 4\,M, 2\,F) will be allocated for a given patch of sky.
Given the limited amount of observing time of exceptional quality that is needed to achieve the depth of F plates, 
only a finite number of designed F plates are expected to get observed during the five-year survey lifetime. 
Targets of all designed plates are selected, assigned and checked in advance.

Based on the above principles, there are five steps in the LSS-GAC survey design and target selection.
Firstly, a clean sample of targets is generated from the XSTPS-GAC catalogues.
Secondly, targets are assigned with different priorities and fed to the LAMOST Survey Strategy System (SSS; Yuan et al. 2008), 
which allocates fibres to targets for each designed plate scheduled for observation.
Thirdly, field centres of all designed plates, which have to be on a star brighter than $V \leq 7.3$\,mag, are 
carefully selected and adjusted to maximize 
the spatial uniformity of sampling over the whole survey area.
Fourthly, for each chosen centre of FoV, the SSS is run to assign targets 
(along with sky fibres) for all designed plates. The designed plates are then ready for observational scheduling. 
Finally, to account for the field rotation of LAMOST, prior to the actually execution of scheduled observations 
(typically in a week), the SSS has to be rerun and fibres re-assigned for targets of each 
designed plate. A small fraction of targets, less than $\sim 100$ (excluding sky fibres), 
fail to get reallocated a fibre and thus will not get observed.

The XSTPS-GAC (Liu et al. 2014) surveys in SDSS $g$, $r$ and $i$ bands an area of approximately 5,400\,deg$^2$ centred on the 
GAC, from ${\rm RA} \sim 3$ to 9$^{\rm h}$ and ${\rm Dec} \sim -10$ to +60$^{\rm o}$, 
plus an extension of $\sim 900$\,deg$^2$ to the M\,31, M\,33 region, 
with the Xuyi 1.04/1.20\,m Schmidt Telescope equipped with a 4k$\times$4k CCD.
The resulting catalogues archive about a hundred millions stars down to a limiting magnitude of $\sim 19.0$\,mag ($10\sigma$). 
The catalogues are astrometrically calibrated using the PPMXL (Roeser et al. 2010) as reference and globally flux-calibrated 
(using the ubercal algorithm, Padmanabhan et al. 2008) 
against the SDSS DR8 (Aihara et al. 2011) using the overlapping fields. The calibration has achieved an astrometric accuracy 
of 0.1\,arcsec (Zhang et al. 2014) and a global photometric accuracy of 2 per cent (Yuan et al., in prep.). The 
catalogues are used to generate a clean sample of targets for the LSS-GAC, with the requirements that: 
1) The stars are detected in at least two bands, including $r$ band, and have magnitudes  $ 14.0 < r \le 18.5$\,mag; 
2) Positions of stars measured in different bands agree within 0.5 arcsec; 
3) They are not flagged as galaxies nor star pairs in either $r$ or $i$ band; 
4) They have no neighbors within 5 arcsec that are brighter than ($m$ + 1) mag., 
where $m$ is magnitude of the star concerned in each of the three bands; 
and 5) They have local sky background no more than three times brighter than the sky background of the whole frame, 
i.e., they are not around extremely bright stars.
Fig.\,\ref{r18.5} shows the distribution of stellar number densities per square arcmin of the thus constructed clean photometric sample 
within the LSS-GAC footprint.
For comparison, a reddening map for the same region that shows values of \ebv~of the interstellar extinction from SFD is shown in Fig.\,\ref{SFD}.
The dusty Galactic disc is clearly visible, and exhibits number densities over 10,000 stars per\,deg$^2$ after imposing the above cuts.
The stellar number densities decrease rapidly toward high Galactic latitudes, down to about 3,000 per deg$^2$ at 
$|b| \sim 30\degr$. Patchy or filamentary dark lanes or bands caused by obscuration of dark clouds are abundant.
In the direction of GAC, more dark clouds are found in 
the southern Galactic hemisphere than in the northern hemisphere, as can be clearly seen in Fig.\,\ref{SFD}. 
There are a few roughly rectangle regions in Fig.\,\ref{r18.5} around longitude $l = 173\degr$ and latitude $b= 4.5\degr$,
with a total area of about 30\,deg$^2$, that have fewer stars than their surroundings due to the 
shallower limiting magnitudes caused by poor observing conditions when those regions were observed.
The black corner in the top right of Fig.\,\ref{r18.5} 
has declinations higher than $63\degr$, falling outside the XSTPS-GAC footprint.     

Before target priority assignment, we first need to estimate values of $m_1$ and $m_2$, the
border magnitudes separating B, M and F plates.  
Due to the spatial variation of the stellar populations, the survey depths of XSTPS-GAC,  as well as the patchy 
distribution of the interstellar extinction,
$m_1$ and $m_2$ vary slightly over the survey area, even within a single LAMOST field.
Fixing $m_1$ and $m_2$ to constant values for the whole survey area will lead to discontinuities 
of sample star number density in the colour-magnitude diagrams
at the border magnitudes separating B, M and F plates. To avoid such undesired features, 
we have allowed values of $m_1$ and $m_2$ to float and determined their appropriate values for 
individual patches of sky of 1\,deg. by 1\,deg. in size.
The whole survey area is first divided into boxes of 1\,deg. side in RA and Dec. 
For stars in each box, ($r$, $g - r$) and ($r$, $r - i$) Hess diagrams 
are constructed from the clean sample. Stars of extremely blue
colours, $(g - r)$ or $(r - i) \leq -0.5$, and of extremely red colours, $(g - r)$ or
$(r - i) > 2.5$, are first selected and removed from the diagrams.  Stars in the remaining colour space are then
selected using a Monte Carlo method. First two uniformly distributed random numbers, $r$ in the range
(14, 18.5] and $c$ in ($-1$, 5], are generated. If $-1 < c \leq 2$, then 
the ($r$, $g - r$) Hess diagram is used to select the next star, assuming $g - r = c + 0.5$. Otherwise the
($r$, $r - i$) Hess diagram is used instead, assuming $r - i = c
- 2.5$.  For a given colour-magnitude set ($r$, $g - r$) or ($r$, $r - i$), if there are stars in
the appropriate Hess diagram in a box centred on the simulated set of colour-magnitude
and of length 0.2\,mag in magnitude and 0.3\,mag in colour, then the star of
colour-magnitude values closest to the simulated set is chosen and removed from
the pool. If not, the process is repeated until a total of 1,000 stars,
including those of extreme colours, per deg$^2$ are selected\footnote{Note that 
the individual boxes have areas $cos(\delta)$ deg$^2$, i.e. smaller than 1 deg$^2$ for declination $\delta \neq 0$\,deg.}. 
The stars are then sorted in magnitude from bright to faint. Then $m_1$ and $m_2$ are set to the 
faint end magnitudes of the first 40\% and 80\% sources, respectively. 
The above procedure applies to the survey area of $|b| > 3.5^{\circ}$, where
the LSS-GAC plans to sample 1,000 stars per deg$^2$. For $|b| \leq 3.5^{\circ}$,
a similar procedure is applied except that 2,000 stars per deg$^2$are selected and the selections are carried out 
in ($l$, $b$) space instead of (RA, Dec).

\begin{figure}
\includegraphics[width=90mm]{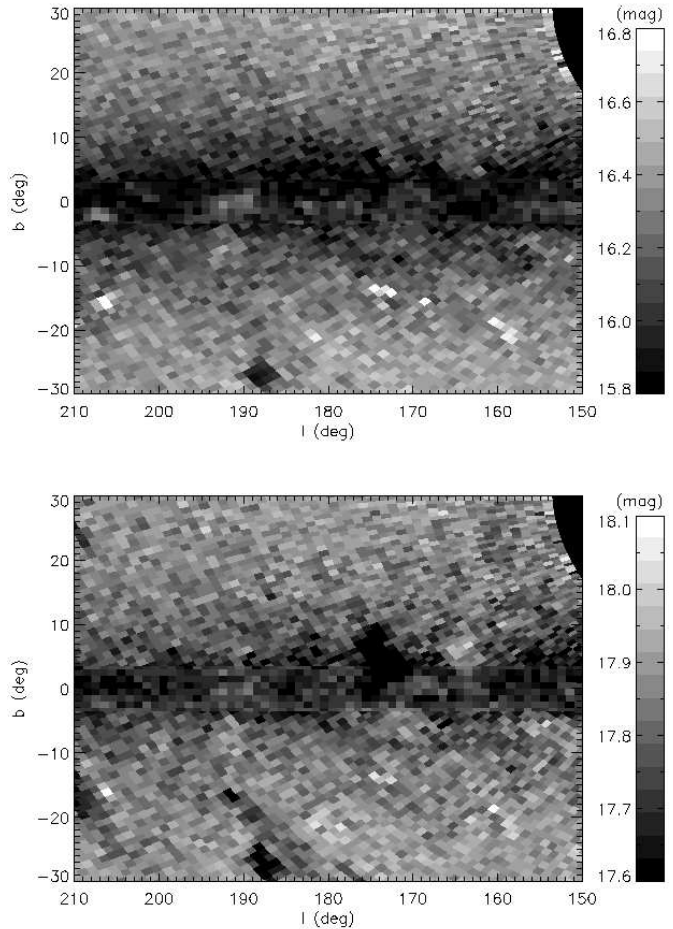}
\caption{
Spatial variations of $m_1$ (top panel) and  $m_2$ (bottom panel) within the footprint of the LSS-GAC main survey.
}
\label{ds9_m1}
\end{figure}

Fig.\,\ref{ds9_m1} shows the spatial variations over the survey footprint of $m_1$ and $m_2$.
$m_1$ has a typical value of 16.3\,mag and varies between 15.8 -- 16.5\,mag for most regions, whereas
$m_2$ has a typical value of 17.8\,mag and varies between 17.6 -- 18.0\,mag for most regions.
Both $m_1$ and $m_2$ decrease toward the Galactic mid-plane.
In regions where the survey depths of XSTPS-GAC are shallower than the designed limiting magnitude of LSS-GAC, 
such as around ($l,b$) = ($173\degr, 4.5\degr$) and ($188\degr,-27\degr$), the values of $m_1$ and $m_2$ 
are smaller than those of the surrounding areas.

\begin{figure*}
\includegraphics[width=120mm]{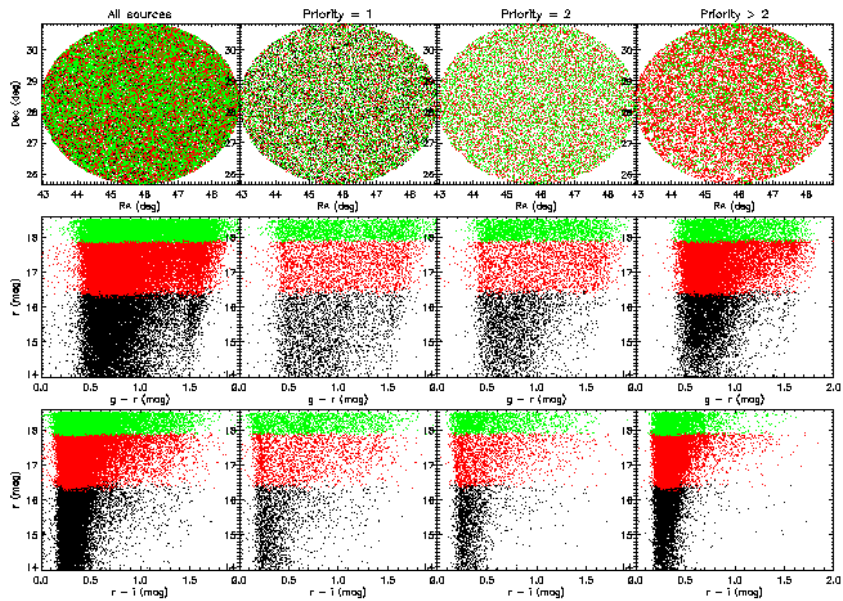}
\caption{Distributions of stars in (RA, Dec) (top panels) and in the ($r$, $g-r$) (middle panels) and ($r$, $r-i$) (bottom panels) 
colour-magnitude diagrams for all stars (first column), stars of first (second column), second (third column) and 
lower (fourth column) priorities 
in the clean photometric sample, for a field centred around RA = 3$^{\rm h}$ and Dec = 28$^{\circ}$. 
Black, red and green dots represent targets of B, M and F plates, respectively.
Note for the 2nd -- 4th columns in the middle panels, only stars selected in the ($r$, $g-r$) diagram are plotted.       
While for the 2nd -- 4th columns in the bottom panels, only stars selected in the ($r$, $r-i$) diagram are plotted.
}
\label{target_selection}
\end{figure*}

Once the values of $m_1$ and $m_2$ are set, stars of B, M and F plates are
selected and assigned priorities, in batches of 200 stars per deg$^2$. For each category of plates, up to
10 batches are selected pending on the availability of stars. The first
batch of stars are assigned the highest priority, the second one
priority lower, and so forth. When selecting each batch of stars, if a star of a
given coordinate set, (RA, Dec) or ($l$, $b$), is selected, then all other stars within 2\,arcmin of that star 
are removed from the pool to avoid small scale clustering. Those removed stars are however put back
to the pool when selecting the next batch of stars.
As an example, Fig.\,\ref{target_selection} shows the spatial distributions in (RA, Dec) and in the ($r$, $g-r$)
and ($r$, $r-i$) Hess diagrams for all stars and stars of different priorities in the clean sample, for a field of 
relatively high Galactic latitude centred  
around ${\rm RA} = 3^{\rm h}$ and ${\rm Dec} = 28^{\circ}$. 
Note for the 2nd -- 4th columns in the middle panels, only stars selected in the ($r$, $g-r$) Hess diagram are plotted. 
While for the 2nd -- 4th columns in the bottom panels, only stars selected in the ($r$, $r-i$) Hess diagram are plotted. 
As expected, for stars of a given priority, their
spatial distribution is fairly uniform and clustering in 
the ($r$, $g-r$) and ($r$, $r-i$) diagrams is much reduced.  
Note also that for a given priority, given the broader colour range of $g - r$,
more targets are selected from the ($r$, $g-r$) diagram than that of ($r$, $r-i$), 
especially for bright plates. 

\begin{figure*}
\includegraphics[width=120mm]{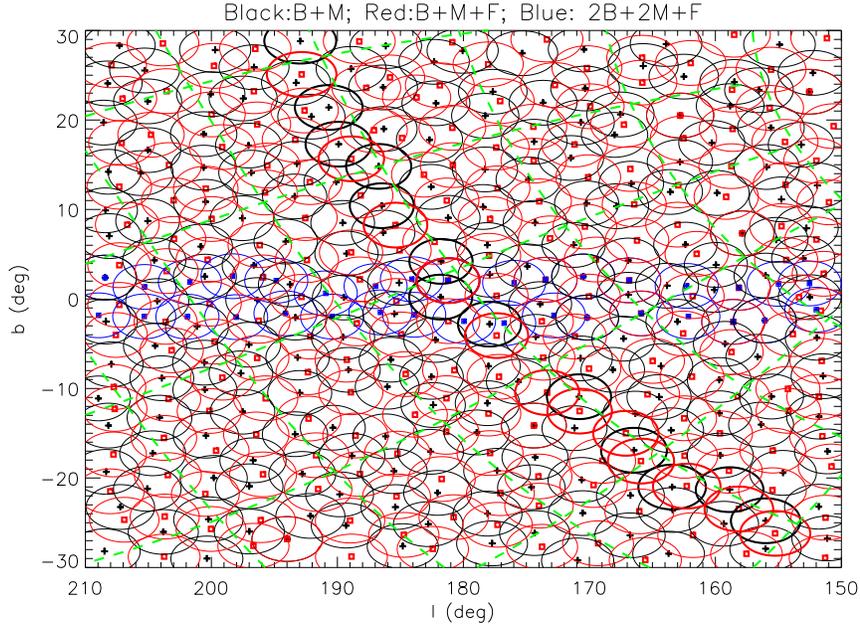}
\caption{Distribution of field centers, which have to be on stars brighter than $V = 7.3$\,mag,  
for G\,1 (black), G\,2 (red) and G\,3 (blue) fields of the LSS-GAC main survey. 
The green dashed lines stretching from top-left to bottom-right and from bottom-left to top-right 
delineate grids of constant Declination and Right Ascension, respectively.  
The thicker circles distributed roughly along Declination 29$^\circ$ represent fields designed for observations during
the Pilot Survey of LSS-GAC main survey.
}
\label{central_stars}
\end{figure*}

After priority assignment, 
we define the field centers, which, as stated earlier, must be centred on  
bright stars in order for the LAMOST active optics to operate. 
We use the Hipparcos and Tycho Catalogs (Perryman \& ESA 1997),
which contain 118,218 stars with accurate positions, proper motions and parallaxes and are complete to $V-$band magnitudes between 7.3 and 9.0\,mag.
We require that the stars must be single and have magnitudes $4 \le V \le 7.3$\,mag in order to make the active optics work
even under unfavourable conditions such as bright moon or partially cloudy nights. 
A couple of stars with a companion of separations less than 0.5 arcsec are however retained in places of few bright stars. 
The LAMOST active optics works well under such small separations.
To maximize the uniformity of spatial sampling over the whole LSS-GAC survey area and achieve the designed sampling density,
three groups (G\,1, G\,2 and G\,3 hereafter) of fields are defined, with the first two groups covering the whole survey footprint
and G\,3 for the Galactic plane ($|b|$ $\le$ 3.5$\degr$) only.
For the G\,1, G\,2 and G\,3 groups of fields, two (1 B and 1 M), three (1 B, 1 M and 1 F) and five (2 B, 2 M and 1 F) plates are planned, respectively.
We first partition the survey area in ($l, b$) by equilateral triangles of length $2.5\degr \times \sqrt3$,
assuming that the celestial sphere is flat in ($l, b$), not a bad assumption
given that the survey extends to a latitude of only 30$^\circ$.
Bright stars from the Hipparcos and Tycho catalogues that satisfy the aforementioned criteria and are closest to the 
centres of those equilateral triangles are selected as G\,1 field central stars.
Central stars of G\,2 fields are then selected by shifting the equilateral triangles in order to 
fill up as much as possible the gaps that are not covered by the G\,1 fields. 
Central stars of G\,3 are selected in a similar way, except that the survey area 
($|b| \leq 3.5^\circ$, $150 \leq l \leq 210^\circ$) are partitioned by 
squares of length  $2.5\degr \times \sqrt2$  such that the central stars must have a latitude $|b|$ smaller than $2.5^\circ \times \sqrt{2}/2$.
In total, 216, 216 and 34 central stars are defined as the G\,1, G\,2 and G\,3 field centers, respectively,
yielding a total of 1,250 plates (500 B, 500 M and 250 F).
Their spatial distributions are displayed in Fig.\,\ref{central_stars}.
The field central stars are not necessarily unique and two fields may share the same star. 
There are 5, 4 and 6 common stars shared by G\,1 and G\,2,
G\,1 and G\,3, and G\,2 and G\,3 fields, respectively. 

\begin{figure*}
\includegraphics[width=100mm]{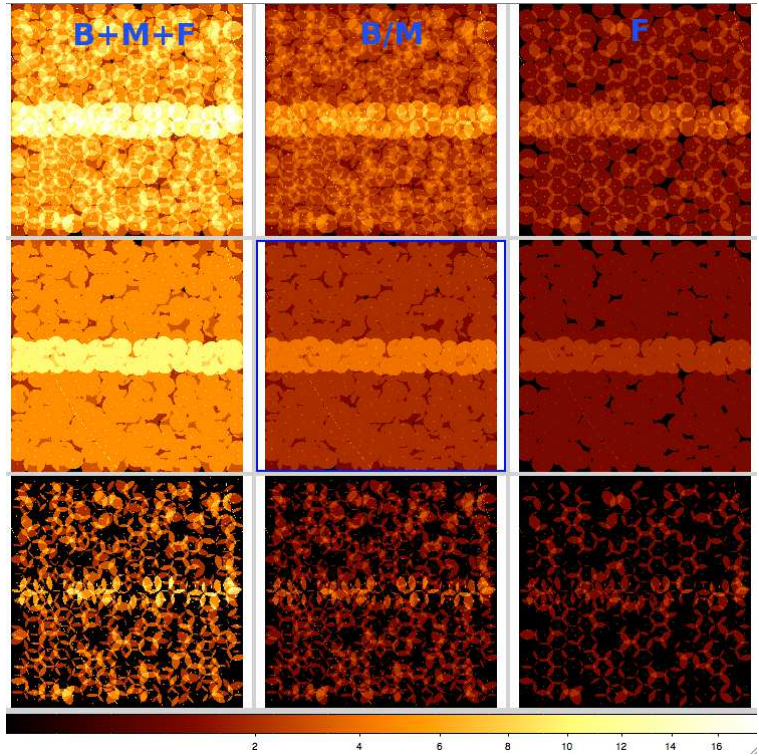}
\caption{Spatial distributions of plate sampling density (top panels), that after discounting overlapping plates 
from the same field group (middle panels), and that for the overlapping areas for
time-domain spectroscopy (bottom panels) for the LSS-GAC main survey. 
From left to right the panels represent all, B (or M) and F plates, respectively.
Each panel represents the footprint of LSS-GAC main survey, and covers 
$l$ from 210$\degr$ to 150$\degr$ from left to right, and $b$ from $-30$$\degr$ to $30\degr$ from bottom to top. 
The effects of the five holes in the LAMOST focal plane have been ignored.
A colourbar is plotted at the bottom.} 
\label{sampling_density}
\end{figure*}

Since the LAMOST FoV is circular, field overlapping can not be avoided in order to achieve a contiguous sky coverage.
In the ideal cases where the centres of partition fall on qualified bright stars, 
the fraction of overlapping areas is about 20 per cent for G\,1 and G\,2 fields, and more for G\,3 fields.
There are two options to make use of the overlapping areas of adjacent fields. 
One is to increase the sampling density in the overlapping regions by observing different targets in different plates.
The other is to make duplicate observations for some sources in the overlapping areas for time-domain spectroscopy. The latter is 
preferred considering that a) the sampling density is already high and the margin gained by increasing the sampling density is
not huge; b) the sampling density will be more uniform over the whole survey footprint; 
and most important of all, c) it opens up the possibility of time-domain 
spectroscopy that may lead to new discoveries (e.g., Yang et al. 2014). 
To implement such an approach, we require that there are no common targets amongst plates of  
different groups of fields, but plates belonging to the same group, target selections of overlapping plates are independent, i.e. 
fibres are assigned to stars regardless whether the stars have been assigned a fibre or not in the adjacent plates. 
The procedure guarantees that adjacent plates belonging to the same group target, with few exceptions, 
the same sources in their overlapping areas.    

The top panels of Fig.\,\ref{sampling_density} display the spatial distributions of plate density of the LSS-GAC main 
survey for all, B (or M) and F plates. The middle panels show the same after discounting  
overlapping regions of adjacent plates from the same field group. The distributions become much more uniform than those
displayed in the top panels.
Detailed fractions of survey footprint covered by different plate densities as 
illustrated in the second row of Fig.\,\ref{sampling_density} 
are listed in Table\,\ref{sampling_density_table}.
About three quarters of the survey area
are sampled by five plates (2 B, 2 M and 1 F) for a given patch of the sky out of the Galactic plane.
For the highly sampled Galactic plane ($|b| \leq 3.5^\circ$), essentially all the region, which accounts for about 
13.5 per cent of the whole survey footprint is sampled by ten plates
(4 B, 4 M and 2 F) for a given patch of the sky. 
About 10 per cent of the survey area is covered only by plates of fields of either G\,1 or G\,2.
Only 0.4 per cent of the survey footprint is left uncovered.
The bottom panels of Fig.\,\ref{sampling_density} display the plate density distributions 
for overlapping areas of adjacent fields for time-domain spectroscopy.
About one quarter of all targets will be observed twice and about 2 per cent will be observed three times. 
Note that in the above calculations, the effects of the four holes in the LAMOST focal plane have not been taken into account.

\begin{table}{}
\small
\centering
\begin{minipage}{90mm}
\caption{Sampling density of the LSS-GAC main survey.}
\label{sampling_density_table}\end{minipage}
\tabcolsep 1mm
\begin{tabular}{lccll}
  \hline\noalign{\smallskip}
Plate   & Area fraction &  Area   & Plate & Plate  \\
density &  (\%) &  (deg$^2$)   & type& group  \\
  \hline\noalign{\smallskip}
0& 0.4 & 134  &          &             \\
2& 5.1 & 175 &  B+M     & G\,1\\
3& 5.8 & 200 &   B+M+F   & G\,2\\
5& 74.3 & 2554 &   2B+2M+F & G\,1, G\,2\\
7& 0.45& 15  &  3B+3M+F & G\,1, G\,3\\
8& 0.43& 15  &  3B+3M+2F& G\,2, G\,3\\
10&13.5 & 464 &   4B+4M+2F& G\,1, G\,2, G\,3\\
\noalign{\smallskip}\hline
\end{tabular}
\flushleft
\end{table}

Once the target priorities have been assigned, the plate centres and categories as well as field groups  fixed,
the next (fourth) step in the LSS-GAC survey design and target selection is to run 
the SSS to allocate fibres to targets of individual plates.
Usually the SSS requires three input catalogues: targets, 
standard stars for flux calibration and positions of blank sky for sky subtraction.
Given the unknown extinctions, which are likely to be significant, to individual disc stars targeted by the LSS-GAC, 
it is not feasible, as in the case of SDSS which primarily targets sources at high Galactic latitudes, to
pre-select F-turnoff stars as flux calibration standards based on the measured photometric colours. 
On the other hand, as described earlier, our target selection algorithm is specifically designed 
to target stars of all colours (thus spectral types), and  
as shall be shown in the next subsection, our targets include a large fraction of F stars.
Consequently, we have used a modified version of the SSS for which no  
standard stars need to be supplied for its operation. 
A separate pipeline that differs from the LAMOST default 2D pipeline  
has also been been developed at Peking University to flux-calibrate plates collected for  
the LSS-GAC using an iterative approach (Xiang et al. 2014a).
Catalogs of positions of blank sky contain a sub-sample of positions of 
empty sky that have no 2MASS, GSC2.3 (Lasker et al. 2008) and 
USNOB1.0 (Monet et al. 2003) sources in a square box of size 36\,arcsec centred on the position.
This sub-sample of blank sky positions is constructed such that the positions are more or less 
uniformly distributed spatially and has a number density of about 100 per deg$^2$.
Typically 320 sky fibres are allocated per plate for sky background measurements.
This gives 20 sky fibres per spectrograph, or about 16 sky fibres per deg$^2$, which is about 3.5 times denser than the SDSS.
In addition to the input catalogues, to run the SSS, an observing date has to be 
specified and the date when the field passes the prime meridian at midnight is chosen.
In running the SSS, four guiding stars are first selected from the GSC2.3 catalogues to 
allow for corrections of fibre positioning errors produced by the expansion/contraction (due to changes of the focal length), 
shift and rotation of the focal plane.
If four guiding stars can not be located, the field centre is shifted to the nearest qualified bright star.

The SSS assigns LSS-GAC targets according to their priorities.
Small numbers of targets, such as emission line objects from the INT/WFC Photometric H$\alpha$ Survey (IPHAS; Drew et al. 2005) 
of the northern Galactic plane (Witham et al. 2008), 
candidates of young stellar objects and white dwarfs can be added\,{\it ad hoc} to 
the input target catalogues, and assigned the highest priorities.
This ensures that adding those special targets will not jeopardize the overall spatial uniformity of the LSS-GAC targets. 
Ideally, a LAMOST plate should make full use of all the 4,000 fibres, with 320 of them targeting blank sky and 
the remaining targeting science objects. In reality, only 3700 -- 3930 fibres per plate are assigned 
a target (including blank sky) for most plates designed for the LSS-GAC main survey. 
The unassigned fibres are either dead or have no targets left in their reachable areas.

Finally the SSS has to be rerun in order to update the 
fibre allocation prior to the actual observation taking place (usually within a week), based on the 
output target list (including blank sky) generating by the first run of the SSS.
This is because the difference between the date specified when the SSS is first run and the actual observing 
date can lead to loss or change of guiding stars originally selected. In most cases, 
no loss or change of guiding stars occurs, and most 
targets assigned a fibre in the first round of runs of SSS get re-assigned a fibre.
Only a few, if not nil targets, fail to get reallocation of a fibre.
The occasional loss of a small number of targets is caused by the small 
differences in aberration corrections for fibre positions at different dates.
This ensures that all LSS-GAC plates can be designed, targets assigned 
and checked well in advance of actual observation.
In some rare cases, one or two of the selected guiding stars get changed.
The change of one guiding star can lead to a loss of 
200 -- 300 targets originally assigned a fibre.
If two guiding stars are changed, about 500 -- 600 original targets can get lost.
This problem has however been fully solved since November, 2013 by 
supplying the SSS when it is rerun lists of guiding stars 
that contain only those selected in the first run of SSS, instead of the whole GSC2.3 catalogues. 

It is at this very last stage that add-on targets are assigned fibres if possible.
Add-on targets usually have the highest priorities.  
They are assigned to fibres in a way that is very similar to blank sky (or standard stars 
in the case of high Galactic latitude plates targeted by the spheroid survey for example).
No more than 3 add-on targets per clip of 25 fibres and no more than 20 per spectrograph 
are assigned to ensure that the overall homogeneity and completeness of the main sample are 
not affected. 

\subsection{Simulation with the Besan\c{c}on model}

\begin{table*}{}
\small
\centering
\begin{minipage}{180mm}
\caption{Survey completeness of the LSS-GAC main survey for different stellar types at different Galactic latitudes 
for a one-degree wide stripe of $|b| \le 30\degr$ centred at $l = 180\degr$.
\label{Tabcompleteness} } \end{minipage}
\tabcolsep 2mm
\begin{tabular}{lcccccccc}
  \hline\noalign{\smallskip}
 &       \multicolumn{2}{c}{$-2.5\degr \le b \le 2.5\degr$} &     \multicolumn{2}{c}{$10\degr \le b \le 15\degr$}  &  \multicolumn{2}{c}{$25\degr \le b \le 30\degr$} 
 &  \multicolumn{2}{c}{$-30\degr \le b \le 30\degr$}        \\
Type     &  Num.  & Completeness                   &  Num. & Completeness              &  Num. & Completeness                 &  Num.  & Completeness        \\
 \hline\noalign{\smallskip}
K-(sub)giant  & 2,005  & 0.55   & 174 &0.56   &  14 & 0.67  & 4,285   & 0.47   \\
G-(sub)giant  & 677  & 0.37   & 129 &0.24   &  62 & 0.48  & 2,252   & 0.27    \\
A-dwarf  & 1,573  & 0.32   & 12 &0.75   &  0 & -  & 2,989   & 0.39    \\
F-dwarf  & 1,644  & 0.07   & 1,180 &0.20   &  778 & 0.57  & 13,393   & 0.14    \\
G-dwarf & 1,277  & 0.07   & 994 &0.07   &  1,081 & 0.38  & 12,451   & 0.08    \\
K-dwarf & 1,346  & 0.15   & 1,557 &0.14   &  1,692 & 0.46  & 18,251   & 0.19    \\
M-dwarf & 917  & 0.60   & 810 &0.53   &  740 & 0.63  & 9,328   & 0.55    \\
\noalign{\smallskip}\hline
\end{tabular}
\flushleft
\end{table*}

\begin{table}{}
\small
\centering
\begin{minipage}{90mm}
\caption{Survey depths of the LSS-GAC main survey of different types of star for a limiting magnitude $r = 14$ and 18.5\,mag, 
respectively, assuming zero extinction.\label{Tabvolume} } 
\end{minipage}
\tabcolsep 1mm
\begin{tabular}{lccccc}
  \hline\noalign{\smallskip}
Type &  $M$($g$)   & $M$($r$) & $M$($i$)  & Dist. (kpc)    & Dist. (kpc)       \\
                  &  &   &  & ($r=14$\,mag)  &  ($r=18.5$\,mag)                   \\
  \hline\noalign{\smallskip}
A0V & 1.29  & 1.53   & 1.75 &3.12   &  24.77   \\
F0V & 2.35  &   2.27 &  2.33  & 2.22   &   17.62   \\
G0V & 4.47  &  4.08  & 4.01 & 0.96    &  7.66      \\
K0V & 6.28  &  5.62  & 5.45 & 0.55    &  3.77      \\
M0V &  10.93  &  9.62  & 8.85  & 0.08    &  0.60  \\
K0III& 1.28   &  0.41  & 0.15 & 5.23   &   41.50  \\
M0III& $-$0.22  &  $-$1.71 & $-$2.59 & 13.9  &   110.15  \\
\noalign{\smallskip}\hline
\end{tabular}
\flushleft
\end{table}

\begin{figure*}
\includegraphics[width=180mm]{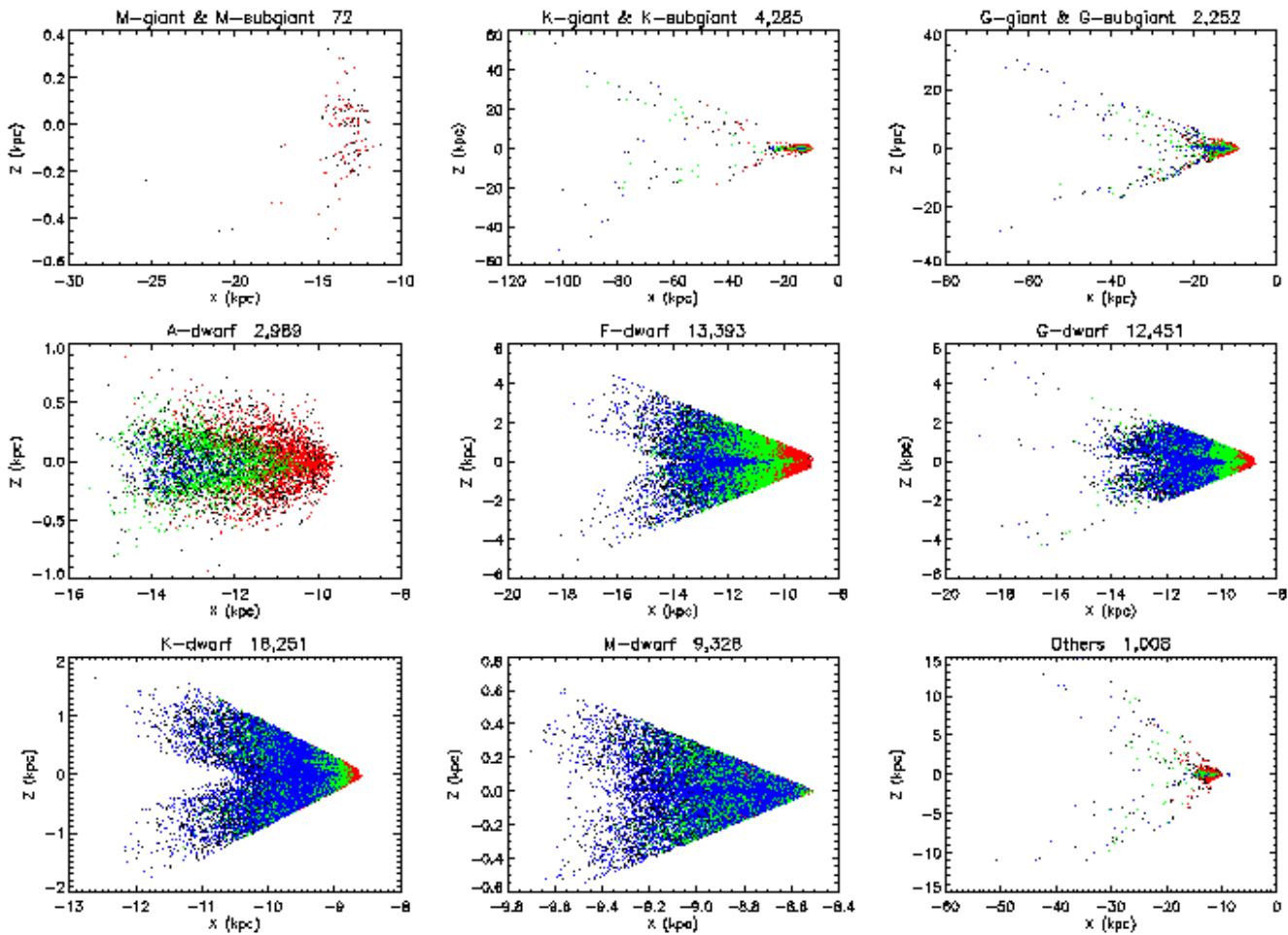}
\caption{
Spatial distributions of targets simulated with the Besan\c{c}on model of the LSS-GAC main survey in the (X, Z) 
plane of Galactic radius and height for a one-degree wide stripe of $|b| \le 30\degr$ centred at $l = 180\degr$. 
The numbers of different types of star targeted are labelled. 
Red, green and blue dots represent sources of B, M and F plates, respectively (c.f. Fig.\,\ref{target_selection}).
Note the Sun is located at (X, Z) = ($-$8,500, 15) pc and the disc is truncated at a radius of 14 kpc in the Besan\c{c}on model.
\label{simulation}
}
\end{figure*}

\begin{figure}
\includegraphics[width=90mm]{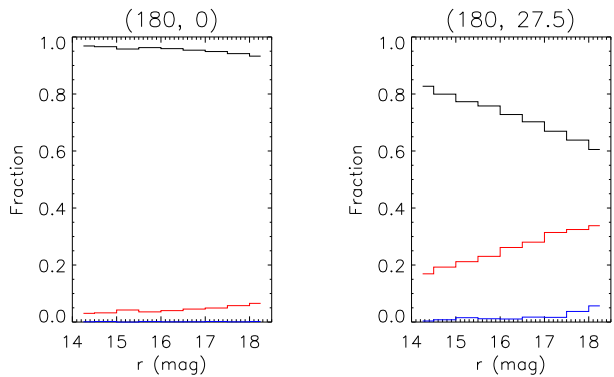}
\caption{
Relative fractions of targets from the Galactic thin disk (black lines), thick disk (red lines) and halo (blue lines)
plotted against $r$ band magnitude for two sightlines of the LSS-GAC main survey toward  
(l, b) = ($180\degr, 0\degr$) (left) and (l, b) = ($180\degr, 27.5\degr$)  (right), simulated with the Besan\c{c}on model.
\label{fraction}
}
\end{figure}

The above survey strategy and target selection algorithm are tested using mock catalogues generated from the Besan\c{c}on 
Galactic model (Robin et al. 2003). As an example, for a $1\degr$ wide stripe of 56.7 deg$^2$ centred at $l = 180\degr$ and 
stretching from $b$ = $-30\degr$ to +30$\degr$, the model yields 396,075 stars of $14 \le r \le 18.5$\,mag. 
Amongst them, 0.018, 2.3 and 2.1 per cent are respectively M, K and G subgiants/giants, 
and 1.9, 24, 39, 24 and 4.3 per cent are respectively A, F, G, K and M dwarfs. 
Note that $u,g,r,i,z$ magnitudes delivered by the Besan\c{c}on model refer to the CFHT/MegaCam filter system, and 
have been transformed to the SDSS photometric system using the calibration of the SNLS 
group\footnote{http://www.astro.uvic.ca/$\sim$pritchet/SN/Calib/ColourTerms-2006Jun19/index.html.} for the $g, r, i$ and $z$ 
bands and that from the CFHT website\footnote{http://cfht.hawaii.edu/Instruments/Imaging/MegaPrime/\\generalinformation.html.} 
for the $u$ band, respectively.
Photometric errors are also assigned to magnitudes yielded by the model according to their values  
to simulate the error-magnitude relations of the XSTPS-GAC photometry.
Running this mock catalogue through the target selection procedure outlined above, 
one finds that in total 64,029 of all stars, 99, 47 and 27 per cent of all M, K and G subgiants/giants, 39, 14, 8, 19 and 55 per cent 
of all A, F, G, K and M dwarfs, are selected and targeted by the LSS-GAC, respectively (Table\,\ref{Tabcompleteness}). 
Table\,\ref{Tabcompleteness} also shows that the completeness increases toward high Galactic latitudes for all stellar types. 
At $|b|$ = 25$\degr$ -- 30$\degr$, about half stars of all types get targeted. 
Fig.\,\ref{simulation} shows the spatial distributions of various types of star targeted in the (X, Z) plane of Galactic radius and height.
The LSS-GAC targets range from M dwarfs to M giants, probing volumes of a wide range of depths.
The first three columns of Table\,\ref{Tabvolume} give respectively the absolute magnitudes in $g$, $r$ and $i$ band of 
individual types of star, whereas the last 2 columns list the maximum distances probed by those types of star 
for a limiting magnitude of $r = 14.0$ and 18.5\,mag, respectively.     
At $r=18.5$\,mag, M0V, K0V, G0V, F0V, A0V, K0III and M0III stars
as far as 0.6, 3.8, 7.7, 17.6, 25, 40 and 110 kpc are reached, respectively, assuming zero reddening.
The simulation confirms that our target selection algorithm is well designed to 
meet the scientific goals of LSS-GAC.

Fig.\,\ref{fraction} shows the relative fractions of targets from the Galactic thin disk, thick disk and halo 
for two sightlines of the LSS-GAC main survey, simulated with the Besan\c{c}on model.
In the direction of Galactic anti-center, over 90 per cent of the targets are from the Galactic thin disk.
In the direction of (l, b) = ($180\degr, 27.5\degr$), as the targets get fainter, 
the fraction of thick disk stars increases from 17 to 35 per cent, 
whereas the fraction of halo stars increases from zero to 6 per cent.

\subsection{Target selection for the M31-M33 and VB Surveys} 

\begin{figure}
\includegraphics[width=90mm]{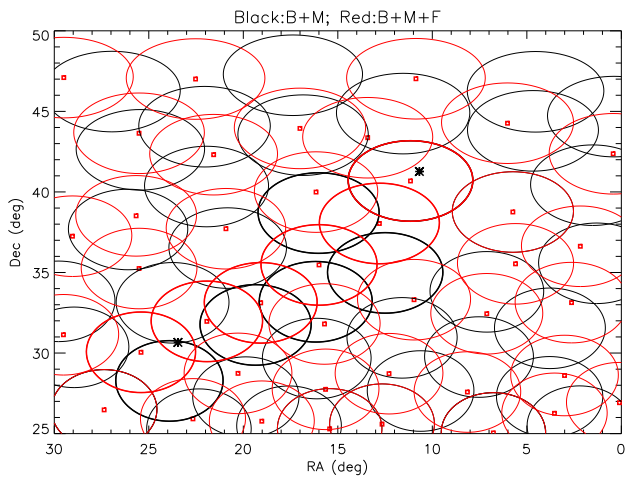}
\caption{Field centres of Group 1 (black) and 2 (red) for the LSS-GAC M\,31-M\,33 survey. 
The positions of M\,31 and M\,33 are indicated by stars.
The thicken fields that run from M\,31 to M\,33 are fields targeted during the Pilot Surveys.
}
\label{m31_central_stars}
\end{figure}

The targets in the M\,31-M\,33 field include known and candidate PNe, H\,{\sc ii} regions, supergiants and globular clusters of 
M\,31 and M\,33, known and candidate background quasars, as well as foreground Galactic stars from the XSTPS-GAC photometric catalogues.
The stars are selected, assigned and observed in the same way as for the LSS-GAC main survey. 
The selection  criteria of various types of target and candidate of interest
are presented elsewhere (e.g., Yuan et al. 2010 for PNe; Huo et al. 2010, 2013 for quasars). 
They are assigned higher priorities than (foreground Galactic) stars.
The field central stars for the M\,31-M\,33 survey are selected similarly to the LSS-GAC main survey in the (RA, Dec) plane.
Two groups of field central stars are chosen to cover the M\,31-M\,33 area 
($0\degr \le$ RA $\le 30\degr$, $25\degr \le$ Dec $\le 50\degr$) independently. 
Each group contains 41 central stars.
Their distributions are displayed in Fig.\,\ref{m31_central_stars}.

To make use of bright nights or nights of unfavourable conditions (large seeing or low atmospheric transparency), 
very bright (VB) plates are designed to target stars brighter than $r=14$\,mag 
in all sky area accessible to the LAMOST ($-10\degr \le$ Dec $\le 60\degr$).
Within the XSTPS-GAC footprint, all stars of $r \le 14.0$\,mag from the XSTPS-GAC catalogues and stars of 
$9 \le J \le 12.5$\,mag from the 2MASS catalogues are selected as potential targets.
Outside the XSTPS-GAC footprint, all stars of $ 10.0 \le b1 \le 15.0$\,mag, or $ 10.0 \le b2 \le 15.0$\,mag, 
or $9.0 \le r1 \le 14.0$\,mag, or $ 9.0 \le r2 \le 14.0$\,mag, or $ 8.5 \le i \le 13.5$\,mag from the PPMXL catalogues (Roeser et al. 2010) and stars 
of $9 \le J \le 12.5$\,mag from the 2MASS catalogues are selected as potential targets.
Stars are targeted with equal priorities. 
The field centres are selected similarly to the LSS-GAC main survey but in the (RA, Dec) plane after dividing the
sky area accessible to the LAMOST into three bins in Dec ($-$10$\degr$ -- 30$\degr$, 30$\degr$ -- 50$\degr$ 
and 50$\degr$ -- 60$\degr$).
Only one group of fields centred on a total of 966 bright stars are chosen.
For each of them, 1 to 9 plates are planned, depending on the number of stars available, 
yielding a total number of 2,594 VB plates.  
Typically 2,000 -- 3,400 stars get assigned a fibre for a VB plate, depending on its 
location (in particular its latitude).
About 10 per cent targets are duplicates targeted by adjacent plates. 
In the overlapping areas of adjacent plates, fibres are assigned to stars of a plate 
regardless whether they have been assigned a fibre or not in the adjacent plates.
No flux standard stars are allocated, as for the LSS-GAC main survey.
Due to relatively low densities of stars available compared to those of the main survey and the fact that 
VB plates are observed under unfavourable observing conditions (bright moon, poor seeing etc.), 
more ($\sim$ 460) sky fibres are allocated for more efficient use of available fibres and better sky subtraction.
The VB survey is supplementary to the LSS-GAC main survey in terms of (wider) sky coverage, (brighter) magnitude limit 
and (simpler) target selection criteria, yielding an excellent sample of local stars to probe the solar neighbourhood.

\section{Observations and data reduction} 

\subsection{Observations}
Being a meridian reflecting Schmidt telescope,
the LAMOST can only observe a given plate between 2h before and after the transit,
thus putting a strong constraint on the range of RA of plates that can be targeted at a given time.
The focal plane has a FoV of 5\degr~in diameter, where the 4,000 auto-positioning fibres are 
nearly evenly distributed, feeding the light to 16 low-resolution spectrographs, 250 fibres each.
The fibres have a diameter of 3.3 arcsec projected on the sky.
Slit marks of width 2/3 the fibre diameter, i.e. 2.2\,arcsec,  are placed in front of the fibre clips 
to increase the spectral resolution to a resolving power of $R \sim 1,800$ 
(Deng et al. 2012; Liu et al. 2014), similar to that of the SDSS. 
The spectra cover a wavelength range from 3700 -- 9000\,\AA~and are recorded in two arms, 
3700 -- 5900\,\AA~in the blue and 5700 -- 9000\,\AA~in the red.
A Shack-Hartmann system is housed at the centre of the FoV for the operation of the active optics.
It takes typically 30 minutes, depending on the pointing of the telescope and 
brightness of the field central star, to bring individual segments of the primary and corrector mirrors into focus.
Four guiding CCD cameras are placed about halfway out from the field centre to monitor  
guide stars during the exposures for accurate fibre positioning. 
The guide stars are also used to measure seeing during the observation.

Following a two-year commissioning phase, the LAMOST Pilot Surveys were initiated in October 2011 and completed in June 2012. 
The Regular Surveys were initiated in October 2012.
In each year, a sufficient number of plates of the LSS-GAC are planned in advance for observations.
For the main survey, we adopt a strategy that 
we start with the stripe along Dec = 29$\degr$ observed during the Pilot Surveys (Fig.\,\ref{central_stars}) 
and then extend both ways to higher and lower Declinations.
The main survey plates are observed in dark/grey nights.
Typically 2 -- 3 exposures are obtained for each plate, with typical integration time per exposure of 
600 -- 1200s, 1200 -- 1800s, 1800 -- 2400s for  B, M and F plates, respectively, depending on the weather.
Some test nights reserved to monitor the telescope performance are also used to observe the LSS-GAC plates.
The seeing varies between 3 -- 4 arcsec for most plates, with a typical value of about 3.5 arcsec.

\begin{figure}\includegraphics[width=90mm]{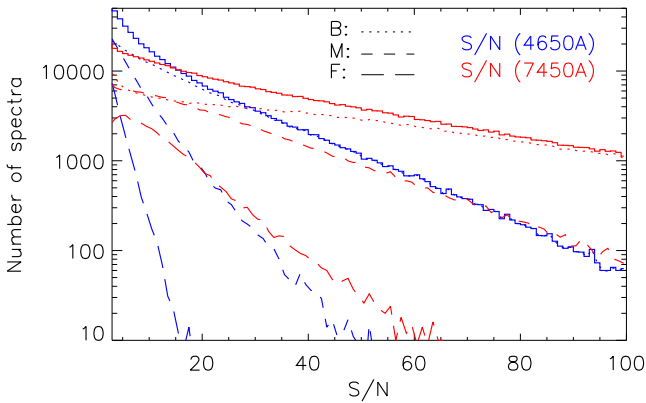}
\caption{Histogram showing the distributions of S/N(4650\AA) (blue) and S/N(7450\AA) (red) of 
targets observed in the LSS-GAC main survey by June 2013
from B (dotted), M (short dashed), F (long dashed) and all (solid) plates during the Pilot and the first year Regular Surveys.
\label{snr_gac}}
\end{figure}

\begin{figure*}\includegraphics[width=140mm]{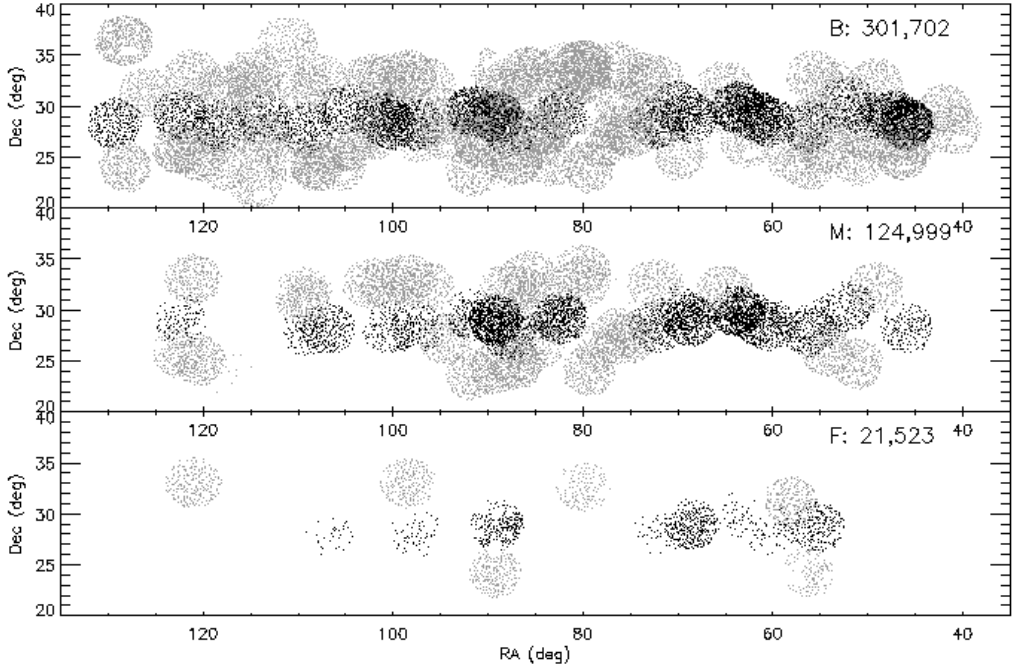}
\caption{
Spatial distributions of targets of spectral S/N(4650\AA) $\ge$ 10 or S/N(7450\AA) $\ge$ 10 
in the LSS-GAC main survey from B (top), M (middle) and F (bottom) plates 
during the Pilot (Oct. 2011 -- Jun. 2012, black dots) and the first year Regular (Oct. 2012 -- Jun. 2013, grey dots) Surveys.
In total, 301,702, 124,999 and 21,523 spectra 
have been collected from B, M and F plates, respectively. To avoid overcrowding, only one-in-ten targets are shown.
\label{spatial_gac_bluered}}
\end{figure*}

\begin{figure*}\includegraphics[width=180mm]{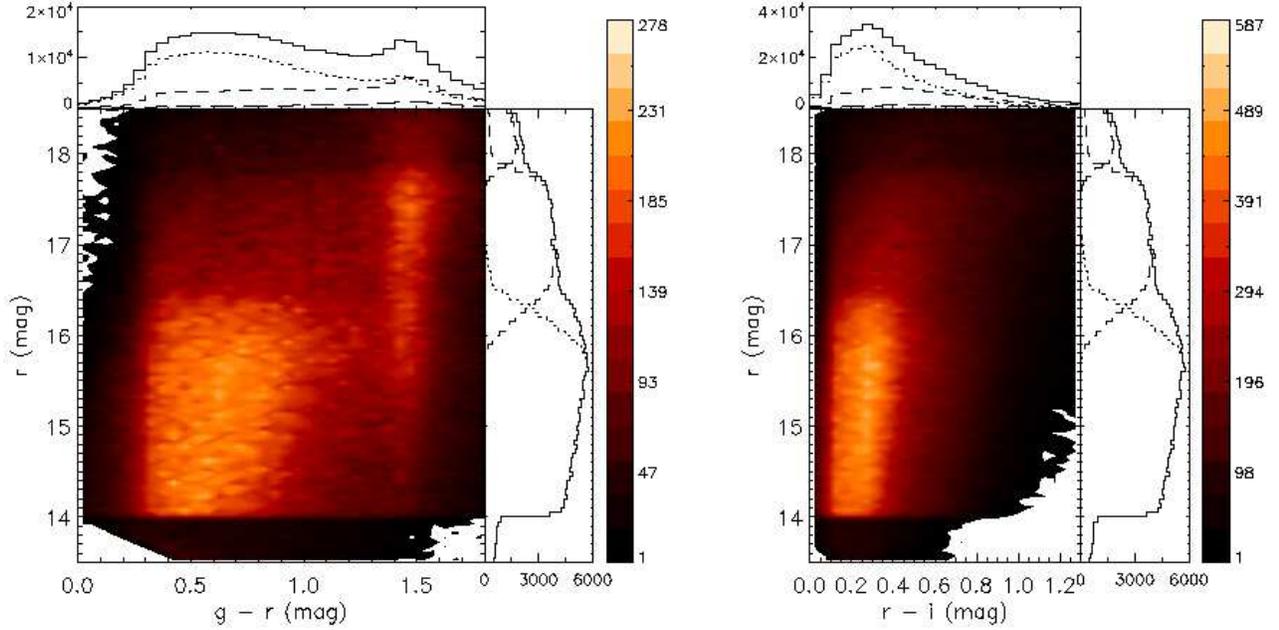}
\caption{
Contour maps of stellar number density in the $(g - r)$ versus $r$ (left) and $(r - i)$ versus $r$ (right) 
planes of spectra of S/N(4650\AA) $\ge$ 10  or S/N(7450\AA) $\ge$ 10 collected in the LSS-GAC main survey  
during the Pilot and the first year Regular Surveys. The magnitudes and colours are from the XSTPS-GAC survey without corrections for the reddening. 
The bin sizes are (0.05, 0.05)\,mag in both panels. Duplicate targets are not accounted. Colour bars are over-plotted by the side.
Histograms of star counts  in $(g - r)$ and $(r - i)$ colours as well as in $r$ magnitude 
are also over-plotted in solid black lines, with dotted, short-dashed and long-dashed lines indicating those from B, M and F plates, respectively. 
Note that the two sequences in $g -r$ colour around 0.6 and 1.4\,mag correspond to solar-type stars and 
M-dwarfs, respectively. That the M-dwarf sequence is not apparent in plots of $r - i$ colour is due to
the fact that M-dwarfs have a very narrow colour range in $g - r$ but spread out in $r - i$ colour.
\label{GAC_CMD_bluered}}
\end{figure*}

\begin{figure*}\includegraphics[width=180mm]{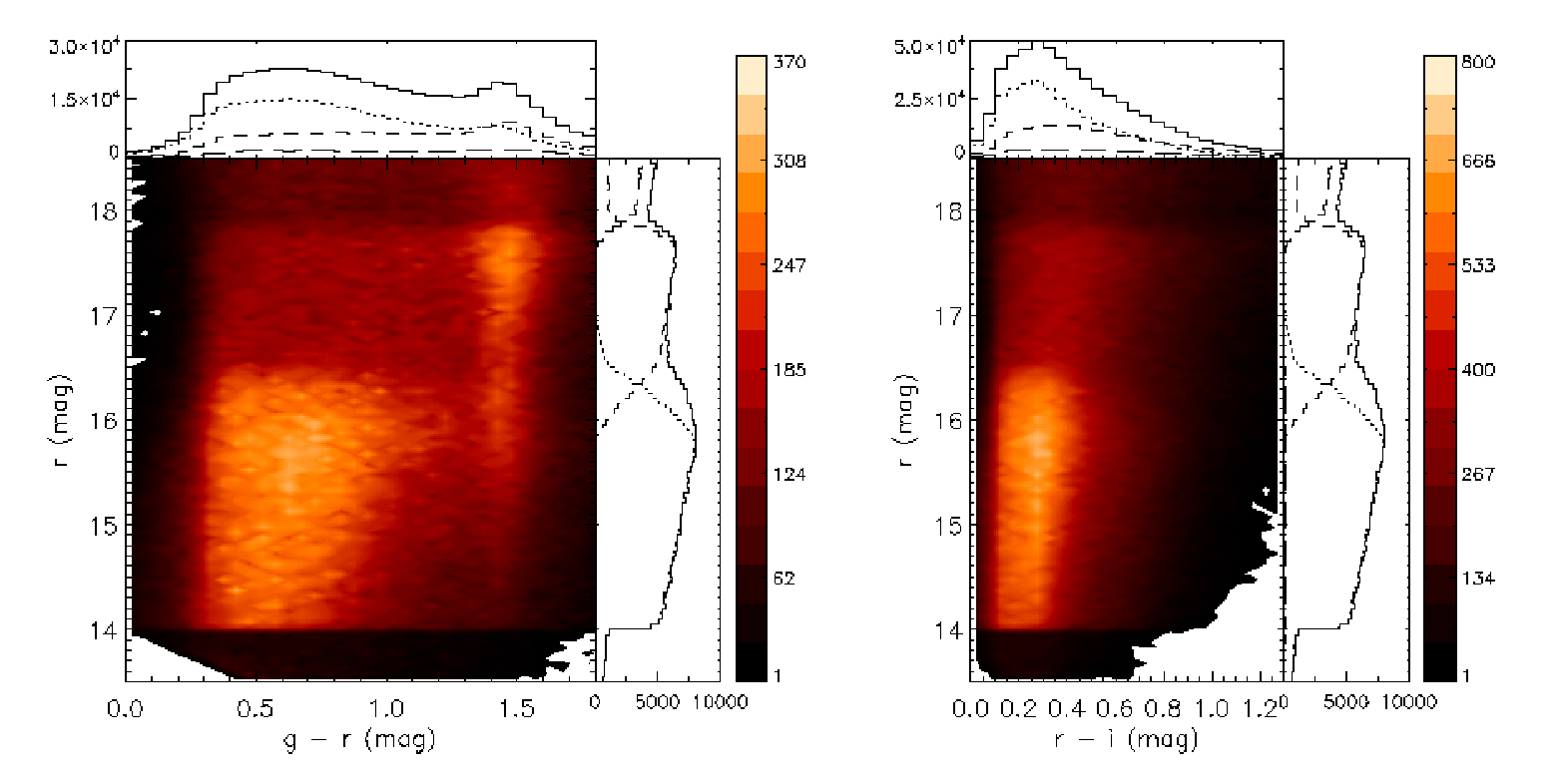}
\caption{
Same as Fig.\,\ref{GAC_CMD_bluered} but for all the LSS-GAC main survey sources targeted
during the Pilot and the first year Regular Surveys, including those having a S/N below 10. 
\label{GAC_CMD_observed}}
\end{figure*}

By June 2013, 142 B, 75 M and 20 F plates were targeted, 
yielding a total number of 727,478 spectra of 536,602 unique targets recorded in 237 plates. 
About 73.4, 20.2, 4.5, 1.3, 0.4, 0.1 per cent targets were observed 1 to 6 times, respectively. 
The duplicate observations came from targets reobserved in the overlapping areas of adjacent plates 
of the same category and group of plates or from targets of observed plates that failed to meet the quality control and then got re-observed.
Fig.\,\ref{snr_gac} shows histogram distributions of S/N(4650\AA) and 
S/N(7450\AA) for the observed targets, which 
roughly follow a power law distribution for S/N's higher than 10. 
A total of 448,224 spectra of 359,433 unique targets of S/N(4650\AA)  
or S/N(7450\AA) higher than 10 are obtained. This 
includes 225,522 spectra of 189,042 unique targets of S/N(4650\AA) higher than 10, 
and 439,560 spectra of 352,775 unique targets of S/N(7450\AA) higher than 10.
Their spatial distribution is shown in Fig.\,\ref{spatial_gac_bluered}. 
About two thirds of the spectra are from B plates, less than one third from M plates and about 5 per cent from F plates.
More quality spectra were obtained in the first year Regular Surveys than in the Pilot Surveys due to more observing time 
available and better observing strategy and quality control. 
The distributions in ($r$, $g-r$) and ($r$, $r-i$) Hess diagrams of those stars are shown in Fig.\,\ref{GAC_CMD_bluered}.
For comparison, ($r$, $g-r$) and ($r$, $r-i$) Hess diagrams of all targets observed in the main survey
during the Pilot and the first year Regular Surveys are shown in Fig.\,\ref{GAC_CMD_observed}.
Note that a few targets brighter than $r=14$\,mag were observed during the early stage of the Pilot Surveys, 
October 2011 and November 2011, when the bright and faint limiting magnitudes were set as $14 \le g < 19$\,mag or 
$14 \le r < 19$\,mag or $14 \le i < 19$\,mag instead of $14 \le r < 18.5$\,mag.  

\begin{figure}\includegraphics[width=90mm]{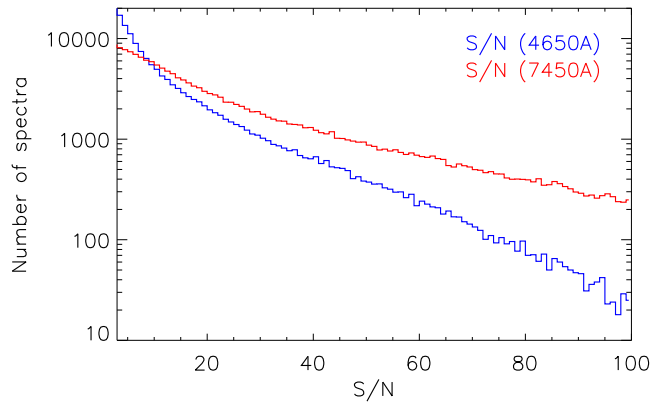}
\caption{Histogram showing the distributions of S/N(4650\AA) (blue) and S/N(7450\AA) (red) of targets observed in
the LSS-GAC M31-M33 fields during the Pilot and the first year Regular Surveys.\label{snr_m31}}
\end{figure}

\begin{figure}\includegraphics[width=90mm]{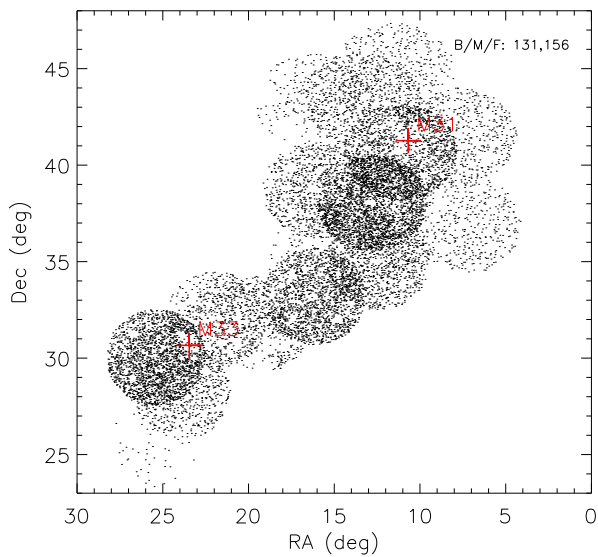}
\caption{
Spatial distribution of targets of spectral S/N(4650\AA) $\ge$ 10 or S/N(7450\AA) $\ge$ 10 
in the LSS-GAC M31-M33 fields during the Pilot and the first year Regular Surveys.
The positions of M\,31 and M\,33 are marked by red pluses.
In total, 131,156 spectra 
have been collected. To avoid overcrowding, only one-in-ten stars are shown.
\label{spatial_m31_bluered}}
\end{figure}

\begin{figure*}\includegraphics[width=180mm]{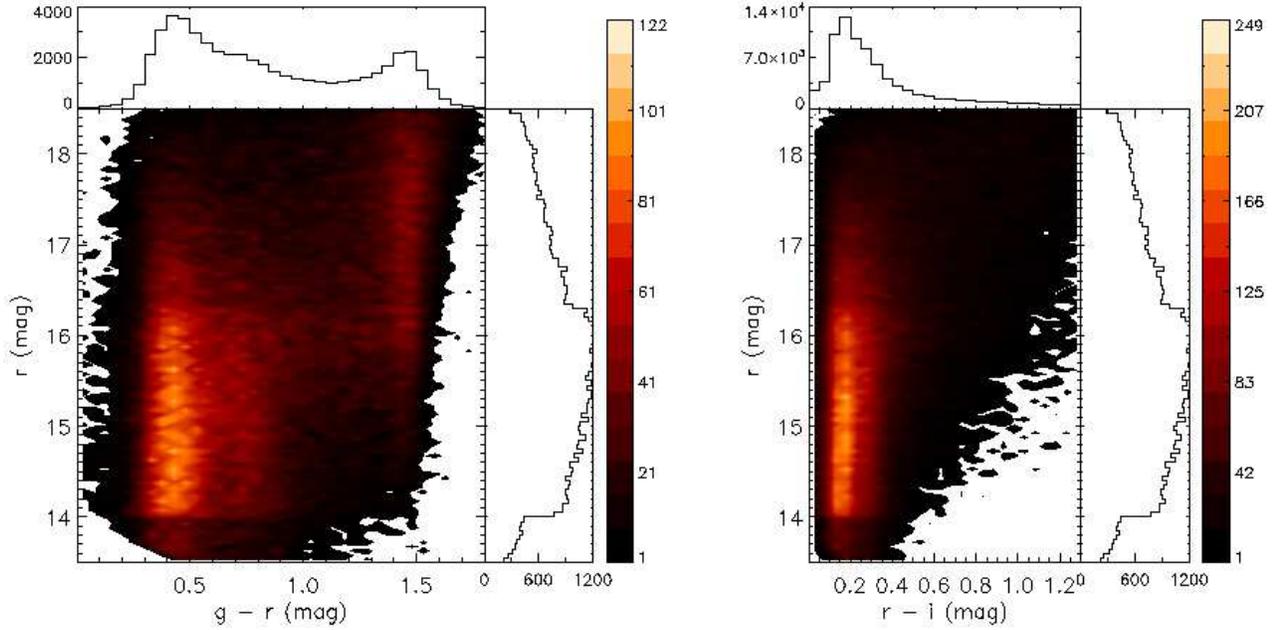}
\caption{
Same as Fig.\,\ref{GAC_CMD_bluered} but for spectra collected 
in the LSS-GAC M31-M33 fields during the Pilot and the first year Regular Surveys.
\label{M31_CMD_bluered}}
\end{figure*}

\begin{figure*}\includegraphics[width=180mm]{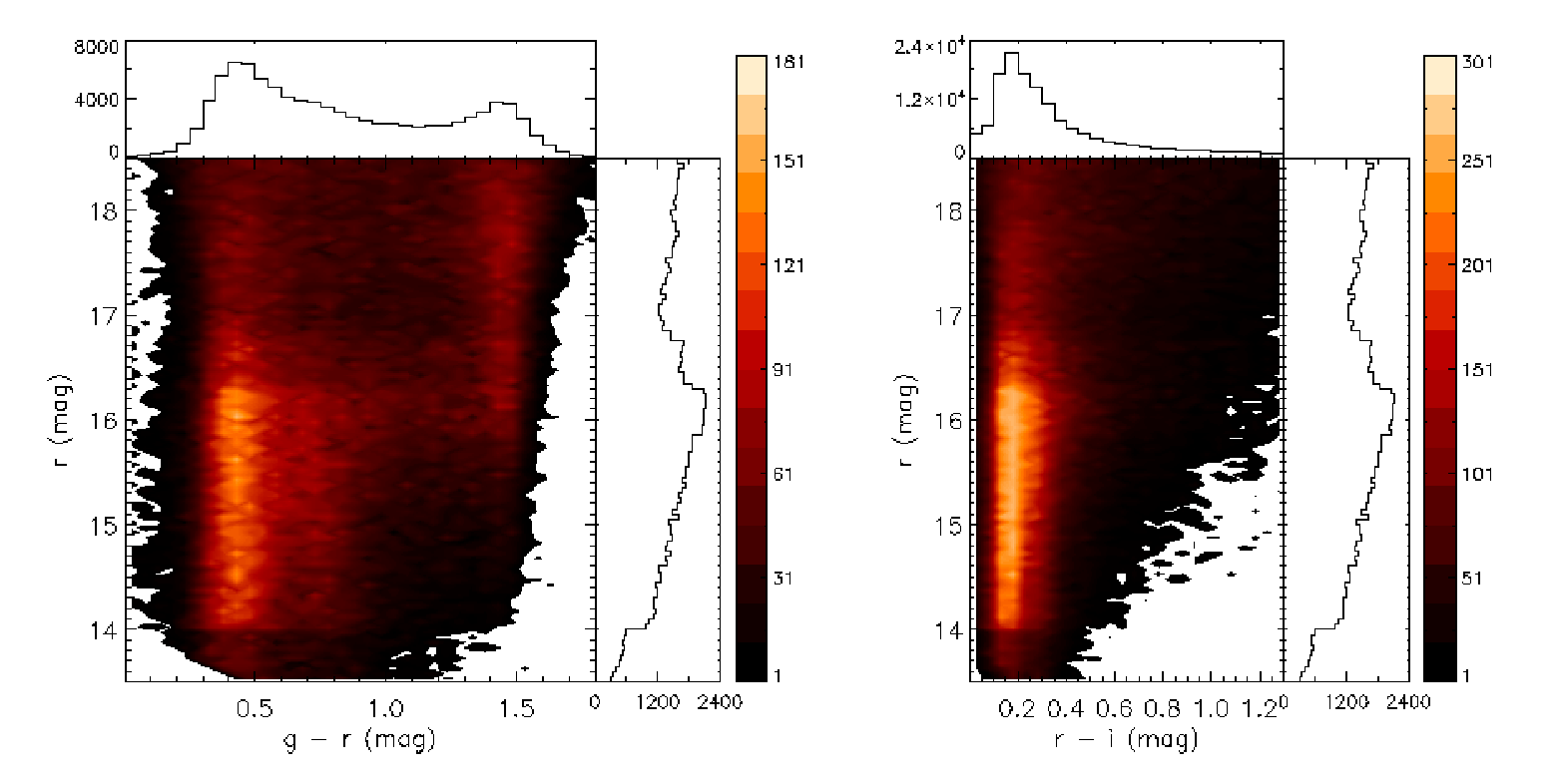}
\caption{
Same as Fig.\,\ref{M31_CMD_bluered} for 
all the sources targeted in the M31-M33 fields during the Pilot and the first year Regular Surveys, 
including those of a S/N lower than 10. 
\label{M31_CMD_observed}}
\end{figure*}

Plates of the M\,31-M\,33 survey were observed in a similar way as the main survey. 
By June 2013, 87 plates were targeted, yielding 265,745 spectra of 154,319  unique stars. 
About 64.0, 17.8, 9.4, 3.9, 2.4, 1.3, 0.7 per cent stars were observed by 1 to 7 times, respectively.
Fig.\,\ref{snr_m31} shows histogram distributions of S/N(4650\AA) and S/N(7450\AA) for those stars.
Again the distributions follow a power law for S/N's higher than 10.
A total of 131,156 spectra of 84,084 unique targets of S/N(4650\AA)  
or S/N(7450\AA) higher than 10 are obtained,
including 67,439 spectra of 46,501 unique targets having S/N(4650\AA)  higher than 10,
and 123,029 spectra of 79,733 unique targets having S/N(7450\AA)  higher than 10.
Their spatial distribution is shown in Fig.\,\ref{spatial_m31_bluered}.
About 80 per cent of the spectra are from B plates. 
The distributions in ($r$, $g-r$) and ($r$, $r-i$) Hess diagrams are shown in Fig.\,\ref{M31_CMD_bluered}.
For comparison, ($r$, $g-r$) and ($r$, $r-i$) Hess diagrams of all stars 
observed in the M\,31-M\,33 fields during the Pilot and the first year Regular Surveys are shown in Fig.\,\ref{M31_CMD_observed}.
Again a few stars brighter than $r=14$\,mag were observed during the early stage of the Pilot Surveys.

\begin{figure}\includegraphics[width=90mm]{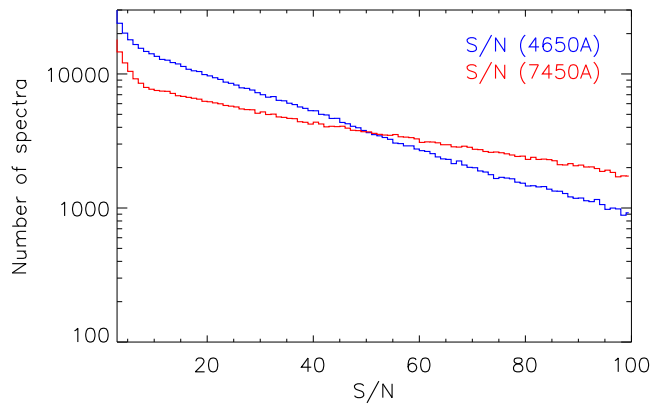}
\caption{Histogram showing the distributions of S/N(4650\AA) (blue) and S/N(7450\AA) (red) of 
LSS-GAC VB spectra collected during the Pilot and the first year Regular Surveys.\label{snr_vb}}
\end{figure}

\begin{figure*}\includegraphics[width=120mm]{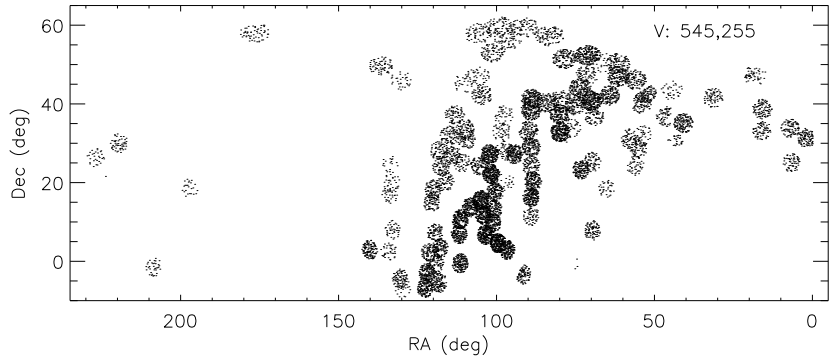}
\caption{
Spatial distribution of LSS-GAC VB targets of spectral S/N(4650\AA) $\ge$ 10  or S/N(7450\AA) $\ge$ 10  
observed during the Pilot and the first year Regular Surveys.
In total, 545,255 spectra of either S/N(7450\AA) $\ge$ 10  or S/N(4650\AA) $\ge$ 10  have been collected.
To avoid overcrowding, only one-in-fifty stars are shown.
\label{spatial_vb_bluered}}
\end{figure*}

\begin{figure}\includegraphics[width=90mm]{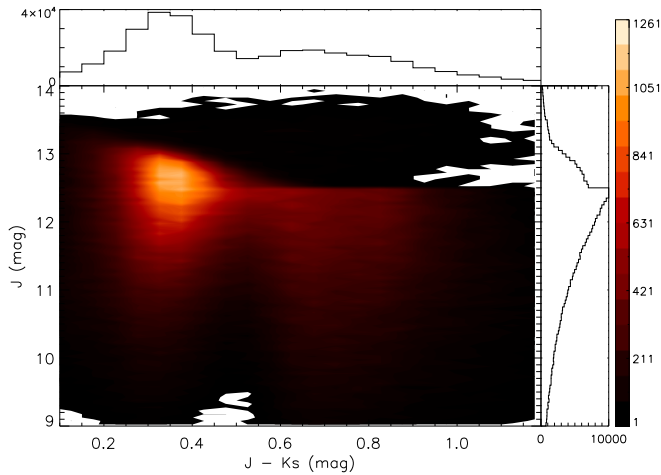}
\caption{
Contour map of stellar density in the $(J - K_{\rm s})$ versus $J$ 
plane for LSS-GAC VB targets observed during the Pilot and the first year Regular Surveys that have 
spectral S/N(4650\AA) $\ge$ 10  or S/N(7450\AA) $\ge$ 10. 
The magnitudes are from the 2MASS without corrections for the reddening.
The bin sizes are (0.05, 0.05)\,mag. Duplicate targets are not accounted. 
A colourbar is over-plotted by the side.
Histograms of star counts in $J - K_{\rm s}$ colour and in $J$ magnitude are also plotted.
\label{VB_CMD_bluered}}
\end{figure}

\begin{figure}\includegraphics[width=90mm]{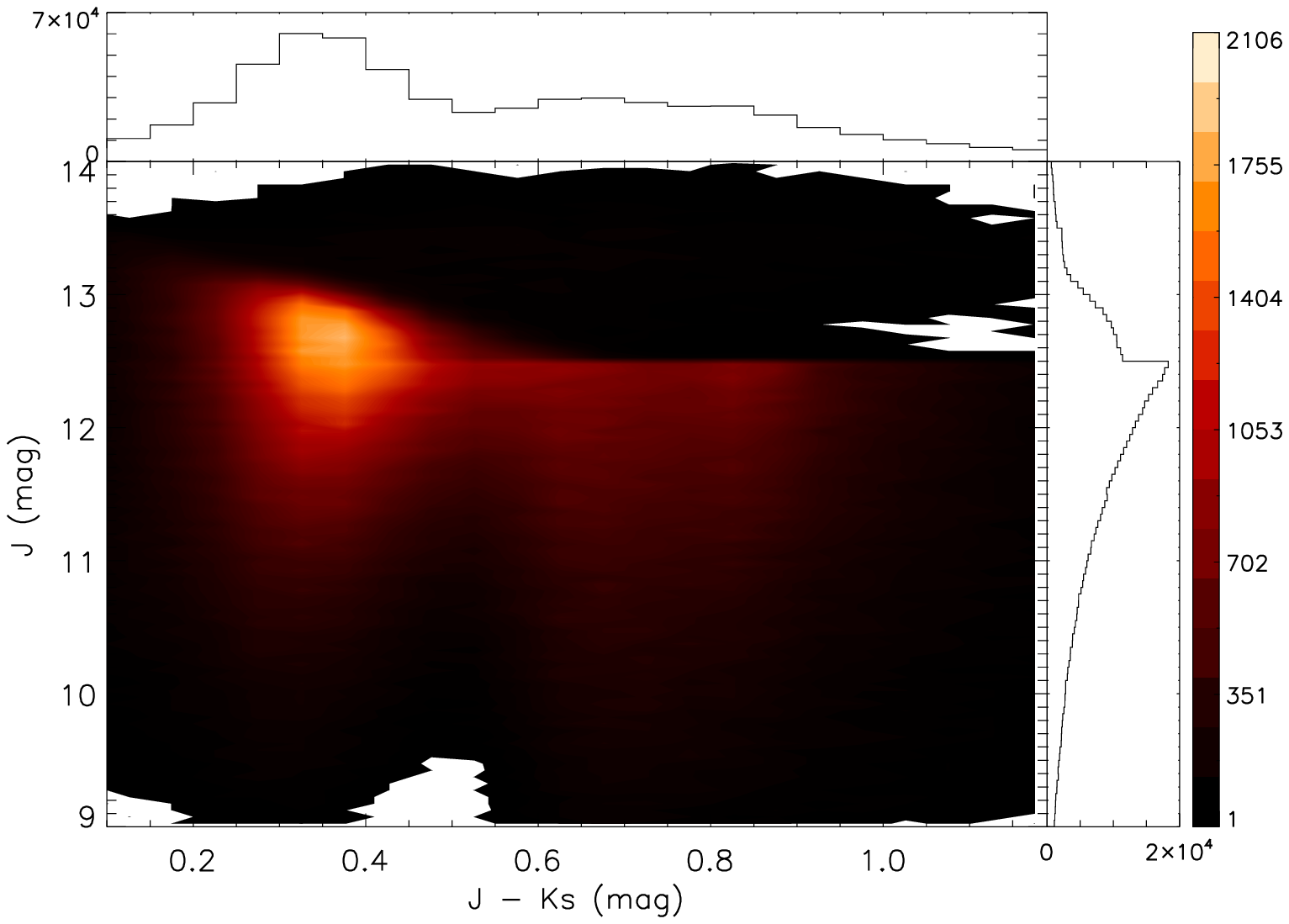}
\caption{
Same as Fig.\,\ref{VB_CMD_bluered} for all 
LSS-GAC VB targets observed during the Pilot and the first year Regular Surveys,
including those of S/N's below 10. 
\label{VB_CMD_observed}}
\end{figure}

VB plates were observed in bright nights or nights of poor observing conditions. 
The observations of VB plates began on January 5, 2012.
Typically each plate was observed with 2 $\times$ 600s.
Plates sharing the same central star are observed as many times as possible in one pointing in order to increase 
time-on-target efficiency of the telescope.
By June 2013, 259 plates were observed, yielding 791,530 spectra of 638,836 unique stars.
About 81.1, 15.1, 2.9, 0.7 and 0.1 per cent of the targets were observed by 1 to 5 times, respectively.
Fig.\,\ref{snr_vb} shows histogram distributions of S/N(4650\AA) and S/N(7450\AA) for the observed stars.
A total of 545,255 spectra of 452,758 unique targets having S/N(4650\AA)  
or S/N(7450\AA)  higher than 10 are obtained,
including 457,906 spectra of 385,672 unique targets having S/N(4650\AA)  higher than 10,
and 479,997 spectra of 398,381  unique targets having S/N(7450\AA)  higher than 10.
Their spatial distribution is shown in Fig.\,\ref{spatial_vb_bluered}. 
Most VB stars are within the footprint of LSS-GAC main survey.
The distribution of those stars in ($J$, $J-K_{\rm s}$) Hess diagram is shown in Fig.\,\ref{VB_CMD_bluered}.
For comparison, a ($J$, $J-K_{\rm s}$) Hess diagram of all stars
observed in the VB plates during the Pilot and the first year Regular Surveys is shown in Fig.\,\ref{VB_CMD_observed}.
Red bright giants stars are clearly visible.   

\begin{figure*}\includegraphics[width=140mm]{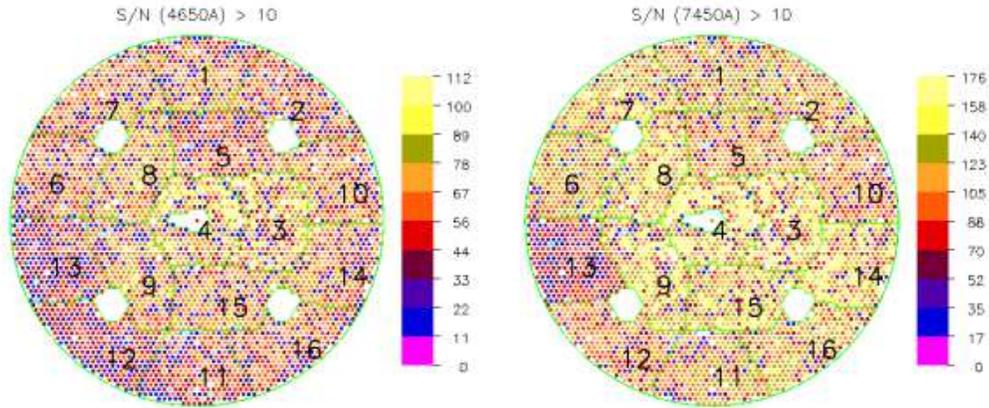}
\caption{Distributions of numbers of spectra of S/N(4650\AA) $\ge$ 10 (left) or S/N(7450\AA) $\ge$ 10 (right) 
in the focal plane collected by individual fibres for targets of the LSS-GAC main survey observed 
during the Pilot and the first year Regular Surveys.  
Individual spectrographs are separated by green lines. The spectrograph IDs are labelled in black.  
\label{snr_focal_gac}}
\end{figure*}

\begin{figure*}\includegraphics[width=140mm]{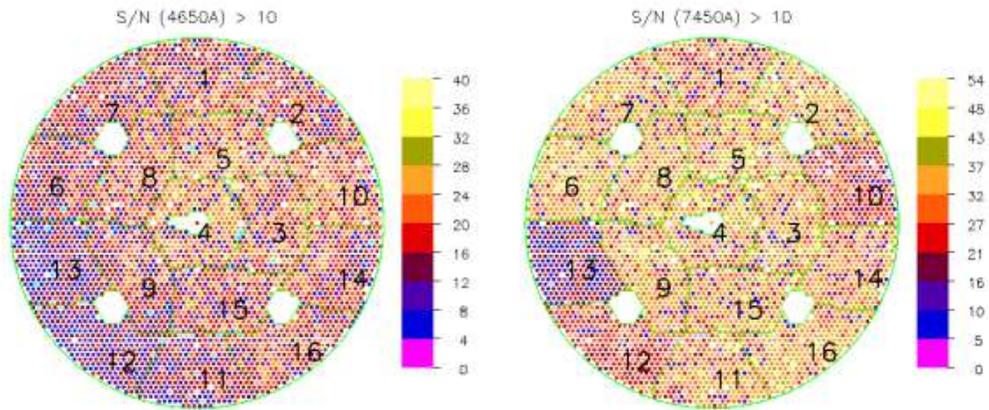}
\caption{Same as Fig.\,\ref{snr_focal_gac} for LSS-GAC targets in the M31-M33 fields.
\label{snr_focal_m31}}
\end{figure*}

\begin{figure*}\includegraphics[width=140mm]{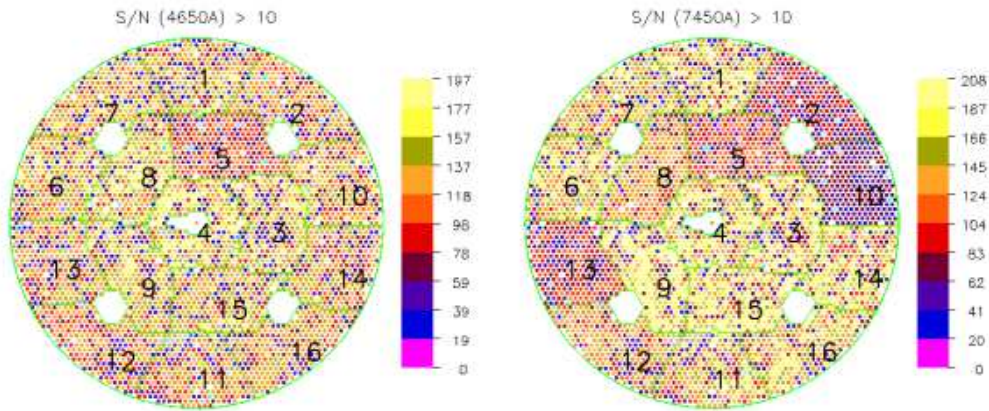}
\caption{Same as Fig.\,\ref{snr_focal_gac} for LSS-GAC VB targets.
\label{snr_focal_vb}}
\end{figure*}

The instrumental throughputs differ spectrograph by spectrograph and fibre by fibre. 
Fig.\,\ref{snr_focal_gac} shows distributions of spectra that have 
S/N(4650\AA) $\ge$ 10 or S/N(7450\AA) $\ge$ 10 in the focal plane for 
the LSS-GAC main survey targets observed during the Pilot and the first year Regular Surveys.
Large fibre-by-fibre variations are seen clearly.
Spectrographs close to the centre have much higher success rates than those near the edge, 
for both the blue and red arms.
Clearly, when defining the selection effects of the samples, one needs to keep such variations in mind.
Within a given spectrograph, a small fraction of fibres have very low success rates, 
probably due to their relatively larger fibre positioning errors. 
Similar but not exactly the same behaviors are found for the M\,31-M\,33 fields (Fig.\,\ref{snr_focal_m31}) 
and for VB plates (Fig.\,\ref{snr_focal_vb}). 

\subsection{Data reduction}

The raw data were reduced with the LAMOST 2D pipeline (Version 2.6) (Luo, Zhang \& Zhao 2004; Bai et al. 2014), 
including steps of bias subtraction, cosmic-ray removal, 1D spectral extraction, flat-fielding, wavelength calibration, and sky subtraction. 
As described earlier, the LAMOST spectra are recorded in two arms, 
3700 -- 5900\,\AA~in the blue and 5700 -- 9000\,\AA~in the red. 
The blue- and red-arm spectra are processed separately in the 2D pipeline and 
joined together after flux calibration. 
No scaling or shifting is performed in cases 
where the blue- and red-arm spectra are not at the same flux level in the overlapping region, 
as it is unclear whether the misalignment is caused by poor flat-fielding or sky subtraction, or both.

Flux calibration for the LSS-GAC observations is not straight-forward, 
as flux standard stars can not be selected based on (optical) photometric colours alone 
due to the unknown extinction towards individual stars in the Galactic disk.
However, the target selection algorithm of LSS-GAC ensures that there are plenty of F-type stars targeted by each spectrograph. 
To flux-calibrated the LSS-GAC spectra, Xiang et al. (2014a) have developed an iterative algorithm.
For a given spectrograph, the spectra are first flux-calibrated using the 
nominal spectral response curve (SRC) in order to derive the initial stellar atmospheric parameters with the LSP3 (Xiang et al. 2014b). 
Then based on the initial estimates of stellar parameters, F-type stars targeted in that spectrograph are selected as flux standards to deduce an 
updated SRC after taking into account the interstellar reddening. 
The reddening is estimated by comparing the observed and synthetic photometric colours. 
The new SRC is then used to re-calibrate the spectra and revise the stellar parameters.
The above process is repeated until a convergence is achieved.
Note that the flux calibration is performed for each spectrograph independently.
Comparisons of stellar colours deduced by convolving the flux-calibrated LSS-GAC spectra 
with the transmission curves of photometric bands and actual measurements from photometric surveys,  
as well as comparisons of multi-epoch observations of duplicate targets suggest that 
an accuracy of about 10 per cent has been achieved for the  
whole wavelength ranges of LSS-GAC spectra. Note that 
in the current implementation of flux-calibration of Xiang et al. (2014a), the 
telluric absorptions, including the prominent Fraunhofer A band at 7590\,{\AA} and B band at 6867\,{\AA} have 
not been removed.

When combining spectra from the individual exposures, 
the standard LAMOST 2D pipeline, following the SDSS approach,  first sorts all flux densities by wavelength and 
then performs a high-order spline fitting to obtain the final, combined flux densities.
While the approach is intended to preserve the spectral resolution as much as possible, it is prone to large, 
unpredictable errors. 
This problem has now been fixed (Bai et al. 2014). Considering that LAMOST spectra are grossly over-sampled, 
Xiang et al. (2014a) has adopted linear rebinning when combining spectra from the individual 
exposures with strict conservation of flux, and the results are therefore immune from the problem.  

\section{Radial velocities and stellar atmospheric parameters}

Radial velocities and basic stellar atmospheric parameters (\teff, \logg~and \feh) released in the LAMOST DR1 are determined by 
template matching with the ELODIE spectral library (Prugniel \& Soubiran 2001; Prugniel et al. 2007) using the LASP (Wu et al. 2014).
Considering that the MILES library (S{\'a}nchez-Bl{\'a}zquez et al. 2006; Falc{\'o}n-Barroso et al. 2011) 
is particularly suitable for the determinations of stellar atmospheric parameters from LAMOST spectra,
a separate pipeline, the LSP3, has been developed at Peking University. 
For a given target spectrum, the LSP3 determines 
the radial velocity by cross-correlating the continuum normalized spectrum with the normalized ELODIE template spectra
and values of \teff, \logg~and \feh~ by template matching with the MILES library, using both a weighted mean algorithm and a simplex downhill algorithm.
Results from the simplex downhill method are only used to cross-check and assign warning flags to parameters obtained from the weighted mean method. 
The LSP3 is thoroughly tested and applied to the LSS-GAC spectra. 
Radial velocities and atmospheric parameters yielded by the LSP3 serve as the core data of 
the add-on catalogues presented in the current work. 
A full description of the LSP3 is presented in a companion paper by Xiang et al. (2014b).

The LSP3 attempts to determine radial velocities and atmospheric parameters for all 
spectra of a S/N(4650\AA) $\ge$ 2.76. 
Stars of all spectral types, from the cool M to the hot O, are dealt with. 
However, parameters derived for the hot OBA-type and very cool M-type stars should be 
treated with caution, mainly due to the limitation of parameter space coverage of the available template spectra.
A variety of techniques are used to estimate the errors of parameters delivered by the LSP3, including both the 
systematic and random errors, as a function of the S/N, \teff, \logg~and \feh. 
For FGK stars and a S/N(4650\AA) = 10, it is estimated that the LSP3 has achieved an accuracy of 5\,km\,s$^{-1}$, 150\,K, 0.25\,dex and
0.15\,dex for $V_{\rm r}$, $T_{\rm eff}$, $\log\,g$ and [Fe/H]  determinations, respectively.
A systematic offset of about $−$3.5 \kms~is found for the LSP3 velocity measurements by comparing with external databases. 
A constant of +3.5 \kms~has been added to all radial velocities yielded by the LSP3.
Systematic offsets in \teff~ranging from 100 -- 300\,K are also found for stars hotter than 7000\,K or cooler than 3700\,K
by comparing the LSP3 estimates with those predicted by photometric colours.
Those offsets have been corrected for using a third-order polynomial. 
The LSP3 parameters for $V_{\rm r}$ and \teff~refer to those corrected values hereafter unless otherwise specified.
Stars whose parameters may have suffered from significant boundary effects are marked by flags assigned by the LSP3.

\begin{figure*}
\includegraphics[width=180mm]{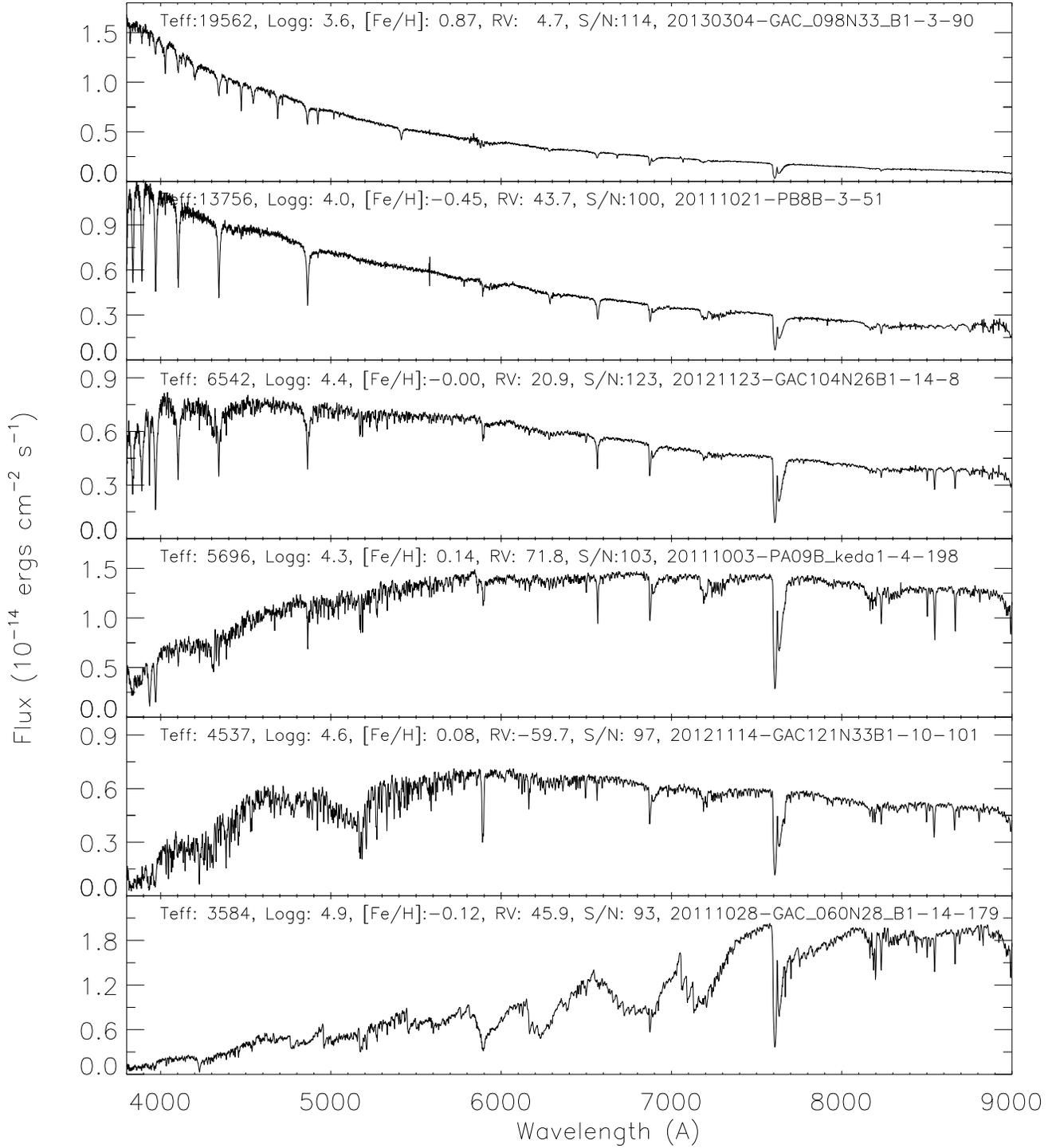}
\caption{Example LAMOST spectra of different type main sequence stars. 
For each star the LSP3 stellar parameters, S/N(4650\AA) and the unique LAMOST spectral ID, in format of 
date-plateid-specgraphid-fibreid, are labelled.} 
\label{spectra1}
\end{figure*}

\begin{figure*}
\includegraphics[width=180mm]{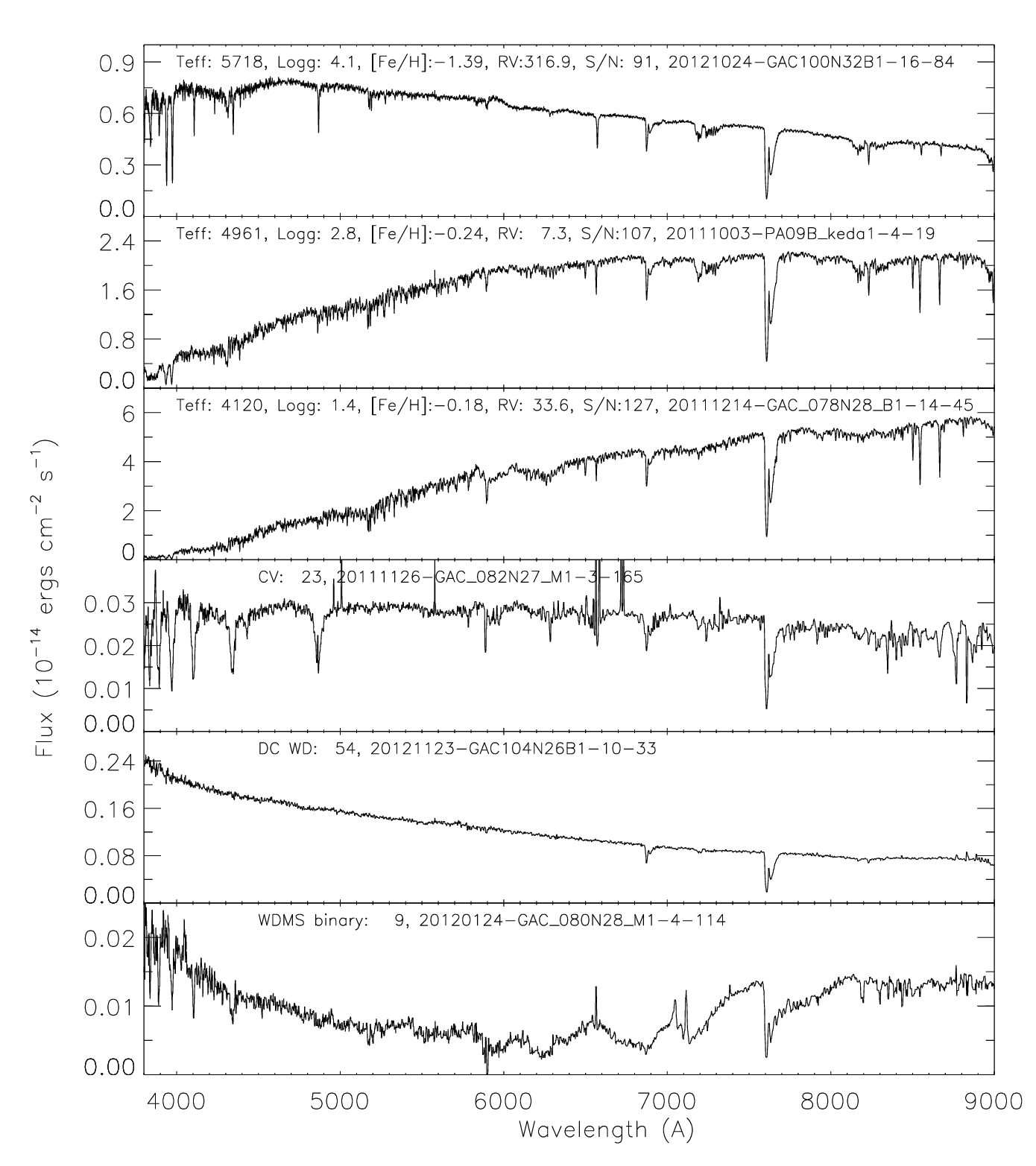}
\caption{
Example LAMOST spectra of stars of special characteristics. From top to bottom 
the figure plots, respectively, spectra of a metal-poor G dwarf, a RC star, a red giant, 
a cataclysmic variable, a DC type WD and a white-dwarf-main-sequence binary. 
For the first three objects, the LSP3 stellar parameters, S/N(4650\AA) and the LAMOST spectral IDs are labelled. 
For the latter three, only object types, S/N(4650\AA) and the LAMOST spectral IDs are labelled.
} 
\label{spectra2}
\end{figure*}

Fig.\,\ref{spectra1} displays six example LAMOST spectra of main sequence stars that range from early the B to
the late M. Examples of a metal-poor star, a RC star and a red giant are given in the top three 
panels of Fig.\,\ref{spectra2}, respectively. For each star, stellar parameters 
delivered by the LSP3, S/N(4650\AA) and the unique LAMOST spectral ID are labelled. 
Stellar parameters of those ``normal" dwarf and giant stars are well determined by the LSP3.  
Since the LSS-GAC adopts a simple yet non-trivial target selection algorithm that targets stars of all colours, 
it also includes many targets of special interests such as white dwarfs (WDs), emission line stars and even quasars. 
Inclusion of those targets significantly increases the discovery space of the LSS-GAC survey.
As described above, 18 spectra of DA type WDs, carbon stars and late-M/L stars 
retrieved from the SDSS database are added to the MILES spectral template library for the purpose of 
classifications of such objects in the LSP3. 
Templates of targets of special spectral characteristics  are far from complete in the current version of LSP3. 
As a result, many sources targeted by the LSS-GAC remain to be properly treated.
The bottom three panels of Fig.\,\ref{spectra2} illustrate example LAMOST spectra of 
a cataclysmic variable star, a DC type WD and a white-dwarf-main-sequence binary, respectively.
At the moment, their stellar parameters can not be reliably determined yet.

\section{Extinction}
The interstellar reddening is a key parameter that its  
accurate determination is vital for reliable derivation of
basic stellar parameters, including effective temperature and distance.
The 2D SFD extinction map provides a robust estimate 
of the (total) line-of-sight Galactic extinction at high latitude regions of 
low and moderate reddening, although there is some evidence that 
it has over-estimated the true values of \ebv~ by about 14 per cent (Schlafly \& Finkbeiner 2011; Yuan, Liu \& Xiang 2013). 
However, the SFD map has the following limitations:
a) The map gives the total amount of extinction, integrated along the line-of-sight to infinite, 
and consequently the value is an upper limit of the real one for a local disc star; b) The map fails at low Galactic latitudes ($|b| \le 5\degr$);
and c) The map has a limited spatial resolution about 6 arcmin, whereas the extinction may vary at smaller scales.
Given that most of the LSS-GAC targets locate at low Galactic latitudes and in the disk,
the SFD map is of limited use.
Determining extinction for individual stars becomes an urgent task for the LSS-GAC.

The SDSS DR9 contains about 0.7 million low-resolution spectra of Galactic stars (Ahn et al. 2012).
The on-going LAMOST Galactic surveys (Deng et al. 2012; Liu et al. 2014)
release over 1 million stellar spectra in its first Data Release (Bai et al. 2014) 
and will take over 5 million spectra when the surveys complete.
With millions of stellar spectra available, observations of stars of essentially identical stellar atmospheric parameters 
in different environments can be easily paired and compared,
enabling a number of studies with the standard pair technique (e.g., Yuan et al. 2014).
Using this technique of pairing a star  with its twins that suffer from  almost nil extinction
but otherwise have almost identical atmospheric parameters, 
Yuan, Liu \& Xiang (2013) have determined the dust reddening in a number of photometric bands for 
thousands of Galactic stars by combining photometric measurements from the far ultra-violet (UV) 
to the mid-infrared (mid-IR) as provided by the GALEX, SDSS, 2MASS and WISE surveys. 
They further derive the empirical, model-free reddening coefficients for those colours.
Their approach has the advantage that the method is straight-forward, model-free and applicable to 
stars of almost all spectral types. 
The same technique has been applied to stars targeted by the LSS-GAC to derive their multi-band reddening values.

\begin{figure*}
\includegraphics[width=180mm]{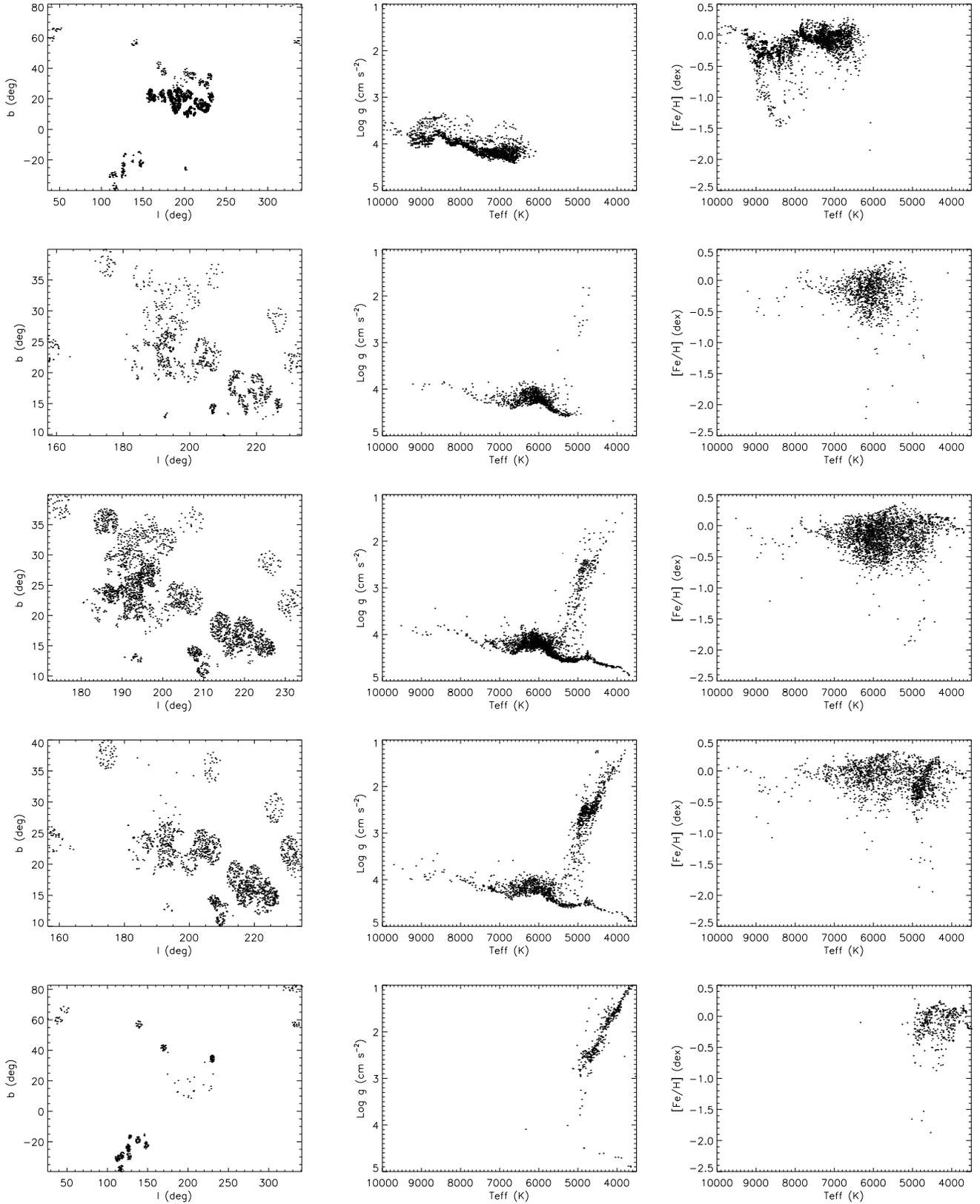}
\caption{Distributions of, from top to bottom, reference control samples R\,1 to 5 in the Galactic coordinates ($l, b$) and 
in the (\teff, \logg) and (\teff, \feh) planes.
To avoid overcrowding, only one in ten stars of R\,2 to 4 are plotted.
} 
\label{ref}
\end{figure*}

The star pair method requires a control sample consisting of stars of nil or extremely low reddening 
in order to estimate the intrinsic colours of target stars.
In order to estimate values of reddening in different colours ranging from the UV to the mid-IR, 5 control samples 
(i.e., R 1 -- 5) are constructed. R\,1, R\,2 and R\,5 are used to determine reddening values of colours 
$FUV - NUV$, $NUV-g$ and $W3-W4$, respectively.
R\,3 is used to determine reddening values of colours $g-r$, $r-i$, $i-J$, $J-H$ and $H-K_{\rm s}$,
and R\,4 for those of colours $K_{\rm s}-W1$, $W1-W2$ and $W2-W3$. 
The criteria of selecting stars of the control samples are listed in Table\,\ref{control}. 
The distributions of control sample stars in Galactic  coordinates, in the 
\teff~ -- \logg~and \teff~ -- \feh~planes are shown in Fig.\,\ref{ref}. 
Note that within the footprint of LSS-GAC, stars in the southern Galactic hemisphere overall suffer from higher extinction than those
in the north (Chen et al. 2014). As a consequence, most control sample stars are from the northern hemisphere. 
All stars in the control samples are reddening corrected using the SFD extinction map and the
reddening coefficients derived by Yuan, Liu \& Xiang (2013). 
The reddening coefficients for colours $W2-W3$ and $W3-W4$ are not available in Yuan, Liu \& Xiang (2013),   
which are adopted to be zero here when dereddening $W2-W3$ and $W3-W4$ colours.
Because the reddening values of the control samples are small, the errors caused by reddening corrections are ignorable.

\begin{table*}{}
\small
\centering
\begin{minipage}[]{180mm}
\caption[]{Control samples for the reddening determinations.}
\label{control}\end{minipage}
\tabcolsep 2mm
\begin{tabular}{lccl}
  \hline\noalign{\smallskip}
Sample & Colours & Num. of stars &  Criteria   \\
  \hline\noalign{\smallskip}
R\,1& $FUV-NUV$ & 3,106 & \ebv~$\le$ 0.1\,mag, S/N$^a$ $\ge$ 20, flags[1]$^b$ $\le$ 1, $FUV$ $\le$ 20\,mag, err($NUV$) $\le$ 0.1\,mag,        \\
    &           &     & $\Delta d^c$(Galex - Xuyi$^d$) $\le$ 2\arcsec, $\Delta d^c$(2MASS - Xuyi) $\le$ 1\arcsec, Q$^e$(2MASS) = `AAA'  \\
    &           &     &          \\
R\,2& $NUV-g$   & 10,330 & \ebv~$\le$ 0.05\,mag, S/N$^a$ $\ge$ 20, flags[1]$^b$ $\le$ 1, err($NUV$) $\le$ 0.1\,mag,     \\
    &           &     &  err($g$) $\le$ 0.02\,mag, $ 12 \le g \le 17$\,mag, $\Delta d^c$(Galex - Xuyi) $\le$ 2\arcsec  \\
    &           &     &          \\
R\,3& $g-r$, $r-i$, $i-J$   & 27,146 & \ebv~$\le$ 0.05\,mag, S/N$^a$ $\ge$ 20, flags[1]$^b$ $\le$ 1, err($g, r, i$) $\le$ 0.02\,mag, \\
    & $J-H$, $H-K_{\rm s}$       &     & $ 12 \le (g, r, i) \le 17$\,mag, Q$^e$(2MASS) = `AAA', $\Delta d^c$(2MASS - Xuyi) $\le$ 1\arcsec    \\
    &           &     &          \\
R\,4& $K_{\rm s}-W1$, $W1-W2$   & 16,985 & \ebv~$\le$ 0.05\,mag, S/N$^a$ $\ge$ 20, flags[1]$^b$ $\le$ 1, Q$^e$(2MASS) = `AAA',  \\
    & $W2-W3$   &     &        ($W1, W2, W3$)  $\ge$ 8\,mag, err($W1, W2$) $\le$ 0.03\,mag, err($W3$) $\le$ 0.2\,mag,   \\
    &           &     &        ext\_flag$^f$(WISE) = `0', cc\_flags$^f$($W1,W2,W3$) = `000', var\_flags$^f$($W1, W2, W3$) $\le$ 6 \\
    &           &     &        ph\_qual$^f$($W1,W2,W3$) = `A/B/C', $\Delta d^c$(WISE - Xuyi) $\le$ 1\arcsec  \\
    &           &     &          \\
R\,5& $W3-W4$   & 452 & \ebv~$\le$ 0.1\,mag, S/N$^a$ $\ge$ 20, flags[1]$^b$ $\le$ 1, 4 $\le$  $W3$ $\le$ 10\,mag,    \\
    &           &     & err($W3$) $\le$ 0.2\,mag, err($W4$) $\le$ 0.4\,mag, $W4$ $\le$7.7\,mag, $W3-W4$ $\le$ 0.3\,mag, \\ 
    &           &     &  ext\_flag$^f$(WISE) = `0', cc\_flags$^f$ = `0000', var\_flags$^f$($W3, W4$) $\le$ 6,      \\
    &           &     &  ph\_qual$^f$($W3,W4$) = `A/B/C', $\Delta d^c$(WISE - Xuyi) $\le$ 1\arcsec       \\
\noalign{\smallskip}\hline
\end{tabular}
\begin{description}
\item[$^a$]  S/N(4650\AA) per pixel.
\item[$^b$]  Second flag assigned to the final LSP3 parameters. See Table 2 of Xiang et al. (2014b). 
\item[$^c$]  Angular distance between the source positions as measured by the two surveys in parentheses. 
\item[$^d$]  Stands for the XSTPS-GAC survey. 
\item[$^e$]  Quality flags of the 2MASS photometry in $J$, $H$ and $K_{\rm s}$ bands. 
\item[$^f$]  Flags from the WISE photometric catalogues. 
\end{description}

\flushleft
\end{table*}

For a given star targeted by the LSS-GAC, its control stars are selected from the control samples as
those having values of \teff, \logg~and \feh~that differ from 
the target values by less than 5 per cent, 0.5 dex and 0.3 dex, respectively. 
For a given colour, if the number of selected control stars are no less than four, 
the colour excess is then calculated as the difference between
the observed colour of the target and the intrinsic colour of a pseudo star of atmospheric parameters identical to those of the target. 
The intrinsic colour of the pseudo star is derived from the observed colours of the selected control sample stars assuming that 
the colour varies linearly with \teff, \logg~and \feh, which is likely to be valid given the narrow ranges of values of 
\teff, \logg~and \feh~ being considered. If the number of selected control stars is less than four, the colour excess of the target is not calculated. 

\begin{figure*}
\includegraphics[width=180mm]{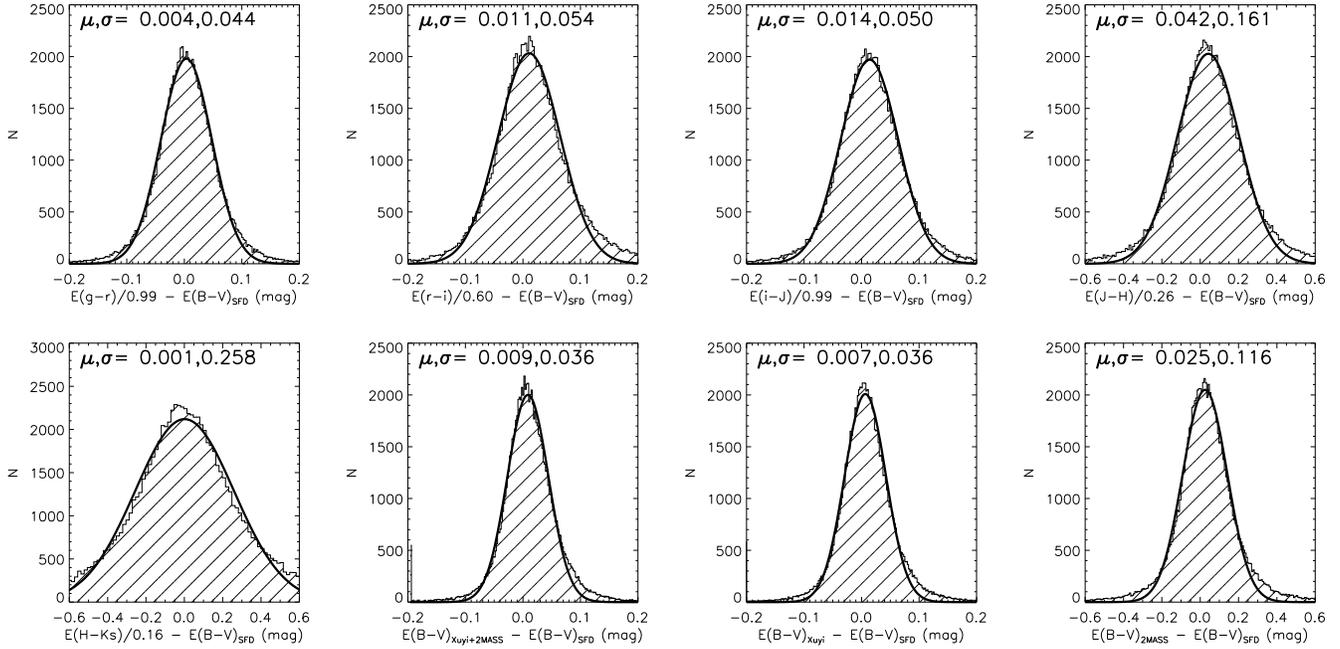} 
\caption{Comparison of \ebv~values derived from the star pair method,
as given by $E(a - b)$ divided by the reddening coefficient taken from Yuan, Liu \& Xiang (2013) for colour $a- b$,
with those from the SFD map for a selected LSS-GAC sample of stars  
of high S/N's [S/N(4650\AA) $\ge$ 10], high Galactic latitudes ($|b| \ge$ 15\degr) and low extinction [$E(B-V)_{\rm SFD} \le$ 0.3\,mag].} 
\label{check_with_SFD}
\end{figure*}

To estimate the errors of reddening values derived from the star pair method, 
we select a sample of stars from the LSS-GAC main survey with the following criteria:
a) S/N(4650\AA) $\ge$ 10;
2) $|b| \ge$ 15\degr;
3) The $E(B - V)$ values from the SFD map, $E(B-V)_{\rm SFD} \le 0.3$ mag; 
and 4) The stars are well detected in $g$, $r$ and $i$ bands by the XSTPS-GAC and 
in $J$, $H$ and $K_{\rm s}$ bands by the 2MASS.
In total, about 90,000 stars are selected.
We then convert the colour excess $E(a-b)$ of colour $a-b$ derived from the pair method for band $a - b$ into \ebv~ 
using the empirical reddening coefficients of Yuan, Liu \& Xiang (2013) and compare the result with $E(B-V)_{\rm SFD}$. 
The results are displayed in Fig.\,\ref{check_with_SFD}, with the mean difference and standard deviation labelled at the top of each panel.
The mean differences are small except for colour $J-K_{\rm s}$, suggesting that the empirical reddening 
coefficients derived by Yuan, Liu \& Xiang (2013) are applicable to the current sample. For $J-K_{\rm s}$, 
the coefficient deduced by Yuan, Liu \& Xiang (2013) may have been underestimated for the current sample of stars, 
due to, say the spatial variations of the extinction law.
The standard deviations for colours $g-r$, $r-i$, $i-J$, $J-H$ and  $H-K_{\rm s}$ 
are 0.044, 0.054, 0.050, 0.161 and 0.259 mag, respectively.
Note that the deviations include contributions from the photometric errors of the XSTPS-GAC and 2MASS measurements, 
the random errors of the LSP3 stellar parameters and possible spatial variations of the extinction law. 
The dispersions have only a weak dependence on S/N for S/N(4650\AA) $\ge$ 10, indicating that the random 
errors of LSP3 parameters are likely to be small even at S/N(4650\AA) = 10. The large values of dispersion for the   
$J-H$ and $H-K_{\rm s}$ colours are simply because the two colours are not as  sensitive to extinction 
as the $g-r$, $r-i$ and  $i-J$ colours of shorter wavelengths.  

To reduce errors of reddening derived from the star pair method, results from different colours are combined. 
If a target has been detected in $g$, $r$ and $i$ bands by the XSTPS-GAC and in $J$, $H$ and $K_{\rm s}$ bands by the 2MASS,
the adopted \ebv~is the weighted mean of values yielded by colours $g-r$, $r-i$ and $i-J$ weighted by 
the corresponding reddening coefficients. 
Values yielded by colours $J-H$ and $H-K_{\rm s}$ are excluded given their low sensitivity to extinction. 
If a star is only detected in the optical, then 
the weighted mean of values yielded by colours $g-r$ and $r-i$ is adopted.
Similarly, if a star is only detected in the near IR by the 2MASS, 
the weighted mean of values given by colours $J-H$ and  $H-K_{\rm s}$ is used. 
Here all reddening coefficients are again taken from Yuan, Liu \& Xiang (2013).
Comparisons between the mean values of extinction derived respectively from the optical plus the near-IR data, 
from the optical or from the near-IR data only, 
and those given by the SFD map are also displayed in Fig.\,\ref{check_with_SFD}. 
The dispersions are 0.035, 0.036 and 0.116 mag, respectively.
Clearly, optical photometry is crucial for accurate determinations of the reddening.

\begin{figure*}
\includegraphics[width=180mm]{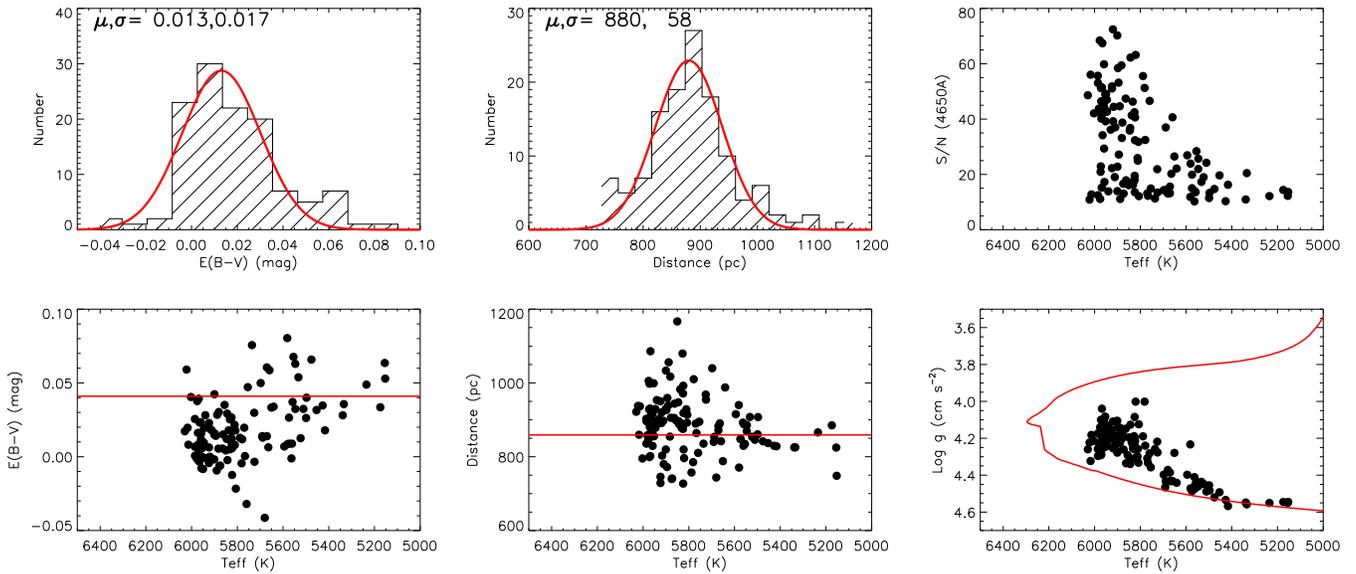}
\caption{Upper left: Distribution of \ebv~values derived from the star pair method for members of open cluster M\,67.
The red curve is a Gaussian fit, with the mean and dispersion labelled. 
Lower left: The values of \ebv~ are plotted as a function of \teff. The red line denotes value given by the SFD map. 
Upper middle: Distribution of distances derived from the empirical relations of absolute magnitudes and 
stellar atmospheric parameters [cf. Eq.(1)] for members of  M\,67. 
The red curve is a Gaussian fit, with the mean and dispersion marked.                     
Lower middle: The distances are plotted as a function of \teff. The red line represents the distance given by 
Pancino et al. (2010). 
Upper right: Values of S/N(4650\AA) of the spectra are plotted as a function of \teff.
Lower right: An H-R diagram of M\,67 member stars constructed from the LSP3 parameters. 
The red line represents a theoretical isochrone 
(age = 4.5 Gyr, \feh = 0\,dex, [$\alpha$/Fe] = 0\,dex, Y = 0.2696) from Dotter et al. (2008).} 
\label{m67}
\end{figure*}

The accuracy of extinction derived from the star pair method is further examined 
by applying the method to member stars of open cluster M\,67 targeted by the LAMOST.  
The list of member stars is from Xiang et al. (2014b).
The upper left panel of Fig.\,\ref{m67} shows that a mean value of \ebv~of 0.013 mag is obtained, is
consistent within the errors with the value of 0.041\,mag given by the SFD map and 
the value from the literature, 0.04 $\pm$ 0.02\,mag (Pancino et al. 2010).
The small dispersion of values yielded from individual stars, 0.017\,mag,  suggests that the star pair method 
is capable of determining reddening to a high accuracy.
A weak dependence of the extinction derived on \teff~ is observed (cf. the lower left panel of Fig.\,\ref{m67}). 
This is probably caused by the low S/N's of spectra of those cool late type stars (cf. the upper right panel of  Fig.\,\ref{m67}). 
    
Reddening can also be determined by comparing the observed and synthetic colours from stellar  model atmospheres.
To calculate the model predicted colours, a grid of synthetic colours is constructed by convolving the synthetic spectra
(Castelli \& Kurucz 2004) with the transmission curves of the SDSS and 2MASS photometric systems.
For a star of given atmospheric parameters,  \teff, \logg~and \feh, the synthetic colours are derived by 
linearly interpolating the grid.
Values of extinction derived from individual colours are then averaged as in the case of star pair method.  
Given that the stellar colour loci for main sequence stars are well described by a single parameter,
i.e., the effective temperature, reddening can also be determined from multi-band photometry alone 
by fitting the stellar SED from the optical to the near-IR (e.g., Berry et al. 2012). 
Using this technique and combining the photometry of XSTPS-GAC in the optical, 2MASS and WISE in the near to mid-IR, 
Chen et al. (2014) has derived values of extinction  
and distances for about 15 million stars surveyed by the XSTPS-GAC, including 
most stars targeted by the LSS-GAC. 

\begin{figure}
\includegraphics[width=90mm]{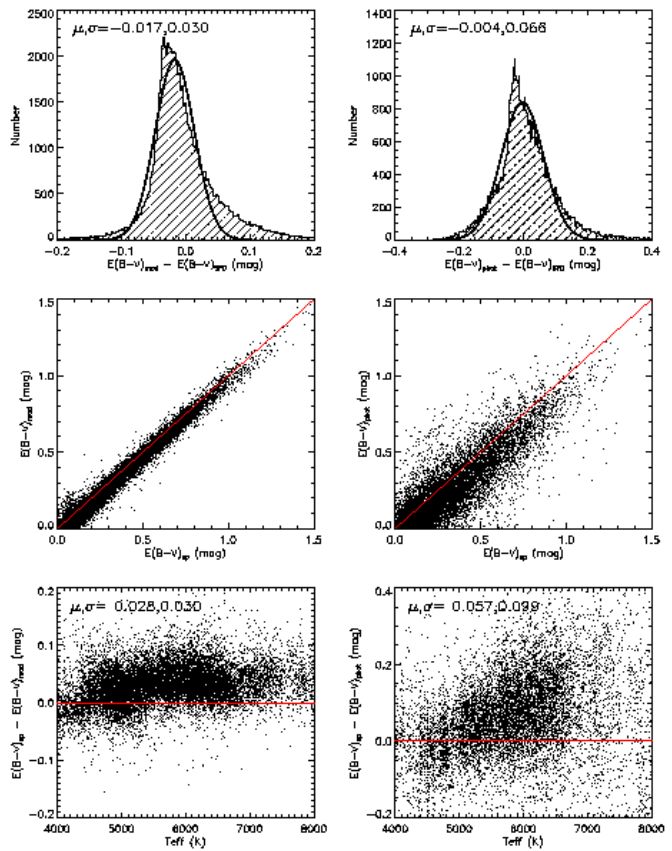} 
\caption{Top: Comparisons between values of $E(B - V)_{\rm mod}$ (left) and $E(B - V)_{\rm phot}$ (right) 
with those from the SFD map for a selected sample of stars of high Galactic latitudes ($|b| \ge$ 15\degr) 
and low extinction [$E(B-V)_{\rm SFD} \le 0.3$\,mag].
Middle: Comparisons between values of $E(B - V)_{\rm sp}$ and $E(B - V)_{\rm mod}$ (left) and 
$E(B - V)_{\rm phot}$ (right).
Bottom: Differences $E(B - V)_{\rm sp}$ $-$ $E(B - V)_{\rm mod}$ (left) and      
$E(B - V)_{\rm sp}$ $-$ $E(B - V)_{\rm phot}$ (right) plotted as a function of \teff. 
To avoid overcrowding, only 10 per cent of targets, randomly selected, are shown in the middle and bottom panels.
}
\label{compare_ebv}
\end{figure}

Errors of reddening derived from the observed and synthetic multi-band colours 
are estimated by comparing the results with those from the  SFD map for the same selected sample of 
LSS-GAC targets used above to estimate the errors of extinction derived from the star pair method. 
The results are shown in the top panels of Fig.\,\ref{compare_ebv}. 
The \ebv~values derived from the star pair method, $E(B - V)_{sp}$, are also compared 
with those derived by comparing the observed and synthetic colours, $E(B - V)_{\rm mod}$, 
and those derived by fitting multi-band photometry to the empirical stellar loci (Chen et al. 2014), $E(B - V)_{\rm phot}$, 
in the middle and bottom panels of Fig.\,\ref{compare_ebv}.
Values of $E(B - V)_{\rm sp}$ and $E(B - V)_{\rm mod}$ have comparable 
errors, and are about two times smaller than those of $E(B - V)_{\rm phot}$. 
A tight correlation between $E(B - V)_{\rm sp}$ and $E(B - V)_{\rm mod}$  
is expected given that both methods use the same set of stellar parameters yielded by the LSP3. 
The small systematic discrepancy between $E(B - V)_{\rm sp}$ and $E(B - V)_{\rm mod}$ suggests that 
there are some systematics between the model predicted colours and the observed ones.
Values of $E(B - V)_{\rm phot}$ are systematically smaller than those of $E(B - V)_{\rm sp}$ 
by about 0.06 mag. The differences are possibly caused by the 
spatial variations of the extinction law as well as by the uncertainties of the photometric method which tends to underestimate 
the true values of reddening (Chen et al. 2014).

\section{Distances}
Studying the six-dimensional phase-space distribution of a large sample of Galactic 
stars plays an essential role in understanding the true nature of the Galaxy and galaxies in general. 
It requires reliable estimates of distance to individual sample stars. 
Distance determinations are coupled with reddening corrections. 
Corresponding to the three methods of reddening determinations described above, we have implemented three 
sets of distance estimates for the LSS-GAC sample stars.

For stars with spectroscopically measured atmospheric parameters, their distances are commonly estimated with the aid of stellar evolutionary models. 
The evolution and properties, including luminosity,  of a (single) star are fully determined by its initial mass, chemical composition and age. 
For a given set of initial mass and chemical composition, the stellar evolutionary models of structure and atmosphere 
yield \teff, \logg, \feh~and absolute magnitudes of specified photometric bands as a function of age.
The absolute magnitudes have a tight relation with \teff, \logg~and \feh. Therefore, it is possible to estimate 
absolute magnitudes, thus distances, from the values of \teff, \logg~and \feh~  
yielded by spectroscopy when combined with reddening corrected photometry of apparent magnitudes. 
Note that \feh~is often used as a proxy of the chemical composition, the effects of 
possible variations in individual elemental abundances, such as the 
$\alpha$-element enhancement have been neglected. 
For stars in the MILES library, the absolute magnitudes can be directly calculated using the Hipparcos parallaxes (Anderson \& Francis 2012). 
Together with the atmospheric parameters, the stars thus provide an excellent sample to establish empirical relations between 
the absolute magnitudes and \teff, \logg~and \feh, enabling  us to determine distances from \teff, \logg~and \feh~directly.  
Compared to the theoretical relations from the stellar evolutionary models, the approach has the advantage of being 
model-independent and straight-forward to use.
Considering that the atmospheric parameters of LSS-GAC targets are derived by template matching with the MILES library, 
it is quite natural to apply the empirical relations of absolute magnitudes as 
a function of stellar atmospheric parameters as derived from the MILES stars to the LSS-GAC targets. 
  
We use polynomials to construct the empirical relations between the absolute magnitudes 
in $g, r, J, H$ and $K_{\rm s}$ bands and \teff, \logg~and \feh~for the MILES stars.
The apparent magnitudes in $g$ and $r$ bands of the stars are transformed from the observed, dereddened $B$ and $V$ magnitudes using 
relations, $g = V + 0.60(B-V)-0.12$\,mag and $r = V + 0.42(B-V)+0.11$\,mag (Jester et al. 2005).
Errors of the derived absolute magnitudes are estimated by combining the photometric and distance (parallax) errors. 
When fitting the data, we only include those MILES stars whose estimated errors of absolute magnitudes are smaller than 0.4\,mag, 
and weight the remaining data points by the inverse square of the absolute magnitude errors.
A $3\sigma$ clipping is used to exclude outliers. 
The MILES library includes stars of a wide range of spectral types and luminosity classes, from the O to the M and from dwarfs to giants. 
Fitting all the stars with a single polynomial  will require a very high order polynomial.
The result is also likely to be unstable and liable to large errors near the periphery of parameter space coverage.
To solve the problem, we divide the stars into four groups of spectral type and luminosity class 1)
the hot OBA stars; 2) FGK dwarfs; 3) KM dwarfs and 4) GKM giants. Each group is fitted  independently.
The groups are defined in atmospheric parameter space with some overlapping.
The definitions are:

\begin{description}
\item[OBA stars:] $T_{\rm{eff}} > 7,000 $~K;
\item[FGK dwarfs:] $ 4,500 < T_{\rm{eff}} < 7,500$~K, \logg~$>$ 3.2 (cm\,s$^{-2}$);
\item[KM dwarfs:] $T_{\rm{eff}} < 5,000 $~K, \logg~$>$ 3.2 (cm\,s$^{-2}$); 
\item[GKM giants:] $T_{\rm{eff}} < 5,500 $~K, \logg~$\le$ 3.5 (cm\,s$^{-2}$).
\end{description}

A three-order polynomial  of the following form is then used to fit the absolute magnitude of a given band:
\begin{align}
&{ M}(T_{\rm eff}, \log g, {\rm [Fe/H]}) =   \nonumber\\ 
&a_0 +  a_1 \times T_{\rm eff} + 
a_2 \times \log g + a_3 \times  {\rm [Fe/H]} + \nonumber\\ 
&a_4 \times (T_{\rm eff})^2 +
a_5 \times (\log g)^2 +
a_6 \times ([Fe/H])^2 + \nonumber\\
&a_7 \times  T_{\rm eff} \times \log g +
a_8 \times  T_{\rm eff} \times  {\rm [Fe/H]} +  \nonumber\\
&a_9 \times \log g \times {\rm [Fe/H]}  + \nonumber\\
&a_{10} \times (T_{\rm eff})^3  + 
a_{11} \times (\log g)^3 +
a_{12} \times ({\rm [Fe/H]})^3 + \nonumber\\
&a_{13} \times  (T_{\rm eff})^2 \times \log g +
a_{14} \times (T_{\rm eff})^2 \times \log g +\nonumber\\
&a_{15} \times  (\log g)^2 \times T_{\rm eff} +
a_{16} \times (\log g)^2 \times {\rm [Fe/H]} + \nonumber\\
&a_{17} \times  ({\rm [Fe/H]})^2 \times T_{\rm eff}  + 
a_{18} \times ({\rm [Fe/H]})^2 \times \log g  +\nonumber\\
&a_{19} \times T_{\rm eff} \times \log g \times {\rm [Fe/H]}. 
\label{eqn:dev}
\end{align}

\begin{figure*}
\includegraphics[width=180mm]{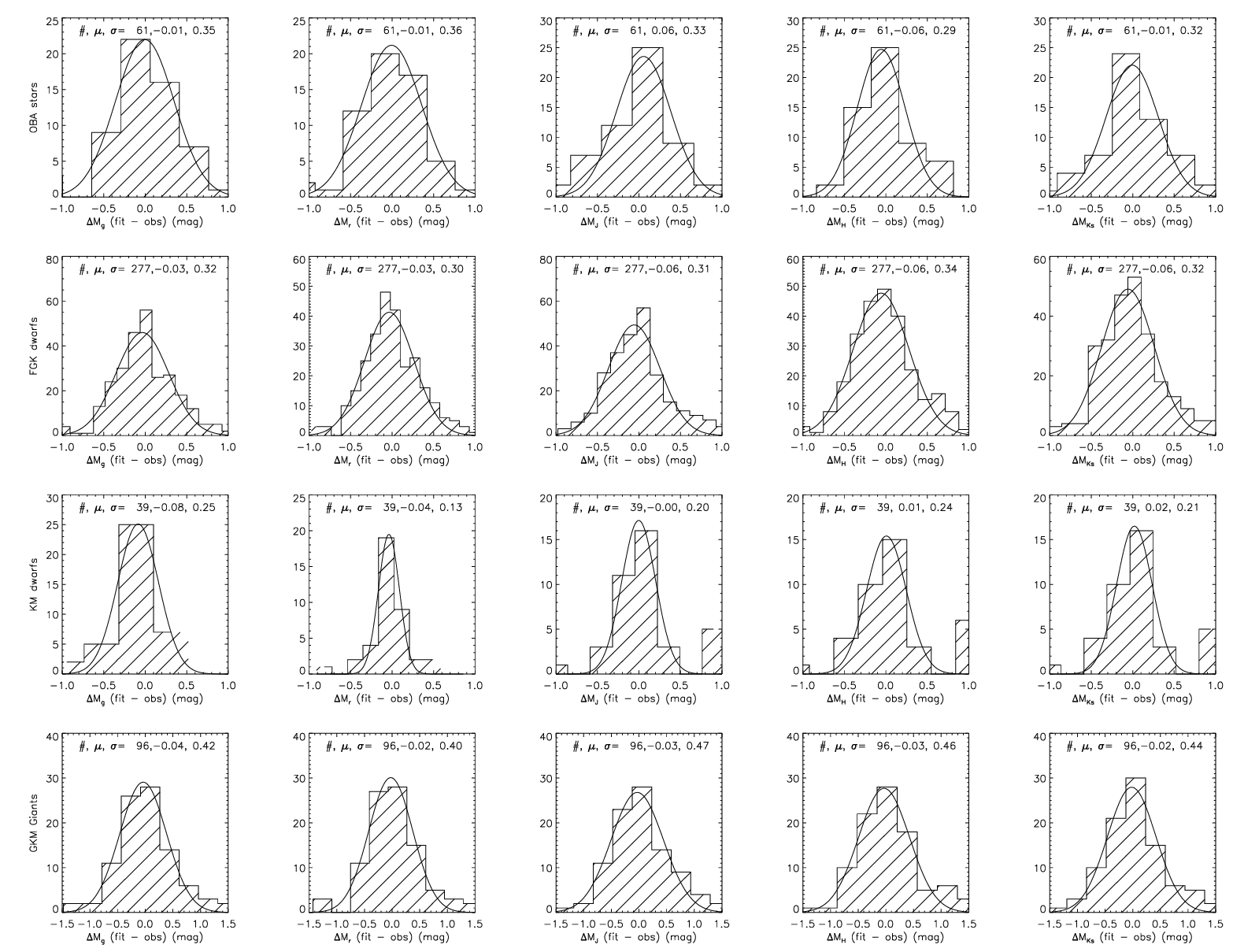}
\caption{
Histograms of fit residuals of absolute magnitudes in $g, r, J, H$ and $K_{\rm s}$ bands 
for the MILES stars of groups, from top to bottom, OBA stars, FGK dwarfs, KM dwarfs and GKM giants, respectively. 
The number of stars, the mean and standard deviation of the residuals are marked in each panel.}
\label{check_miles}
\end{figure*}

The fit coefficients and residuals are listed in Table\,\ref{param}. The residuals are also plotted in Fig.\,\ref{check_miles}
and have dispersion on average of about 0.35, 0.32, 0.22 and 0.44 mag, 
for the four groups of OBA stars, FGK dwarfs, KM dwarfs and GKM giants, respectively, corresponding to a fractional 
distance error of 17, 16, 11 and 22 per cent.
The residuals are partly contributed by errors of the absolute magnitudes as calculated from the apparent magnitudes and Hipparcos parallaxes.

\begin{figure*}
\includegraphics[width=180mm]{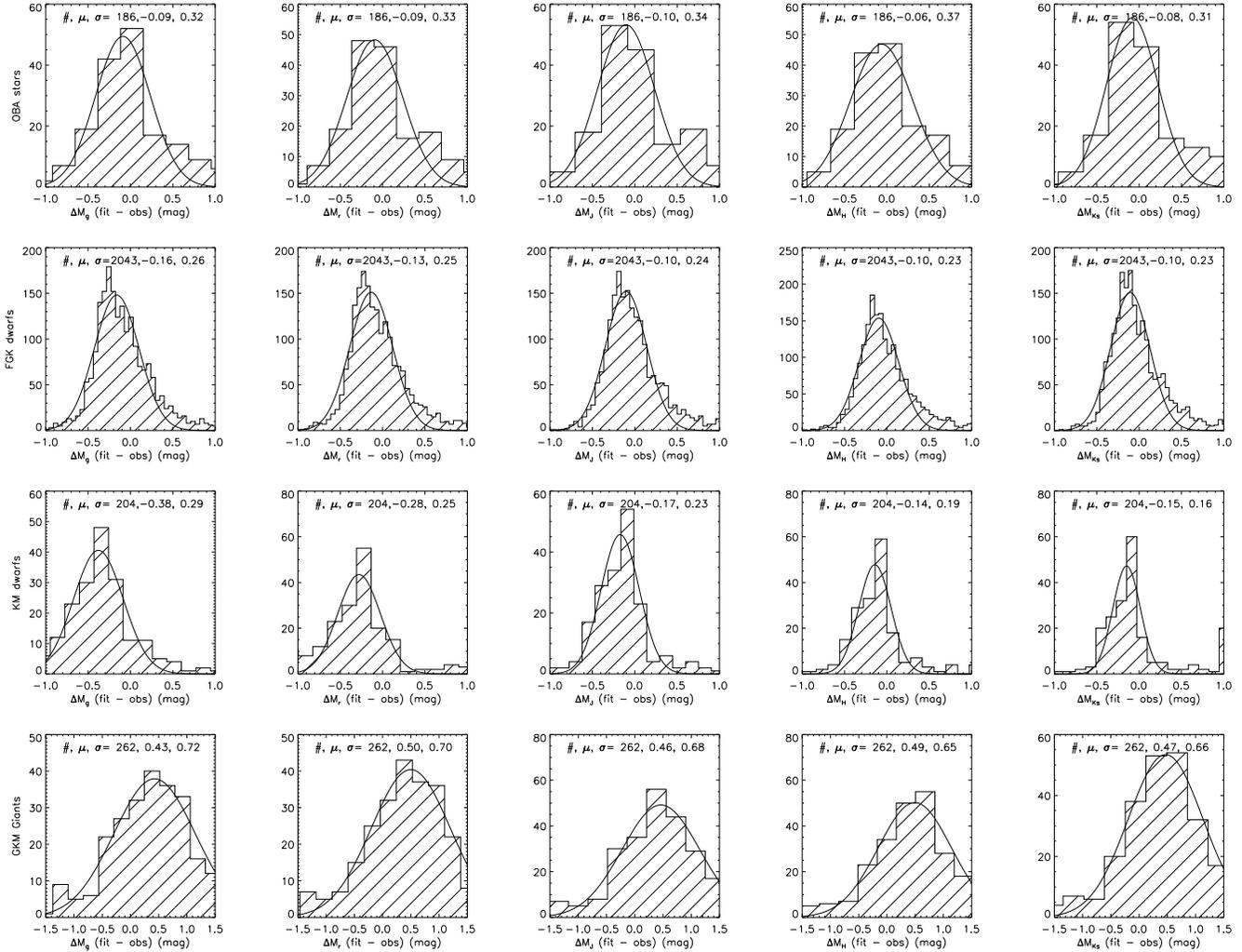}
\caption{From left to right, histograms of differences of absolute magnitudes in $g, r, J, H$ and $K_{\rm s}$ bands 
derived using the empirical relations from the MILES stars and those deduced using the Hipparcos parallaxes for the selected  
PASTEL stars. From top to bottom the four rows of panels correspond to the groups of OBA stars, FGK dwarfs, KM dwarfs and GKM giants, respectively. 
The number of stars, the mean and standard deviation of the residuals are marked in each panel.}
\label{check_pastel}
\end{figure*}

As a robustness check of the relations derived above, 
we apply them to selected stars in the PASTEL database to derive the absolute magnitudes and compare the results 
with those calculated from the Hipparcos distances. 
The results are given in Table\,\ref{param} and plotted in Fig.\,\ref{check_pastel}.
The differences between the two sets of estimates have an average dispersion of about 0.34, 0.25 and 0.25 mag
for the groups of OBA stars, FGK dwarfs and KM dwarfs, respectively, comparable to the fit residuals of the MILES stars.
The results suggest that the fits are robust and applicable to most types of stars, especially the dwarfs. There are, 
however, some small systematic  differences between the two sets of estimates of absolute magnitudes for the selected PASTEL stars, on the level of 
0.09, 0.12 and 0.20 mag for the groups of OBA stars, FGK dwarfs and KM dwarfs, respectively.
For GKM giants, the differences have a dispersion of 0.70 mag, significantly larger than the fit residuals.
In addition, the average absolute magnitudes calculated from the fitted relations are systematically larger by 0.48 mag 
than the ones calculated from the measured parallaxes.
The results suggest that there are some systematic  discrepancies 
between stellar parameters given by MILES library and the PASTEL database, in particular of \logg~for giants.


\begin{table*}{}
\small
\centering
\begin{minipage}[]{180mm}
\caption[]{Fit coefficients and residuals of the relations between absolute magnitudes and stellar atmospheric parameters.}
\label{param}\end{minipage}
\tabcolsep 1mm
\begin{tabular}{lrrrrrrrrrrr}
  \hline\noalign{\smallskip}
  & $M_g$ & $M_r$ & $M_J$ & $M_H$ & $M_{K_{\rm s}}$ & & $M_g$ & $M_r$ & $M_J$ & $M_H$ & $M_{K_{\rm s}}$   \\
  \hline
\noalign{\vskip2pt} \multicolumn{6}{c}{OBA stars} & \multicolumn{6}{c}{FGK dwarfs}   \\
a$_0$ &   1.90e+02 &   1.71e+02 &   1.65e+02 &   8.95e+01 &   1.98e+02 &  &  $-$4.11e+01 &  $-$1.24e+01 &  $-$7.58e+01 &  $-$8.22e+01 &  $-$8.27e+01 \\
a$_1$ &   7.34e$-$04 &   8.34e$-$04 &  $-$2.43e$-$03 &  $-$1.24e$-$02 &  $-$2.33e$-$03 &  &   3.90e$-$04 &   3.12e$-$02 &   5.93e$-$03 &   5.98e$-$03 &   8.53e$-$03 \\
a$_2$ &  $-$1.47e+02 &  $-$1.33e+02 &  $-$1.22e+02 &  $-$4.40e+01 &  $-$1.47e+02 &  &   2.89e+01 &  $-$3.82e+01 &   4.41e+01 &   4.78e+01 &   4.48e+01 \\
a$_3$ &   6.70e+00 &   6.83e+00 &   4.28e+00 &   4.94e+01 &  $-$1.15e+01 &  &   2.32e+01 &   1.56e+01 &  $-$4.67e+01 &  $-$5.83e+01 &  $-$4.24e+01 \\
a$_4$ &  $-$5.80e$-$08 &  $-$1.20e$-$07 &  $-$3.73e$-$08 &   4.38e$-$07 &  $-$7.41e$-$08 &  &  $-$6.79e$-$07 &  $-$4.03e$-$06 &  $-$1.44e$-$06 &  $-$1.28e$-$06 &  $-$1.37e$-$06 \\
a$_5$ &   3.82e+01 &   3.40e+01 &   2.96e+01 &   7.18e+00 &   3.56e+01 &  &  $-$5.01e+00 &   1.42e+01 &  $-$9.23e+00 &  $-$9.76e+00 &  $-$8.38e+00 \\
a$_6$ &  $-$3.49e+00 &  $-$4.06e+00 &  $-$6.82e+00 &  $-$2.06e+00 &  $-$6.30e+00 &  &   1.60e+00 &   2.14e+00 &  $-$3.41e$-$01 &  $-$5.93e$-$01 &  $-$1.02e$-$01 \\
a$_7$ &  $-$4.28e$-$04 &  $-$2.07e$-$05 &   1.43e$-$03 &   4.21e$-$03 &   1.57e$-$03 &  &   1.62e$-$04 &  $-$4.32e$-$03 &   3.64e$-$04 &  $-$4.64e$-$05 &  $-$1.07e$-$03 \\
a$_8$ &   3.36e$-$03 &   3.56e$-$03 &   3.80e$-$03 &   1.64e$-$03 &   4.63e$-$03 &  &  $-$5.50e$-$03 &  $-$1.46e$-$03 &   4.94e$-$03 &   6.02e$-$03 &   3.51e$-$03 \\
a$_9$ &  $-$1.14e+01 &  $-$1.21e+01 &  $-$1.15e+01 &  $-$2.96e+01 &  $-$5.61e+00 &  &  $-$2.64e+00 &  $-$4.65e+00 &   1.59e+01 &   2.00e+01 &   1.59e+01 \\
a$_{10}$ &  $-$3.96e$-$12 &  $-$2.46e$-$12 &  $-$1.26e$-$12 &  $-$3.74e$-$12 &  $-$1.02e$-$12 &  &   7.62e$-$11 &   1.55e$-$10 &   1.00e$-$10 &   9.84e$-$11 &   9.80e$-$11 \\
a$_{11}$ &  $-$3.12e+00 &  $-$2.72e+00 &  $-$2.26e+00 &  $-$3.35e$-$01 &  $-$2.73e+00 &  &   2.12e$-$01 &  $-$1.27e+00 &   6.23e$-$01 &   6.12e$-$01 &   4.59e$-$01 \\
a$_{12}$ &   3.58e$-$01 &   3.59e$-$01 &   3.38e$-$01 &   3.57e$-$01 &   3.27e$-$01 &  &   2.91e$-$01 &   3.42e$-$01 &   3.78e$-$01 &   4.27e$-$01 &   4.34e$-$01 \\
a$_{13}$ &   5.85e$-$08 &   5.82e$-$08 &   2.22e$-$08 &  $-$7.88e$-$08 &   2.88e$-$08 &  &  $-$1.04e$-$07 &   3.28e$-$07 &  $-$5.66e$-$08 &  $-$8.94e$-$08 &  $-$6.48e$-$08 \\
a$_{14}$ &  $-$2.21e$-$08 &  $-$2.73e$-$08 &  $-$1.20e$-$08 &  $-$3.75e$-$08 &  $-$2.82e$-$08 &  &   1.31e$-$07 &  $-$5.58e$-$08 &  $-$2.07e$-$07 &  $-$2.41e$-$07 &  $-$1.67e$-$07 \\
a$_{15}$ &  $-$1.05e$-$04 &  $-$1.59e$-$04 &  $-$2.58e$-$04 &  $-$3.72e$-$04 &  $-$2.90e$-$04 &  &   1.30e$-$04 &   3.54e$-$05 &   3.88e$-$05 &   1.39e$-$04 &   2.34e$-$04 \\
a$_{16}$ &   2.27e+00 &   2.41e+00 &   2.52e+00 &   4.05e+00 &   1.93e+00 &  &  $-$3.71e$-$01 &   1.59e$-$01 &  $-$1.53e+00 &  $-$1.89e+00 &  $-$1.67e+00 \\
a$_{17}$ &  $-$1.81e$-$04 &  $-$1.52e$-$04 &  $-$1.26e$-$04 &  $-$8.28e$-$06 &  $-$7.03e$-$05 &  &   1.16e$-$04 &   9.20e$-$05 &   4.09e$-$04 &   4.49e$-$04 &   4.00e$-$04 \\
a$_{18}$ &   1.44e+00 &   1.52e+00 &   2.20e+00 &   7.19e$-$01 &   1.92e+00 &  &  $-$3.61e$-$01 &  $-$4.21e$-$01 &  $-$2.88e$-$01 &  $-$2.54e$-$01 &  $-$3.03e$-$01 \\
a$_{19}$ &  $-$7.24e$-$04 &  $-$7.46e$-$04 &  $-$8.94e$-$04 &  $-$2.16e$-$04 &  $-$1.01e$-$03 &  &   9.03e$-$04 &   4.74e$-$04 &  $-$6.06e$-$04 &  $-$7.74e$-$04 &  $-$3.85e$-$04 \\

$\Delta$$^a$  & $-$0.01  & $-$0.01 & 0.06 &  $-$0.06 & $-$0.01 & & $-$0.03 & $-$0.03 & $-$0.06 & $-$0.06 & $-$0.06 \\
$\sigma$$^b$ & 0.35     & 0.36    & 0.33 &  0.29    & 0.32    & & 0.32    & 0.30    & 0.31    & 0.34    & 0.32 \\
$\sigma$$_M$$^c$  & 0.19  & 0.19 & 0.28 &  0.24 & 0.21 & & 0.13 & 0.13 & 0.24 & 0.22 & 0.18 \\
$\Delta$$^d$  & $-$0.09 &  $-$0.09 & $-$0.10 & $-$0.08 & $-$0.08 &  & $-$0.15 & $-$0.13 & $-$0.10  & $-$0.10 & $-$0.10 \\
$\sigma$$^e$ & 0.32    &  0.33    & 0.34    & 0.37    & 0.32    &  &  0.26   & 0.25    & 0.24     & 0.23    & 0.23 \\
$\sigma$$_M$$^f$  & 0.16  & 0.16 & 0.27 &  0.24 & 0.20 & & 0.17 & 0.17 & 0.17 & 0.17 & 0.15 \\
\hline
\noalign{\vskip2pt} \multicolumn{6}{c}{KM dwarfs} & \multicolumn{6}{c}{GKM giants}   \\
a$_0$ &  $-$1.91e+01 &  $-$4.71e+01 &   1.28e+01 &  $-$9.41e+01 &  $-$2.68e+01 &  &   3.69e+01 &   4.85e+00 &   5.28e+01 &   3.39e+01 &   2.79e+01 \\
a$_1$ &   1.38e$-$01 &   1.33e$-$01 &  $-$1.58e$-$01 &   3.96e$-$02 &  $-$1.57e$-$02 &  &   1.55e$-$02 &   4.02e$-$02 &  $-$2.56e$-$02 &  $-$1.19e$-$02 &  $-$8.33e$-$03 \\
a$_2$ &  $-$7.73e+01 &  $-$5.82e+01 &   1.13e+02 &  $-$4.13e+01 &   1.22e+01 &  &  $-$1.07e+02 &  $-$1.19e+02 &  $-$6.51e+01 &  $-$7.11e+01 &  $-$6.80e+01 \\
a$_3$ &  $-$8.53e+02 &  $-$8.01e+02 &   3.03e+02 &   5.94e+02 &   2.99e+02 &  &  $-$5.41e+01 &  $-$6.60e+01 &  $-$1.63e+02 &  $-$1.58e+02 &  $-$1.04e+02 \\
a$_4$ &  $-$6.08e$-$05 &  $-$5.57e$-$05 &   1.93e$-$05 &  $-$4.64e$-$06 &  $-$5.09e$-$06 &  &  $-$1.58e$-$05 &  $-$2.21e$-$05 &  $-$1.02e$-$09 &  $-$3.34e$-$06 &  $-$3.90e$-$06 \\
a$_5$ &  $-$2.40e+01 &  $-$2.37e+01 &  $-$3.78e+01 &   2.05e+01 &  $-$1.01e+01 &  &  $-$1.92e+01 &  $-$1.94e+01 &  $-$1.90e+01 &  $-$1.86e+01 &  $-$1.66e+01 \\
a$_6$ &  $-$1.21e+01 &  $-$1.31e+01 &  $-$9.90e+00 &   3.33e+01 &  $-$1.33e+00 &  &   1.71e+01 &   2.01e+01 &   3.21e+01 &   3.35e+01 &   3.07e+01 \\
a$_7$ &   6.25e$-$02 &   5.49e$-$02 &   3.69e$-$02 &  $-$1.19e$-$03 &   2.09e$-$02 &  &   6.39e$-$02 &   6.91e$-$02 &   4.43e$-$02 &   4.66e$-$02 &   4.38e$-$02 \\
a$_8$ &   1.72e$-$01 &   1.63e$-$01 &  $-$4.87e$-$02 &  $-$1.07e$-$01 &  $-$4.86e$-$02 &  &   2.27e$-$02 &   2.86e$-$02 &   9.49e$-$02 &   9.20e$-$02 &   5.99e$-$02 \\
a$_9$ &   2.08e+02 &   1.93e+02 &  $-$8.40e+01 &  $-$1.49e+02 &  $-$8.12e+01 &  &   5.95e+00 &   4.79e+00 &  $-$3.75e+01 &  $-$3.60e+01 &  $-$1.92e+01 \\
a$_{10}$ &   2.44e$-$09 &   2.12e$-$09 &   9.84e$-$10 &   1.30e$-$09 &   1.16e$-$09 &  &   2.31e$-$09 &   2.84e$-$09 &   6.85e$-$10 &   9.54e$-$10 &   9.62e$-$10 \\
a$_{11}$ &   6.72e+00 &   6.31e+00 &   1.92e+00 &  $-$2.71e+00 &   8.13e$-$01 &  &  $-$6.67e$-$01 &  $-$6.56e$-$01 &  $-$9.12e$-$01 &  $-$9.35e$-$01 &  $-$7.95e$-$01 \\
a$_{12}$ &   2.12e+00 &   1.93e+00 &   4.33e+00 &   2.97e+00 &   4.30e+00 &  &   2.10e$-$02 &  $-$3.71e$-$01 &  $-$1.66e$-$01 &  $-$2.76e$-$01 &  $-$1.22e$-$01 \\
a$_{13}$ &   5.79e$-$06 &   5.61e$-$06 &  $-$7.09e$-$06 &  $-$2.85e$-$06 &  $-$2.40e$-$06 &  &  $-$9.09e$-$06 &  $-$9.63e$-$06 &  $-$7.03e$-$06 &  $-$7.26e$-$06 &  $-$6.74e$-$06 \\
a$_{14}$ &  $-$8.49e$-$06 &  $-$8.17e$-$06 &  $-$7.22e$-$07 &   2.30e$-$06 &  $-$5.14e$-$07 &  &  $-$2.25e$-$06 &  $-$2.84e$-$06 &  $-$1.40e$-$05 &  $-$1.37e$-$05 &  $-$9.11e$-$06 \\
a$_{15}$ &  $-$1.27e$-$02 &  $-$1.17e$-$02 &   2.25e$-$03 &   2.16e$-$03 &  $-$3.53e$-$04 &  &   5.06e$-$03 &   5.06e$-$03 &   5.41e$-$03 &   5.37e$-$03 &   4.76e$-$03 \\
a$_{16}$ &  $-$1.22e+01 &  $-$1.12e+01 &   3.06e+00 &   6.26e+00 &   3.00e+00 &  &   1.00e+00 &   1.34e+00 &  $-$2.53e+00 &  $-$3.03e+00 &  $-$2.38e+00 \\
a$_{17}$ &  $-$4.05e$-$04 &  $-$2.96e$-$04 &   1.62e$-$03 &   2.18e$-$04 &   8.66e$-$04 &  &  $-$3.25e$-$03 &  $-$3.90e$-$03 &  $-$8.65e$-$03 &  $-$8.66e$-$03 &  $-$7.69e$-$03 \\
a$_{18}$ &   3.53e+00 &   3.58e+00 &   1.03e+00 &  $-$7.23e+00 &   5.10e$-$02 &  &  $-$2.01e$-$01 &  $-$3.67e$-$01 &   3.99e+00 &   3.42e+00 &   2.75e+00 \\
a$_{19}$ &  $-$2.16e$-$02 &  $-$2.02e$-$02 &   1.21e$-$02 &   1.89e$-$02 &   1.16e$-$02 &  &  $-$1.99e$-$03 &  $-$2.15e$-$03 &   1.17e$-$02 &   1.18e$-$02 &   7.50e$-$03 \\

$\Delta$$^a$  & $-$0.08 & $-$0.04  &  0.00  &  0.01  &  0.02 & & $-$0.04 & $-$0.02  & $-$0.03 & $-$0.03 & $-$0.02 \\
$\sigma$$^b$ &  0.25   & 0.13     &  0.20  &  0.24  &  0.21 & &  0.42   &    0.40  &  0.47   & 0.46    & 0.44 \\
$\sigma$$_M$$^c$  & 0.11  & 0.11 & 0.16 &  0.18 & 0.17 & & 0.12 & 0.12 & 0.32 & 0.27 & 0.30 \\
$\Delta$$^d$  & $-$0.38 & $-$0.28  & $-$0.17 & $-$0.14 & $-$0.15 & & 0.43 & 0.50 & 0.46 & 0.49 & 0.47 \\
$\sigma$$^e$ &  0.29   & 0.25     & 0.23    & 0.19    & 0.16    & & 0.72 & 0.70 & 0.68 & 0.65 & 0.66 \\
$\sigma$$_M$$^f$  & 0.16  & 0.16 & 0.13 &  0.15 & 0.13 & & 0.14 & 0.14 & 0.34 & 0.30 & 0.33 \\

\noalign{\smallskip}\hline
\end{tabular}
\begin{description}
\item[$^a$]  Average of fit residuals for the selected MILES stars. 
\item[$^b$]  Dispersion of fit residuals for the selected MILES stars. 
\item[$^c$]  Median error of the absolute magnitudes calculated from the photometry and Hipparcos parallaxes for the selected MILES stars. 
\item[$^d$]  Average difference of absolute magnitudes derived from the fitted empirical relations and those deduced from the Hipparcos parallaxes 
for the selected PASTEL stars. 
\item[$^e$]  Dispersion of differences of absolute magnitudes derived from the fitted empirical relations and those deduced from the Hipparcos parallaxes
for the selected PASTEL stars. 
\item[$^f$]  Median error of the absolute magnitudes calculated from the photometry and Hipparcos parallaxes for the selected PASTEL stars. 
\end{description} 
\flushleft
\end{table*}

\begin{figure}
\includegraphics[width=90mm]{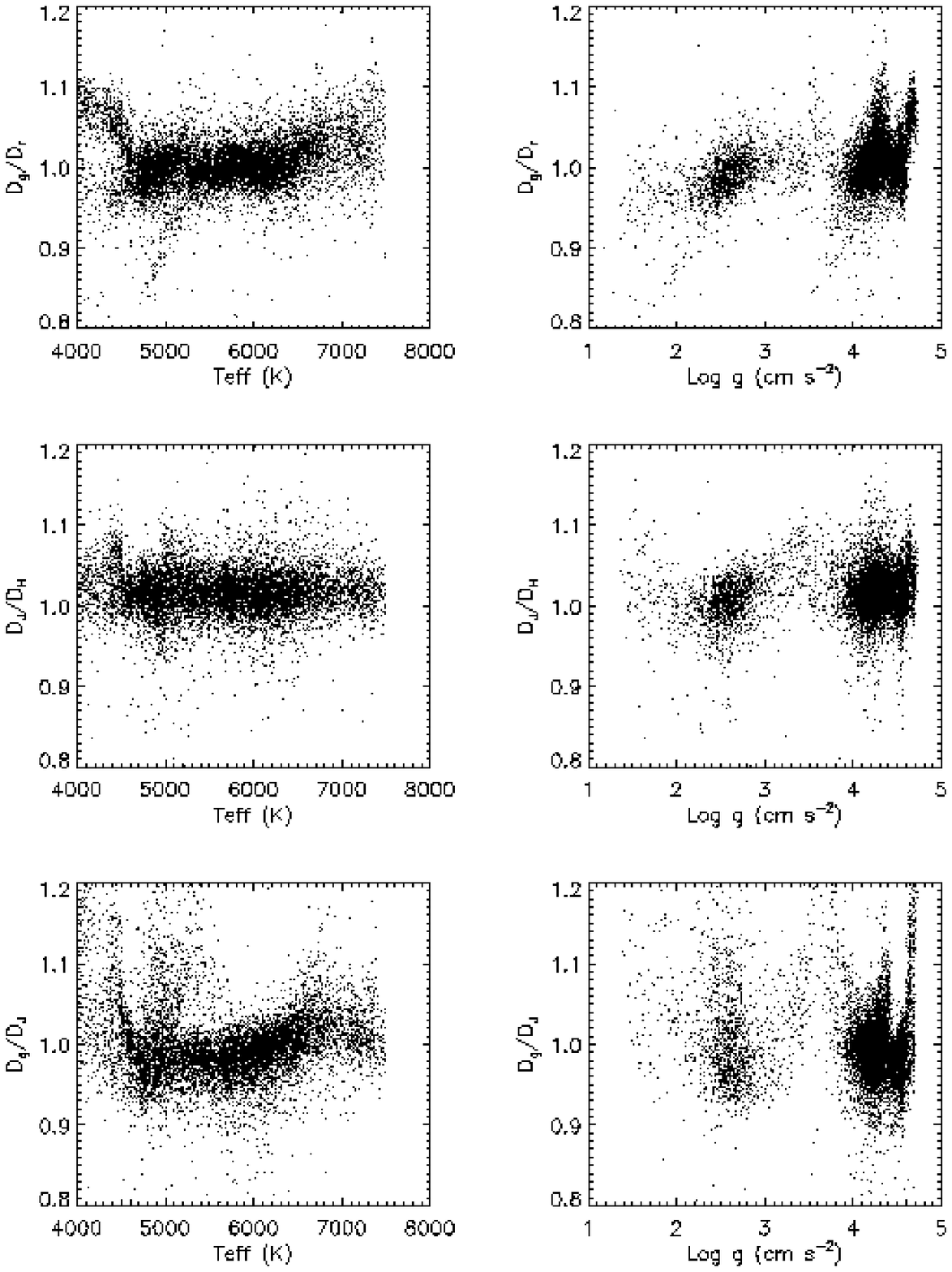} 
\caption{Ratios of distances derived from $g$ and $r$ band magnitudes (top), those from $J$ and $H$ bands (middle) and 
those from $g$ and $J$ bands (bottom) as a function of \teff~(left) and \logg~(right).
}
\label{compare_dist_1}
\end{figure}

Once the empirical relations between the absolute magnitudes and \teff, \logg~and \feh~have been determined,
the following formula is used to calculate distances of the LSS-GAC targets from the photometry of 
a given band $a$ ($a = g, r, J, H$ or $K_{\rm s}$): 
\begin{align}
&D_a = 10^{ \{m_a - E(B-V)_{sp}\times R(a) - M_a(T_{\rm eff}, \log g, \mathrm{[Fe/H]}) + 5\} \times 0.2}, 
\label{eqn:dist}
\end{align}
where $M_a$ is the absolute magnitude of photometric band $a$, $m_a$ the observed apparent magnitude in band $a$, 
$E(B - V)_{\rm sp}$~ the interstellar reddening derived using the star-pair method and $R(a)$ the extinction coefficient from Yuan, Liu \& Xiang (2013).
$M_a$ is calculated from \teff, \logg~and \feh~as derived from the LSS-GAC spectra with the LSP3. 
Note that the \teff~values used here are those before corrected for the systematics as described in Section\,4.
For stars of different groups, the appropriate relations between the absolute magnitudes and stellar atmospheric parameters are used. 
To avoid ambiguities, the defining boundaries between the groups of stars are adjusted slightly  
from those adopted when fitting the relations, and are: 
\begin{description}
\item[OBA stars:] $T_{\rm{eff}} \ge 7,500 $~K;
\item[FGK dwarfs:] $ 4,500\,< T_{\rm{eff}}\,<\,7500$~K, \logg~$>$\,3.5\,(cm\,s$^{-2}$);
\item[KM dwarfs:] $T_{\rm{eff}}\,\le\,4,500$\,K, \logg~$>$\,3.5\,(cm\,s$^{-2}$);
\item[GKM giants:] $T_{\rm{eff}}\,\le\,5,500$\,K, \logg~$\le$\,3.5\,(cm\,s$^{-2}$).
\end{description}
Distances derived from individual bands agree typically with each other within 2 -- 3 per cent for FGK dwarfs, and 5 -- 10 per cent 
for giants, as shown in Fig.\,\ref{compare_dist_1}.
However, the discrepancies become quite large for stars near the boundaries of the four groups of stars such as subgiants. 
The final distances are adopted to be the mean values derived from all available bands.  

RCs constitute a significant fraction of giant stars in the Galactic disc targeted by the LSS-GAC.
RCs have a very narrow luminosity range. Their absolute magnitudes depend very weakly on age and metallicity.
Consequently, they have been widely used as standard candles to measure distances.
In the current work, we have selected stars of 
4,500 $\le$ \teff~$\le$ 5,200\,K  and 2.0 $\le$ \logg~$\le$ 3.0\,dex as RCs.
Their distances are determined assuming a 2MASS $K_{\rm s}$-band  absolute magnitude
of $-1.57\pm0.03$\,mag (Grocholski \& Sarajedini, 2002)
and an SDSS $i$-band absolute magnitude of 0.19$\pm$0.1\,mag, derived from the XSTPS-GAC data using
a technique similar to that of Grocholski \& Sarajedini (2002).
Note that the above simple cuts on \teff~and \logg~ used to select RCs may lead to 
some contamination from red giant stars.

The accuracy of distances derived from the empirical relations is examined using the LAMOST observations of member stars 
of open cluster M\,67. 
The upper middle panel of Fig.\,\ref{m67} shows that the method yields a mean distance of 880 pc for M\,67, 
consistent with the value from the literature (860 $\pm$ 45 pc; Pancino et al. 2010).
The distances for individual stars have a dispersion of 7 per cent, 
consistent with the dispersion of residuals of 
the fits to the MILES stars, as well as with the results of applying the relations to the selected PASTEL stars as discussed above.

\begin{figure*}
\includegraphics[width=180mm]{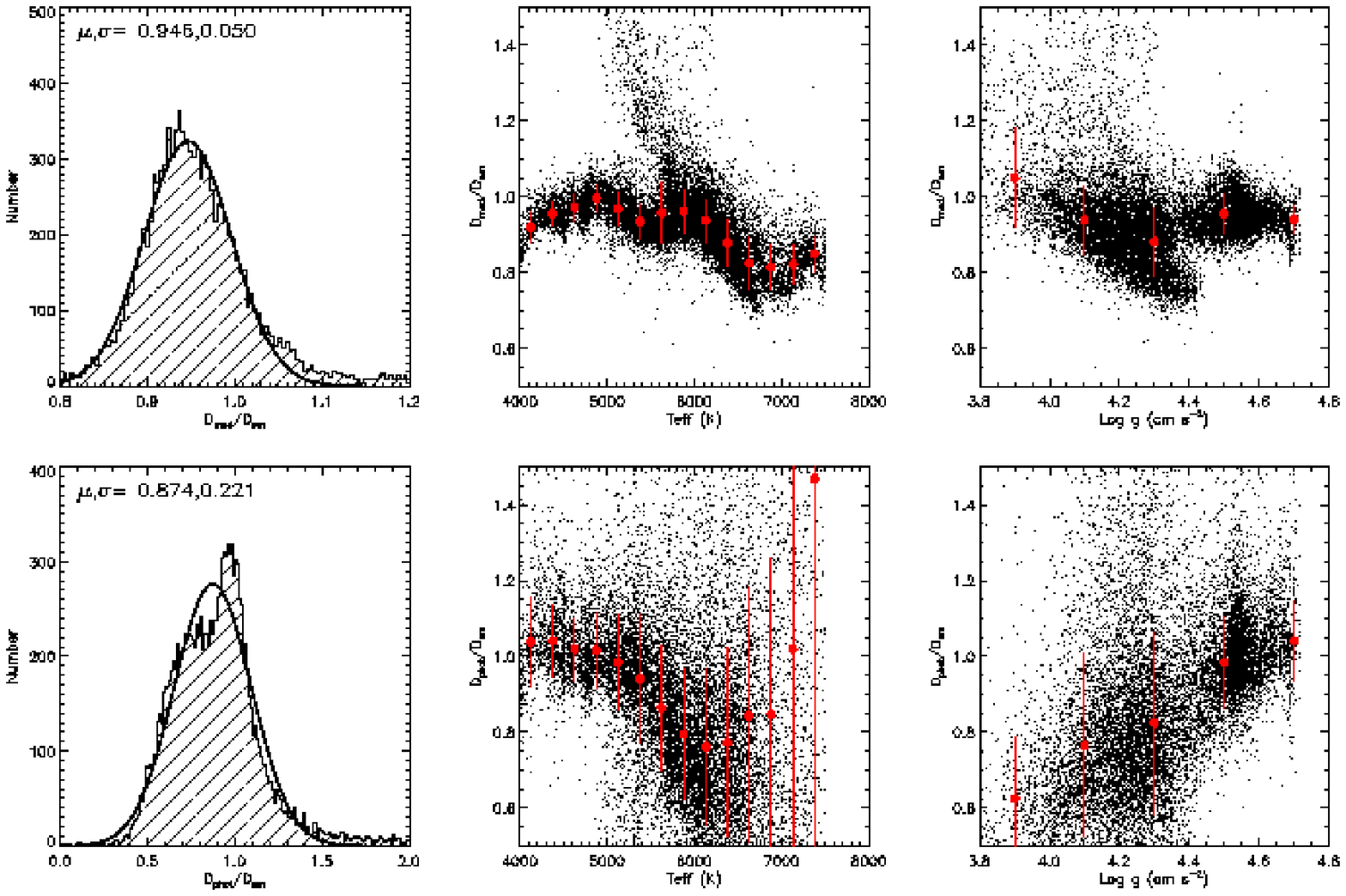} 
\caption{Comparisons of distances derived from the empirical relations 
of absolute magnitudes as a function of stellar atmospheric parameters yielded by the MILES stars with 
those derived from the theoretical isochrones of stellar models (upper panels) and those using the 
photometric parallax method (lower panels) for dwarf stars ($ 4000 \le \teff \le 7500$ K, \logg~$\ge$ 3.8 dex). 
Left: Histogram distribution of the ratios of the distances in comparison.
The mean and standard deviation of the ratios are marked. 
Middle: The ratios plotted against \teff.
Right: The ratios plotted against \logg.
The red dots and error bars in the middle and right panels denote the median values 
and standard deviations of ratios of the individual \teff~or \logg~bins.}    
\label{compare_dist}
\end{figure*}

As we have made use of the synthetic colours predicted by the stellar models to determine the 
extinction of LSS-GAC targets, 
we have also used the absolute magnitudes predicted by the theoretical stellar isochrones to derive the distances.
Once the extinction of a given star has been determined, its 
absolute magnitude is estimated using the isochrones of Dartmouth Stellar Evolution Database (Dotter et al. 2008). 
The isochrones are first linearly interpolated to a dense grid in [Fe/H] and [$\alpha$/Fe], 
with steps of 0.1 and 0.2\,dex, respectively.
Since the current implementation of LSP3 does not provide estimates of [$\alpha$/Fe] yet, we simply 
assign values of [$\alpha$/Fe] to stars according to their [Fe/H] values. We assign 
[$\alpha$/Fe] values of 0, 0.2, 0.4\,dex for stars of [Fe/H] $>-0.5$, 
$-1.0<$ [Fe/H] $\leq -0.5$ and [Fe/H] $\leq -1.0$, respectively to mimic the observed variations of 
[$\alpha$/Fe] as a function of [Fe/H] for Galactic stars (Lee et al. 2011). 
For a star of given \teff, \logg, \feh~and [$\alpha$/Fe], its absolute magnitudes 
are approximated as those of the nearest model of the isochrones.
Since the grid of isochrones is dense enough, the maximum offset 
in parameter due to the simplification is less than 50\,K in $T_{eff}$, 0.05\,dex in [Fe/H], 
0.1\,dex in [$\alpha$/Fe], and 0.05\,dex in log\,$g$, much smaller than the estimated uncertainties of the corresponding parameter.
A comparison of distances estimated using the theoretical absolute magnitudes predicted by the stellar isochrones 
and using those predicted by the empirical relations is given in the 
upper panels of Fig.\,\ref{compare_dist}. For dwarf stars of 4,000 $\le \teff \le$ 7,500\,K and \logg~$\ge$ 3.8\,dex, 
the two estimates agree well, although the distances estimated using absolute magnitudes predicted by the 
theoretical stellar isochrones are about 5 per cent systematically lower than those deduced from the empirical relations. 
The differences depend on \teff~and \logg.  

Both the above two approaches to estimate distances rely on stellar atmospheric parameters 
yielded by spectroscopic observations, thus are limited to stars having high-quality spectra. 
Alternatively, distances can also be estimated using photometric measurements alone. 
Estimates of photometric distances require accurate measurements of photometric colours and reddening.
Compared to spectroscopic distances, photometric distances can be constructed for very large samples 
of stars of different stellar populations. 
Using the reddening values derived from SED fitting by combining 
the XSTPS-GAC, 2MASS and WISE photometry, Chen et al. (2014) have derived photometric distances 
of 15 million stars in the footprint of XSTPS-GAC,   
using the colour-magnitude relation of Ivezi{\'c} et al (2008) for an assumed \feh~=~$-$0.2 dex. 
A comparison of photometric distances obtained by Chen et al. and those 
deduced above using the empirical relations of stellar absolution magnitudes as a function of spectroscopically derived atmospheric parameters
is shown in the lower panels of Fig.\,\ref{compare_dist}.
For dwarf stars of 4,000 $\le \teff \le$ 5,400\,K and \logg~ $\ge$ 4.4 dex, 
the two estimates agree well at a level of about 10 per cent. 
However, for stars hotter than 5,400\,K,  or of surface gravities lower than 4.4 dex, 
the agreement deteriorates, and the discrepancies increase with increasing $T_{\rm eff}$ or decreasing $\log\,g$.
The discrepancies at low surface gravities are possibly caused by the facts that 
1) the colour-magnitude relation of Ivezi{\'c} et al. is not applicable to turn-off stars in the Galactic thin disk;
and 2) the LSP3 may have underestimated \logg~for some FG dwarfs.

\section{Proper motions and orbital parameters}
For convenience, we also provide proper motions of the LSS-GAC stars collected from various sources. 
We have cross-matched the LSS-GAC targets with the 
UCAC4 (Zacharias et al. 2013) and PPMXL (Roeser et al. 2010) catalogues. 
A match radius of 3 arcsec is used.
The UCAC4 is an all-sky, astrometric catalogue that contains 113,780,093 objects complete to $R \sim 16$\,mag. 
Among them over 105 million stars have proper motions, with typical random errors of 4 mas yr$^{-1}$.
The PPMXL is compiled based on the USNO-B1.0, 2MASS, and its previous version PPMX (Roeser et al. 2008). 
It contains over 900 million sources down to $V =20$\,mag, with proper motions for almost all of them. 
The errors of proper motion range from 4 to over 10\,mas yr$^{-1}$, 
depending on the observational history of the sources.

Wu et al. (2011) investigate the systematic and random errors 
of proper motions provided by the PPMXL using a sample of over ten thousand quasars,
which, being distant and point-like objects, should have zero proper motions. 
They find systematic errors on the level of $−$2.0 and $−$2.1 mas\,yr$^{-1}$ 
in  $\mu_{\alpha}\cos\delta$ and $\mu_{\delta}$, respectively.
The errors correlate with source positions but not with 
magnitudes or colours for objects of $16 < r < 21$\,mag and $-0.4 < g - r < 1.2$\,mag.
Carlin et al. (2013) find similar results and fit the systematic errors
as a function of source position using a two-dimensional polynomial. 
We have adopted the fit of Carlin et al. (2013) to correct for systematic errors of the proper motions of PPMXL.
The corrections range from $-$2.7 to 0.7 mas yr$^{-1}$ in $\mu_{\alpha}\cos\delta$ 
and $-$3.9 to 1.1 mas\,yr$^{-1}$ in $\mu_{\delta}$ for the LSS-GAC targets.
For those sources, the random errors vary from  4.0 to  8.0 mas yr$^{-1}$, depending on the magnitudes.

The systematic errors of proper motions of the UCAC4 catalogue are examined by Huang et al. (2014), 
using about 1,700 quasars from the 2nd Release of the Large Quasar Astrometric Catalog (LQAC, Souchay et al. 2012)
that have listed proper motions in the UCAC4.
As in the case of PPMXL, no obvious dependence of the errors on magnitude or colour is found for objects of $11 < R < 16$\,mag and $-0.1 <  B - V < 1.4$\,mag. 
Similar to Carlin et al. (2013), Huang et al. fit the systematic errors as a function of source position 
using a two-dimensional polynomial.
Their results are used to correct the proper motions of UCAC4.
The corrections range between $-$1.1 and 1.0 mas yr$^{-1}$ in $\mu_{\alpha}\cos\delta$ 
and $-$1.8 to $-$1.1 mas yr$^{-1}$ in $\mu_{\delta}$ for the LSS-GAC targets.

Essentially all targets in the LSS-GAC are selected from the XSTPS-GAC (Xuyi hereafter) survey, 
which provides accurate positions (about 0.1\,arcsec) at a typical epoch of 2010 (Zhang et al. 2014).
By combining the measurements of XSTPS-GAC and those of 2MASS observed at a typical epoch of 2000, 
one can obtain independent measurements of the proper motions. The Xuyi-2MASS proper motions are especially 
useful to reject proper motions of large errors from the UCAC4/PPMXL catalogues and 
to search for high proper motion targets, such as nearby hyper-velocity stars. 

\begin{figure}
\includegraphics[width=90mm]{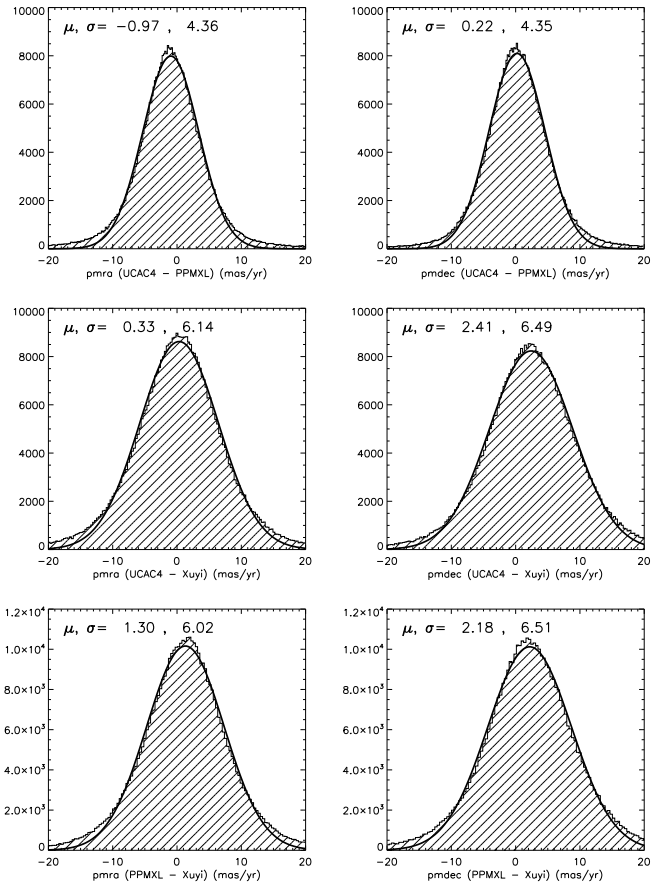}
\caption{
Distributions of the differences of proper motions for a selected sample of  LSS-GAC stars,
as given by the UCAC4 and PPMXL (top), by the UCAC4 and Xuyi-2MASS (middle) and by the 
PPMXL and Xuyi-2MASS (bottom). The black solid line in each panel represents a Gaussian fit to the 
distribution, with the mean and dispersion marked.
}
\label{compare_pm}
\end{figure}

To compare proper motions from the UCAC4, PPMXL and Xuyi-2MASS, we select LSS-GAC targets 
which have LAMOST spectral S/N(4650\AA) higher than 10 and 
proper motions from all the three sources. The results are shown in Fig.\,\ref{compare_pm}.
Compared to the PPMXL, proper motions from the UCAC4 are systematically lower in 
$\mu_{\alpha}\cos\delta$ by $\sim$ 0.97\,mas\,yr$^{-1}$ 
and slightly higher in $\mu_{\delta}$ by $\sim$ 0.22\,mas\,yr$^{-1}$.
Note that the systematic errors in the PPMXL and UCAC4 proper motions have already been corrected for as described above. 
Compared to the values of UCAC4 and PPMXL, the proper motions given by the Xuyi-2MASS measurements are systematically lower 
in $\mu_{\alpha}\cos\delta$ by respectively 0.33 and 1.30 mas yr$^{-1}$, and in 
$\mu_{\delta}$ by respectively 2.41 and 2.18 mas yr$^{-1}$.
If one ignores the small systematic differences, the proper motions from the UCAC4 and PPMXL catalogues 
agree very well, with typical differences of about 4.4 mas yr$^{-1}$ in both $\mu_{\alpha}\cos\delta$  and $\mu_{\delta}$. 
The comparisons suggest that the random errors of the UCAC4 and PPMXL proper motions for 
the selected LSS-GAC targets are typically 3.0 mas yr$^{-1}$, 
assuming that they contribute equally to the observed differences. 
The proper motions of Xuyi-2MASS have larger random errors given the relatively short time span 
between the 2MASS and XSTPS-GAC observational epochs, and are about 6.0 mas yr$^{-1}$ as inferred from 
the differences between the Xuyi-2MASS and UCAC4/PPMXL proper motions. 

\begin{figure}
\includegraphics[width=90mm]{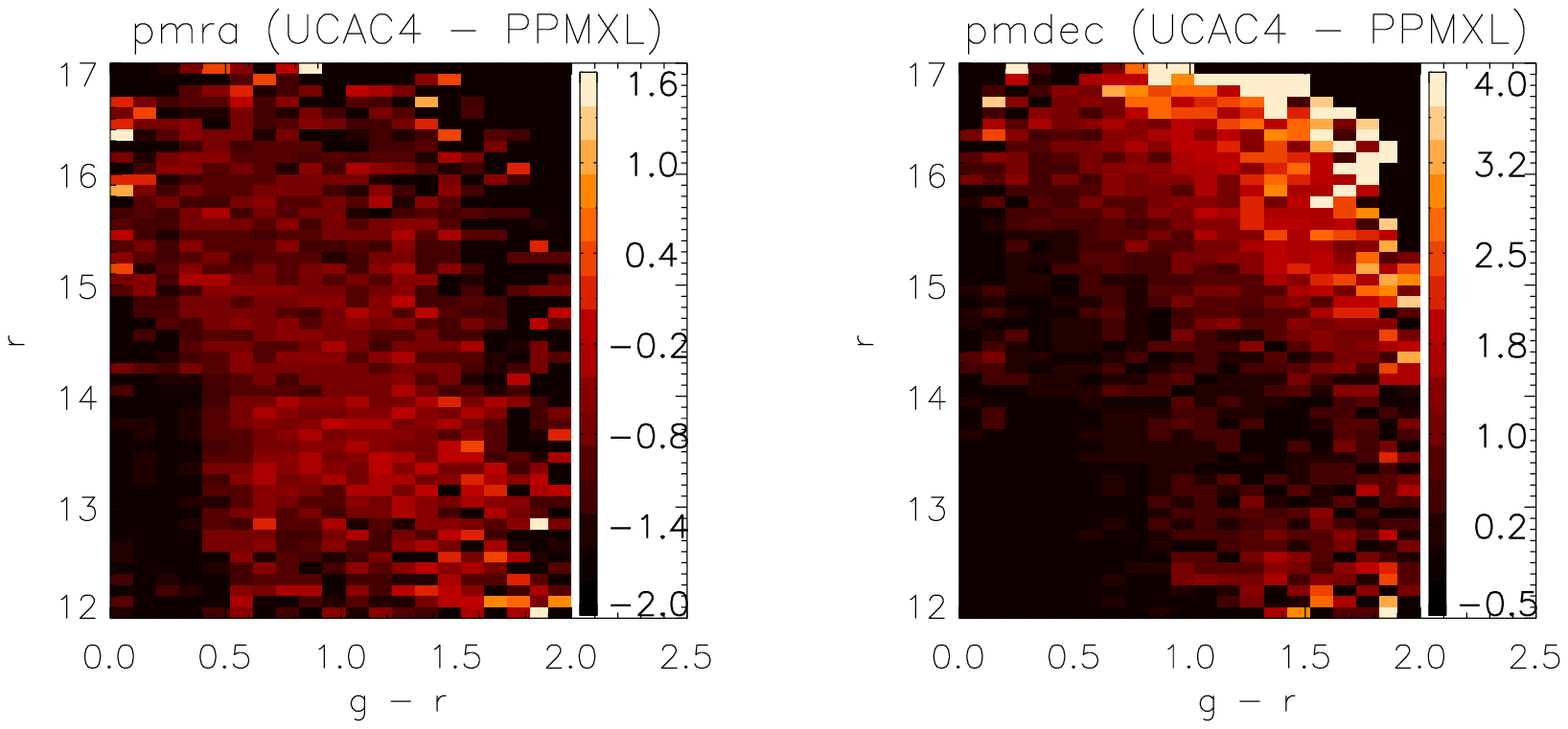}
\caption{
Distributions of the differences of proper motions given by the UCAC4 and by the PPMXL catalogues for 
a selected sample of LSS-GAC targets in the plane
of $g-r$  colour and $r$ magnitude. The colourbars are in units of mas yr$^{-1}$.
}
\label{compare_pm_cmd}
\end{figure}

\begin{figure}
\includegraphics[width=90mm]{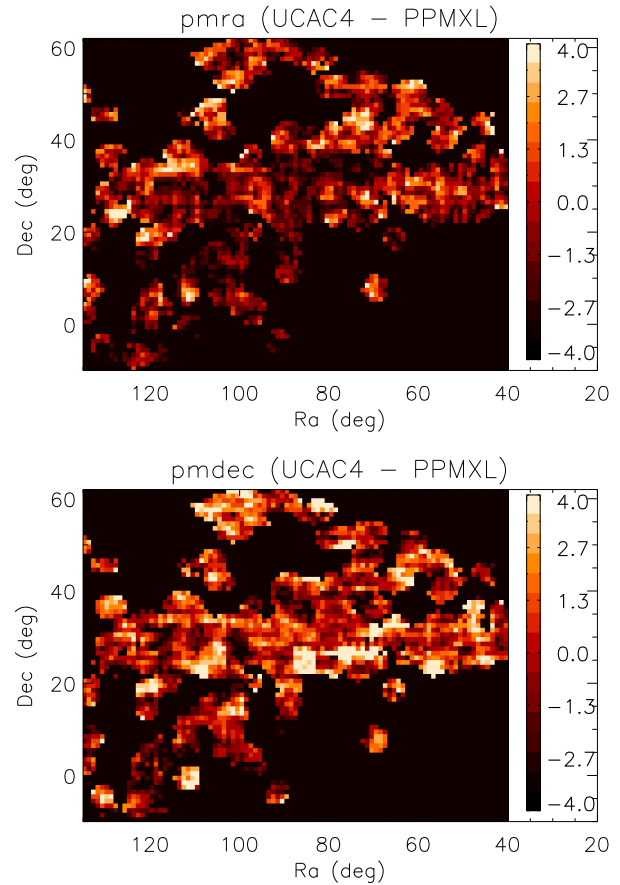}
\caption{
Distributions of the differences of proper motions given by the UCAC4 and by the PPMXL catalogues for a selected sample of 
LSS-GAC targets in the plane of RA and Dec. The colourbars are in units of mas yr$^{-1}$.
}
\label{compare_pm_radec}
\end{figure}

The above selected sample of LSS-GAC targets are divided into bins in colour $g-r$ and in magnitude $r$,
ranging from 0 -- 2 and 12 -- 17\,mag, respectively. The bin size is 0.1 mag  in both quantities.
For bins that have more than 10 stars, the median differences of proper motions given by the 
UCAC4 and PPMXL are calculated and the results are displayed in Fig.\,\ref{compare_pm_cmd}. 
The differences in $\mu_{\alpha}\cos\delta$ are found to be independent of $g-r$  and $r$. 
However, the differences in $\mu_{\delta}$ show a strong dependence on  
$g-r$ and $r$, in the sense that the differences are larger for redder and fainter stars.
The result is inconsistent with the findings of analyses based on the proper motions of quasars. 
One possible explanation is that most quasars discovered so far are from the 
high Galactic latitude regions, the findings from those quasars 
may not be applicable to the low Galactic latitude regions. 
Identifying more quasars at low Galactic latitudes will help clarify the issue.
The LSS-GAC targets are also divided into bins of RA and Dec in steps of 1 degree 
in order to examine possible spatial variations of the differences of proper motions given by the UCAC4 and PPMXL.
The results are shown in Fig.\,\ref{compare_pm_radec}.  
No obvious patterns are found. In addition, the discrepancies 
in $\mu_{\alpha}\cos\delta$ and $\mu_{\delta}$ are found to be independent.  
However, for some randomly distributed patches of sky of 2 -- 5\,deg in size, 
the differences are more than twice the typical values. 
 
For each LSS-GAC target, given its celestial position, distance, radial velocity and proper motions,
we calculate its orbit in the Galactic potential of Gardner \& Flynn (2010).
In the model of Gardner \& Flynn, the Galaxy consists of a disc (Miyamoto \& Nagai 1975), 
a Plummer bulge (Plummer 1911) with an inner core, and a spherical logarithmic dark matter halo. 
The characteristic parameters can be found in their Table 1.
In calculating the orbit, we take the local standard of rest (LSR) values from Huang et al. (2014),
i.e., (U, V, W) = (7.01, 10.13, 4.95) km s$^{-1}$ and the Sun is located at (X, Y, Z) = ($-$8.0, 0.0, 0.0)\,kpc
for a Galactocentric Cartesian coordinate system where the x-axis oriented along a line connecting the Sun and 
the Galactic centre and points towards the Galactic center, the y-axis is in the direction of disc rotation 
and the z-axis points towards the North Galactic Pole.  
The eccentricity of the orbit is defined as:
$e = (R_{\rm apo} - R_{\rm peri})/(R_{\rm apo} + R_{\rm peri})$,
where $R_{\rm apo}$ and $ R_{\rm peri}$ are respectively the maximum and minimum Galactic radii reached by the orbit.
The guiding radius $R_{\rm g}$ is also computed using the z-component of the angular momentum and the rotation curve. 
For low eccentricities, $R_{\rm g}$ is very close to the mean of $R_{\rm apo}$ and $ R_{\rm peri}$.

\section{Samples}
We define three LSS-GAC samples: the main, the M\,31-M\,33 
and the VB samples, corresponding to the LSS-GAC main, the LSS-GAC M\,31-M\,33 and the LSS-GAC 
VB surveys, respectively. In this section, the properties of those samples
are discussed below.
 
\subsection{The main sample}
\begin{figure*}\includegraphics[width=140mm]{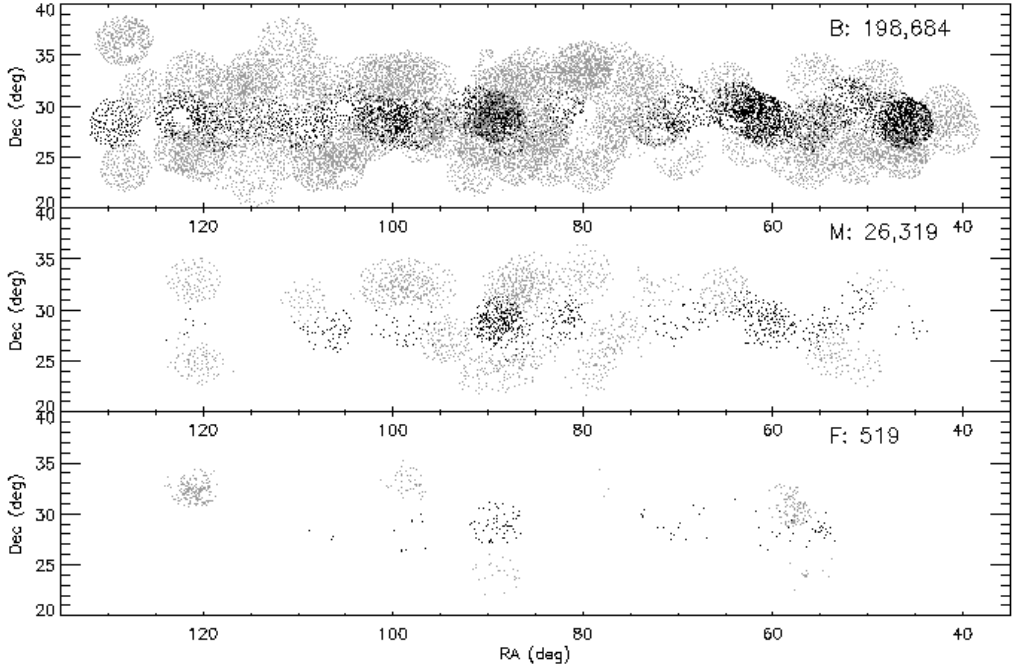}
\caption{
Same as Fig.\,\ref{spatial_gac_bluered} but for the LSS-GAC main sample. 
\label{spatial_gac_b}}
\end{figure*}

\begin{figure*}\includegraphics[width=140mm]{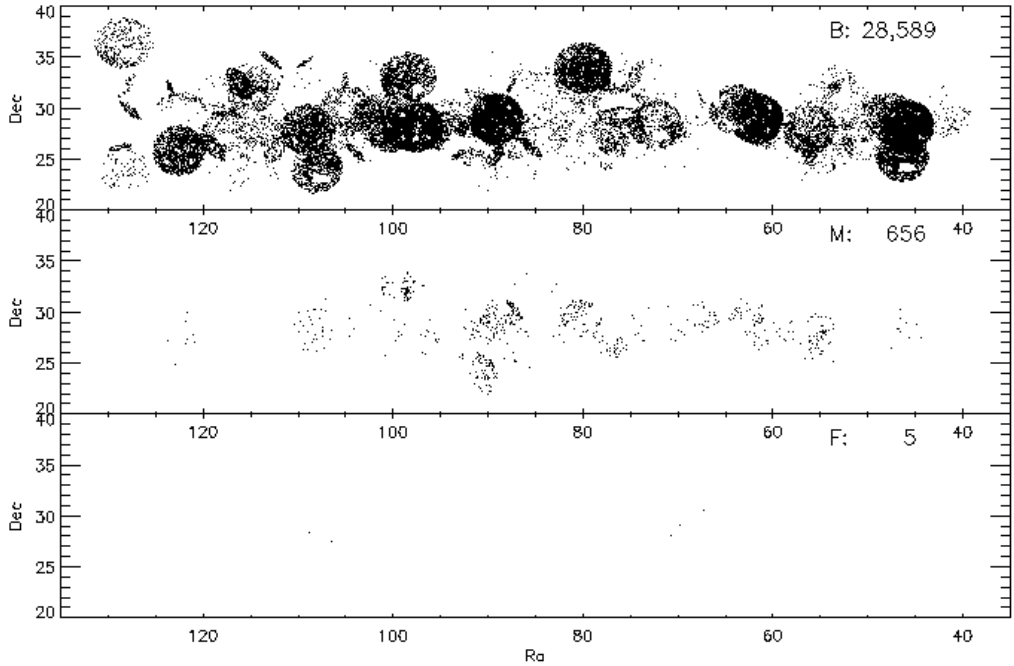}
\caption{
Same as Fig.\,\ref{spatial_gac_bluered} but for the duplicate targets in the LSS-GAC main sample. 
\label{multi_epoch_spatial_gac_b}}
\end{figure*}

\begin{figure*}\includegraphics[width=180mm]{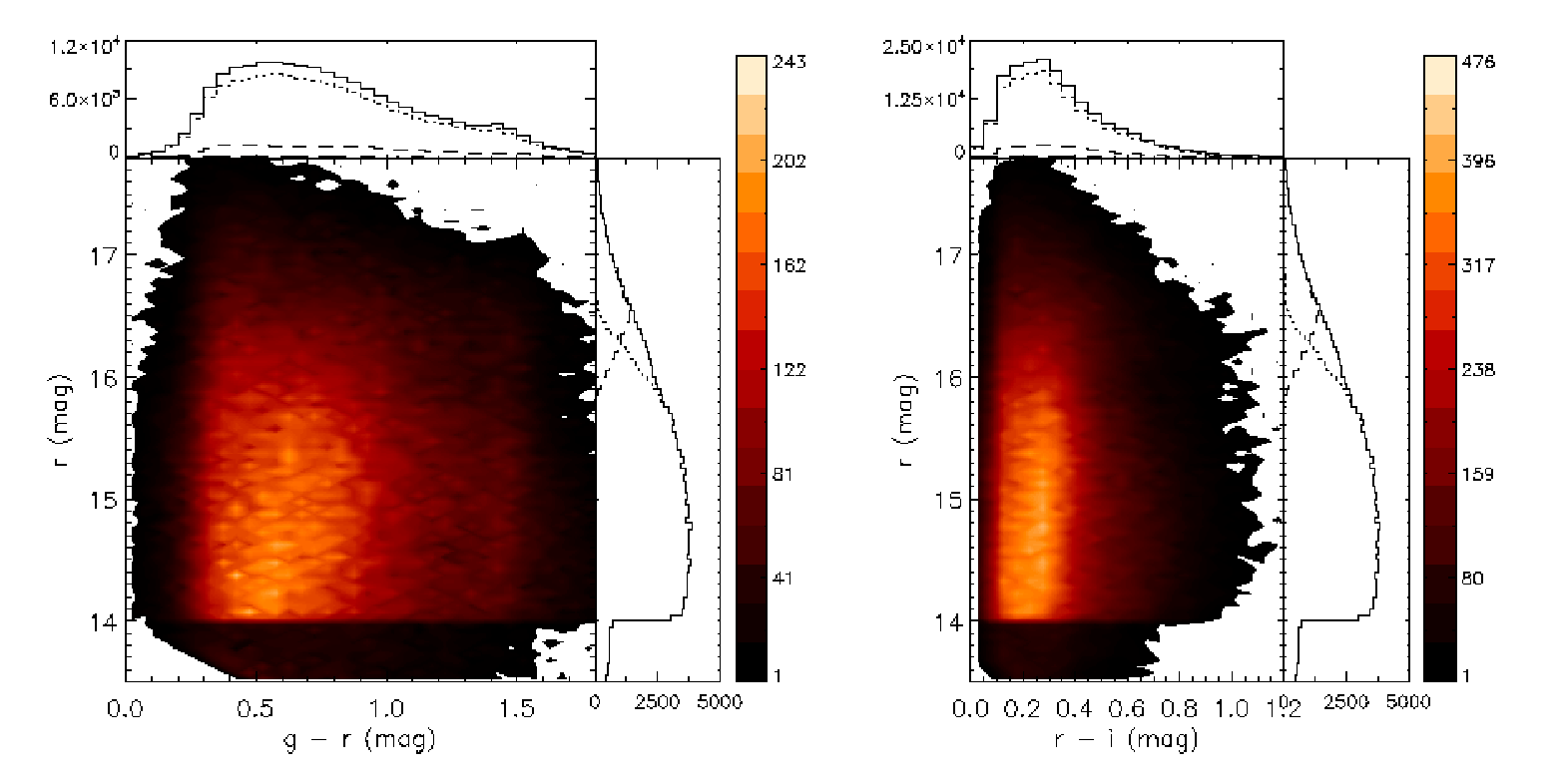}
\caption{
Same as Fig.\,\ref{GAC_CMD_bluered} but for the LSS-GAC main sample.  
\label{GAC_CMD_blue}}
\end{figure*}

By June 2013, 142 B, 75 M and 20 F plates of the LSS-GAC main survey have been observed,
yielding a total of 727,478 spectra of 536,602 unique targets recorded in 237 plates. 
Amongst them, 225,522 spectra of 189,042 unique targets have S/N(4650\AA)  higher than 10, 
and they are defined the LSS-GAC main sample. 
The spatial distribution of the main sample is shown in Fig.\,\ref{spatial_gac_b}.
Most are from the B plates, with the remaining 12 per cent from the M plates and only a few hundred spectra from the F plates.
About 84.5, 12.6, 2.2, 0.5 and 0.1 per cent targets were observed by 1 to 5 times, respectively.
As shown in Fig.\,\ref{multi_epoch_spatial_gac_b}, 
the duplicate observations come from targets in the overlapping areas of adjacent plates of the same category and group
as well as from disqualified plates that get re-observed.
The distributions of the 189,042 unique targets in ($r$, $g-r$) and ($r$, $r-i$) Hess diagrams are shown in Fig.\,\ref{GAC_CMD_blue}.
Compared to the roughly flat but slightly rising distribution in $r$ magnitude of the observed sample of each plate category
(Fig.\,\ref{GAC_CMD_observed}), the distribution in $r$ of the main sample is roughly 
flat but decreases slowly between $14 \le r < 15.8$\,mag, and then drops rapidly at fainter magnitude.
Compared to a nearly flat distribution in $g-r$ colour between 0.4 $\leq g-r \leq$ 1.5\,mag 
for the observed sample (Fig.\,\ref{GAC_CMD_observed}), 
the distribution of the main sample keeps roughly flat between $0.4 < g-r < 0.9$\,mag, but then drops at redder colours.
Compared to the observed sample, the main sample is biased to blue and bright stars.  
Most stars in the main sample are brighter than $g=17.5$\,mag and bluer than $g-r$ = 1.2\,mag and $r-i$ = 0.5\,mag. 
The M-dwarf sequence is almost invisible in Fig.\,\ref{GAC_CMD_blue} because M-dwarfs are red and faint, and thus their spectra 
do not have good SNRs in the blue in general. 
Note the small number of targets brighter than $r=14$\,mag or fainter than $r=18.5$\,mag were observed in 
the early stage of the Pilot Surveys between October 2011 -- November 2011, 
when the bright and faint limits were set 14 $\le$ $g$, $r$ or $i$ $\le$ 19\,mag
instead of $14 \le r < 18.5$\,mag.

\begin{figure*}\includegraphics[width=180mm]{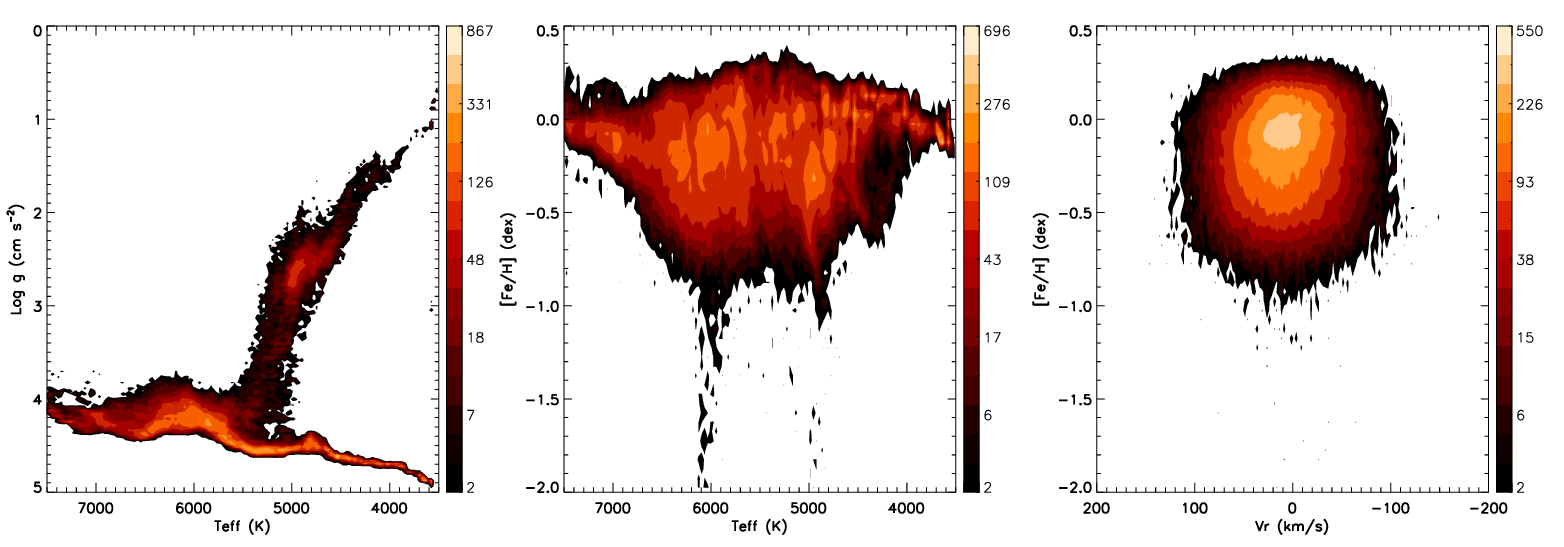}
\caption{
Contour distributions of stellar number density in the  \teff~-- \logg~(left), \teff~-- \feh~(middle) 
and $V_{\rm r}$ --  \feh~(right) planes for the LSS-GAC main sample.
The bin sizes are (30\,K, 0.03 dex), (30\,K, 0.05 dex) and (2.5 \kms, 0.05 dex) in the left, middle and right panels, respectively.
Duplicate targets have been excluded. Colour bars are over-plotted. 
\label{GAC_HR}}
\end{figure*}

Fig.\,\ref{GAC_HR} shows contour distributions of the main sample in the \teff~-- \logg, \teff~-- \feh~ and 
$V_{\rm r}$ -- \feh~planes. The stellar parameters are from the LSP3. 
About 15 per cent of the sample are giants [\logg~$\le$ 3.5\,(cm\,s$^{-2}$)],  
about two thirds of which are possibly RC stars.  
Among the dwarfs [\logg~$>$ 3.5\,(cm\,s$^{-2}$) ], 16.2, 29.5, 37.9, 13.4 and 3.0 per cent of the stars fall in the  
\teff~range of $>$ 7,000\,K, 6,000 -- 7,000\,K, 5,000 -- 6,000\,K, 4,000 -- 5,000\,K and $<$ 4,000\,K,  respectively. 
Most of the giants and stars hotter than 7,500\,K in the main sample are from the Galactic disc and intrinsically luminous, 
and their fractions increase at low Galactic latitudes.
Most stars are metal-rich, only 1.5 per cent of them have \feh~values lower than $-$1.0 dex and 11.7 per cent lower than $-$0.5 dex, 
suggesting that the sample consists mainly of thin disc stars, plus a small fraction of thick disc stars and only few halo stars.  
This result is consistent with the small dispersions of radial velocities of the sample. 
There are very few metal-poor (\feh~$<$ $-$0.5 dex) A- and M-type main sequence stars in the sample. 
One reason is that the MILES library contains few such stars, and as a consequence, the LSP3 
is unable to determine the atmospheric parameters for them. But more importantly, one does not expect many metal-poor main sequence A stars 
given their young ages, nor a lot of metal-poor M\,dwarfs given that they can only be observed from a very small local volume.

\begin{figure}\includegraphics[width=90mm]{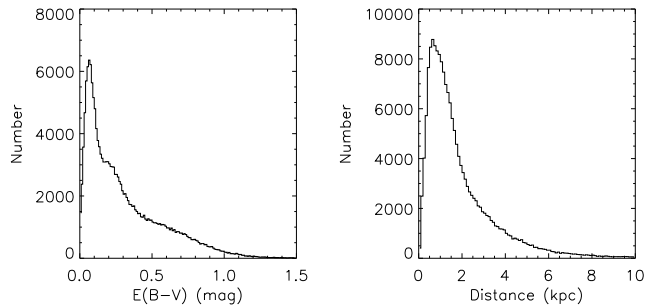}
\caption{
Histograms of values of \ebv~(left) and distances (right) 
for the LSS-GAC main sample. 
The bin size is 0.01\,mag in the left panel and 0.1\,kpc in the right panel.
Duplicate targets have been excluded. 
\label{GAC_ebv_dist}}

\end{figure}

\begin{figure*}\includegraphics[width=180mm]{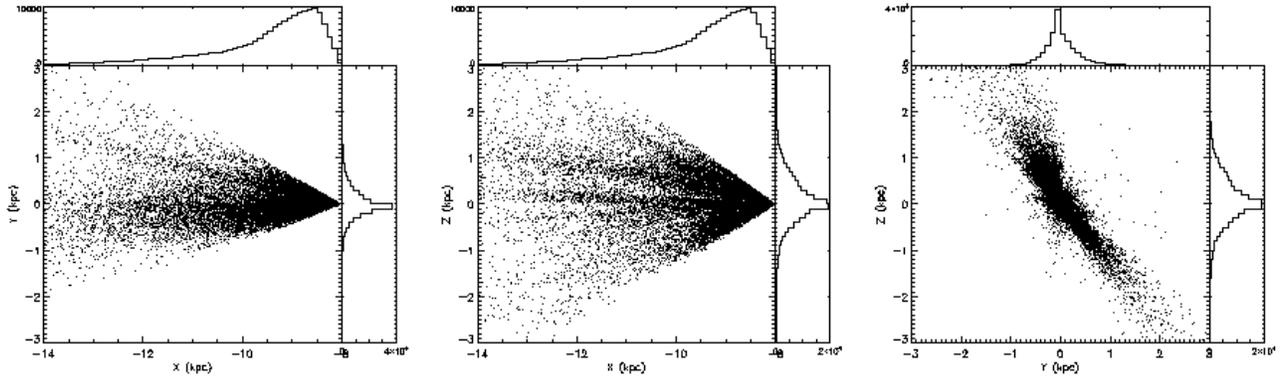}
\caption{
From left to right, distributions of the LSS-GAC main sample in the Galactic 
(X, Y), (X, Z) and (Y, Z) planes.
The Sun is located at (X, Y, Z) = ($-$8.0, 0.0, 0.0)\,kpc. 
Duplicate targets have been excluded. 
To avoid overcrowding, only one-in-ten stars are shown.
Histogram distributions along the X, Y and Z axes are also plotted.
\label{GAC_xyz}}
\end{figure*}

\begin{figure}\includegraphics[width=90mm]{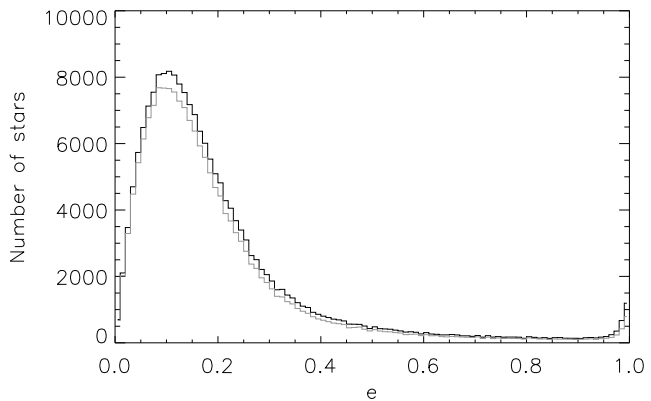}
\caption{
Histograms of eccentricities for the LSS-GAC main sample.
The black and grey lines denote values calculated using proper motions from the PPMXL and the UCAC4 catalogues, respectively.
The bin size is 0.01. 
Duplicate targets and targets without proper motion measurements have been excluded.
\label{eccentricity_gac}
}
\end{figure}

Fig.\,\ref{GAC_ebv_dist} shows histogram distributions of values of \ebv~and distances for the main sample.
The values of \ebv~ are those given by the star pair method and the distances are deduced 
using the empirical relations between absolute magnitudes and stellar atmospheric parameters.
About 44.6, 25.4 and 2.3 per cent stars have \ebv~values higher than 0.3, 0.5 and 1.0\,mag, respectively.
Because most targets in the current LSS-GAC main sample are brighter than $r = 16.5$\,mag, 
the volume that they probe is smaller than designed.
The median distance of the sample is about 1.5 kpc. Only 38.9, 14.8 and 2.7 per cent sample stars 
have distances larger than 2.0, 4.0 and 8.0 kpc, respectively. Most distant targets are giants.
The spatial distributions of the sample in the Galactic (X, Y), (X, Z) and (Y, Z) planes are shown in Fig.\,\ref{GAC_xyz},
assuming that the Sun is located at (X, Y, Z) = ($-$8.0, 0.0, 0.0)\,kpc.
Most stars are from the disk, consistent with the kinematics and chemistry as shown in Fig.\,\ref{GAC_HR}.
Fig.\,\ref{eccentricity_gac} shows histogram distributions of eccentricities for the main sample.
The values calculated using proper motions from the PPMXL and UCAC4 catalogues agree with each other.
The distributions peak at $e = 0.1$ and have median values of 0.15. 
Note that a small fraction of targets have eccentricities close to unity and caused by large 
uncertainties in proper motion measurements or distance estimates.

\subsection{The M\,31-M\,33 sample}

\begin{figure}\includegraphics[width=90mm]{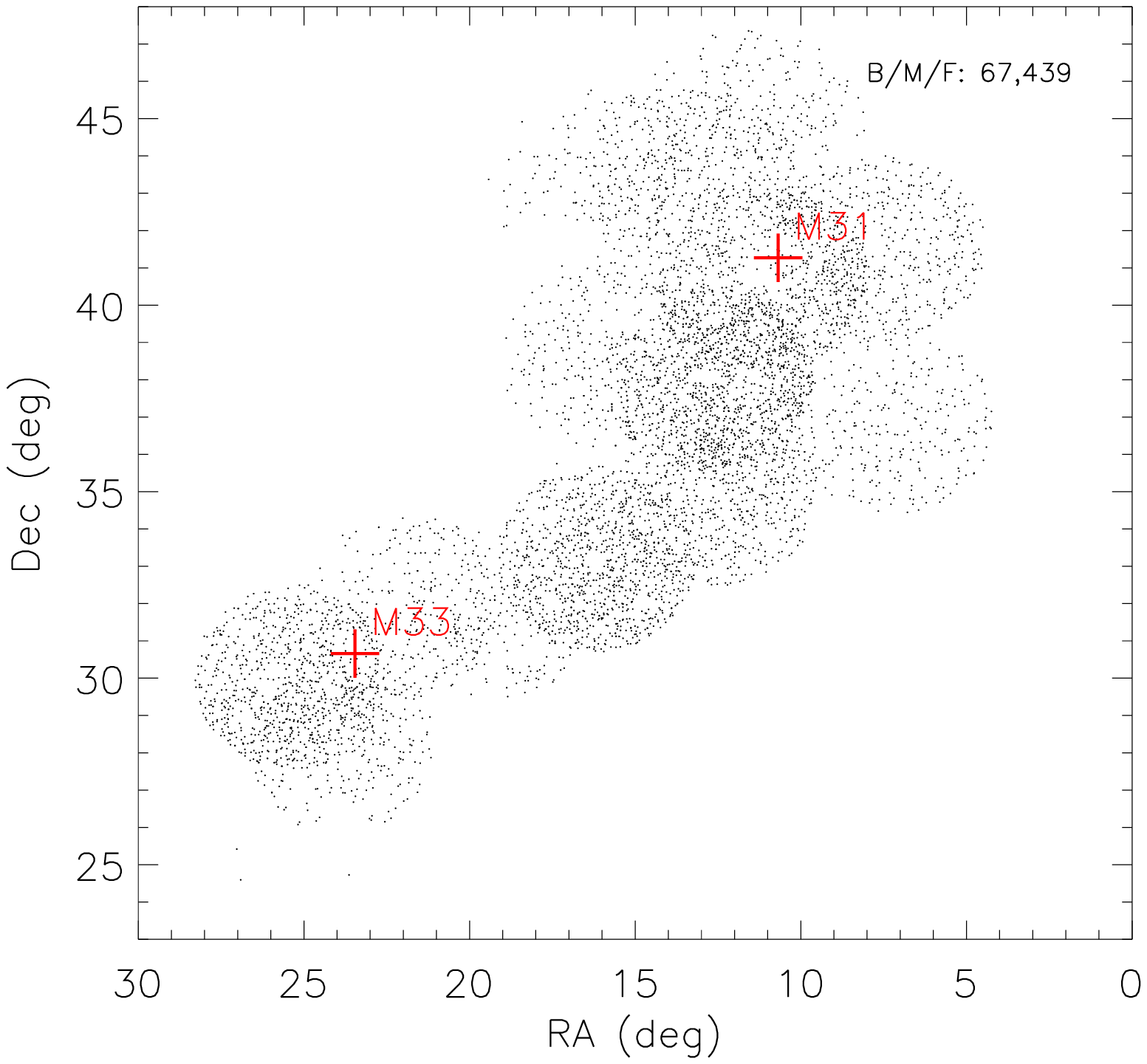}
\caption{
Same as Fig.\,\ref{spatial_m31_bluered} but for the M\,31-M\,33 sample.
\label{spatial_m31_blue}}
\end{figure}

\begin{figure*}\includegraphics[width=180mm]{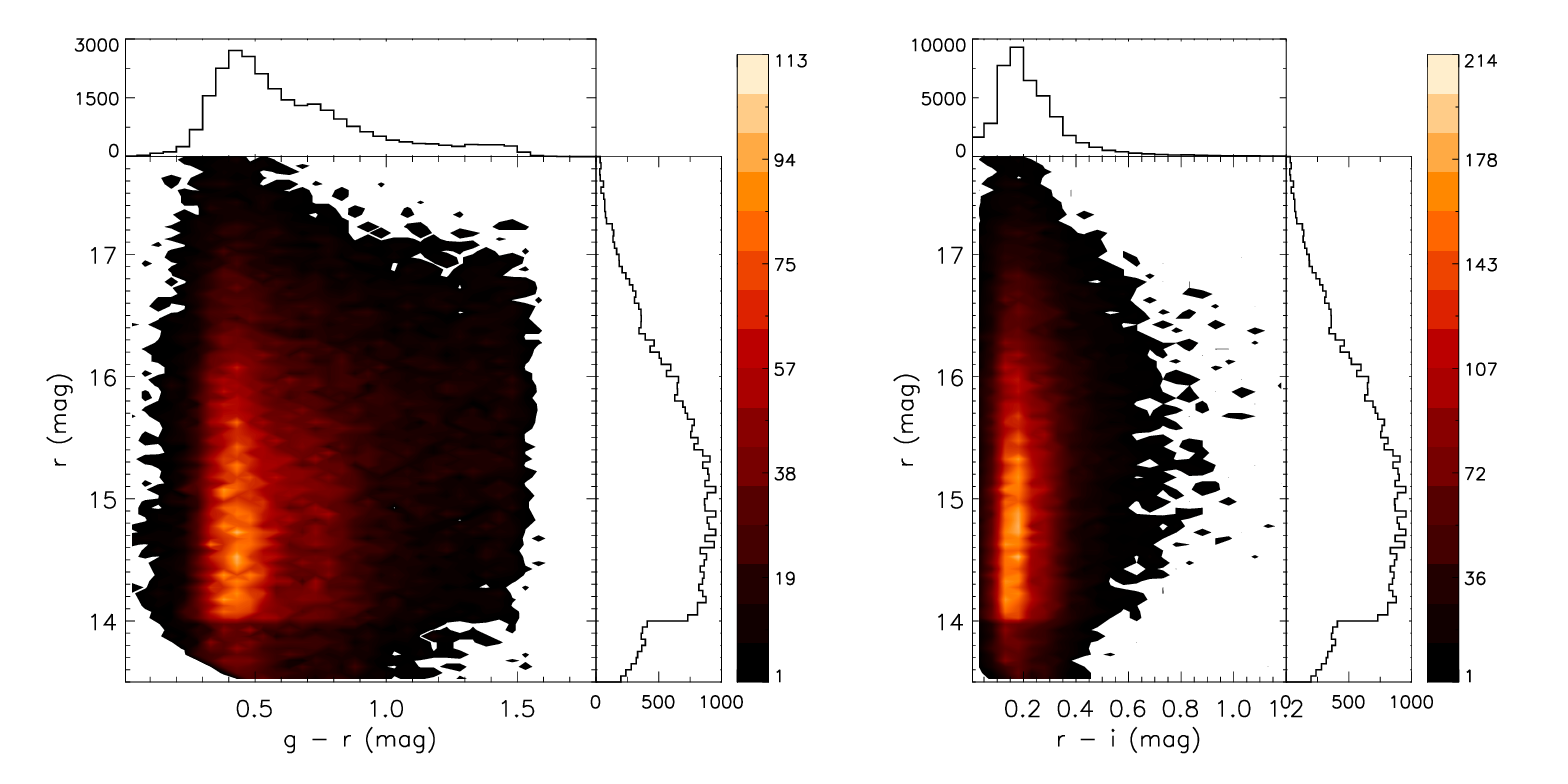}
\caption{
Same as Fig.\,\ref{M31_CMD_bluered} but for the M\,31-M\,33 sample. 
\label{M31_CMD_blue}}
\end{figure*}

By June 2013, 87 plates have been observed in the M\,31-M\,33 fields as parts of the LSS-GAC, 
yielding a total of 265,745 spectra of 154,319 unique stars.
Amongst them, 67,439 spectra of 46,501 unique targets have S/N(4650\AA)  higher than 10
and they are defined as the M\,31-M\,33 sample. 
Their spatial distribution is shown in Fig.\,\ref{spatial_m31_blue}.
About 71.0, 18.9, 6.5, 2.4, 0.8, 0.3 per cent of them are observed 1 -- 6 times, respectively.
Note that compared to the LSS-GAC main sample, 
the fraction of duplicate stars in the M\,31-M\,33 sample are about two times higher.
The distributions in ($r$, $g-r$) and ($r$, $r-i$) Hess diagrams are shown in Fig.\,\ref{M31_CMD_blue}.
As in the case of the main sample, 
compared to the targeted sample (Fig.\,\ref{M31_CMD_observed}),
the current M31-M33 sample that meet the S/N requirement in blue, is 
biased to bluer and brighter stars.
As in the case of the main sample, the M-dwarf sequence is hardly seen in Fig.\,\ref{M31_CMD_blue} either.
As for the LSS-GAC main sample, the sample includes a few stars brighter than $r=14$\,mag or 
fainter than $r=18.5$\,mag observed during the early stage of the Pilot Surveys.

\begin{figure*}\includegraphics[width=180mm]{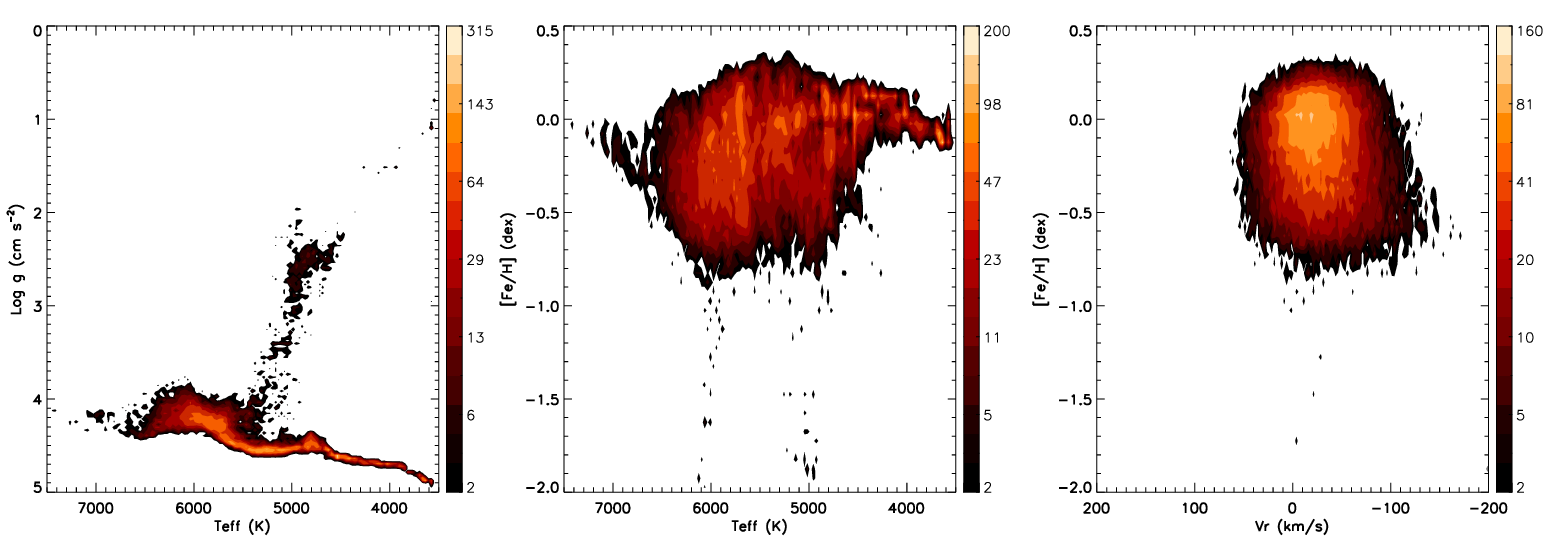}
\caption{
Same as Fig.\,\ref{GAC_HR} but for the M\,31-M\,33 sample.
\label{M31_HR}}
\end{figure*}

The distributions of the M31-M33 sample in the \teff~-- \logg, \teff~-- \feh~ and $V_{\rm r}$ -- \feh~
planes are shown in Fig.\,\ref{M31_HR}. 
Only 8.5 per of the stars are giants. 
Among the dwarfs, 1.4, 18.2, 56.0, 20.8 and 3.6
per cent fall the temperature range of $>$ 7,000\,K, 6,000 -- 7,000\,K, 5,000 -- 6,000\,K, 4,000 -- 5,000\,K and $<$ 4,000\,K,  respectively.
Due to the relatively high Galactic latitudes of the  M\,31-M\,33 fields,
compared to the LSS-GAC main sample, the M\,31-M\,33 sample contains a smaller fraction of giants 
and hot stars but a larger fraction of FGK dwarfs.
The sample is also  metal-rich, with only 2.8 per cent of the stars having a \feh~value lower than $-$1.0 dex and 14.6 per cent lower than $-$0.5 dex.
This is consistent with the small dispersions of radial velocities of the sample.
Note that due to the Galactic rotation, most of the M\,31-M\,33 sample stars have a blue-shifted radial velocity.

\begin{figure}\includegraphics[width=90mm]{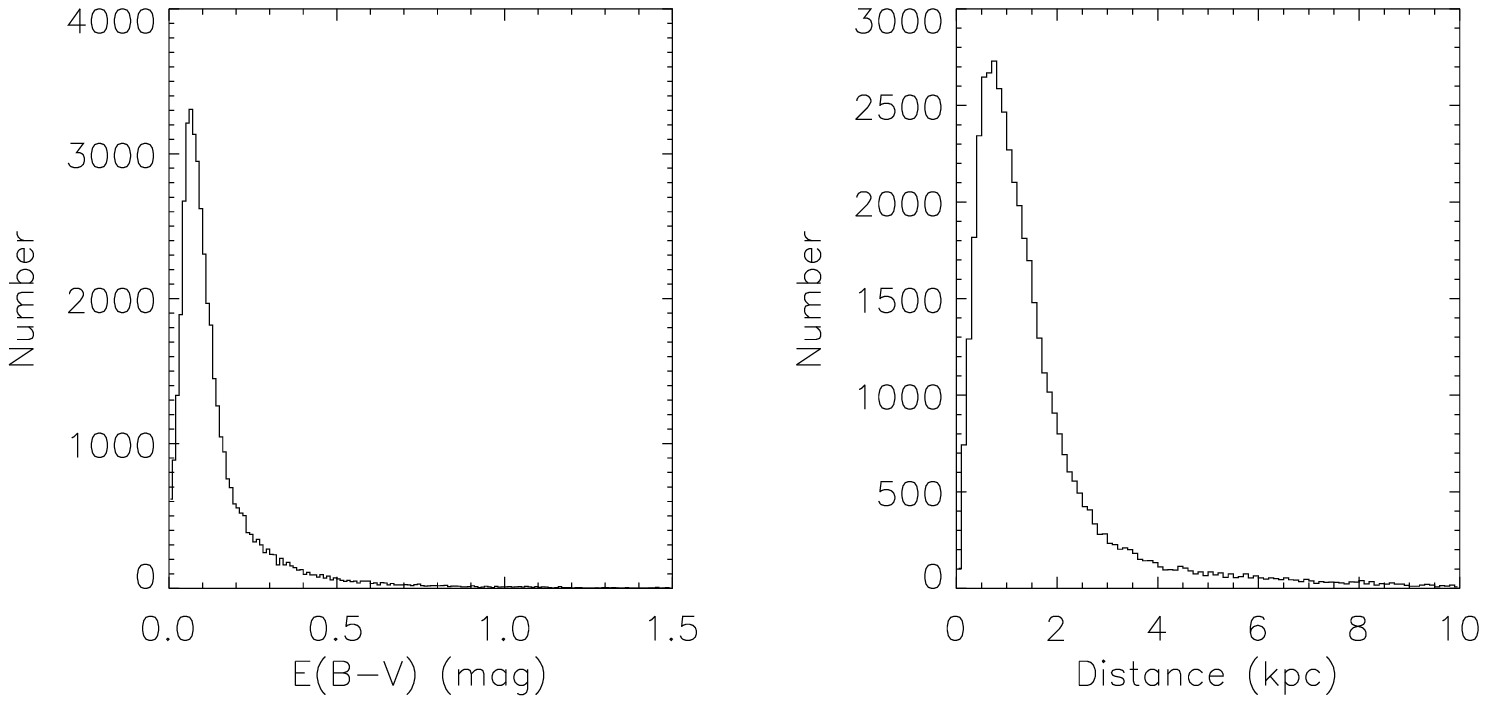}
\caption{
Same as Fig.\,\ref{GAC_ebv_dist} but for the M\,31-M\,33 sample.
\label{M31_ebv_dist}}
\end{figure}

\begin{figure}\includegraphics[width=90mm]{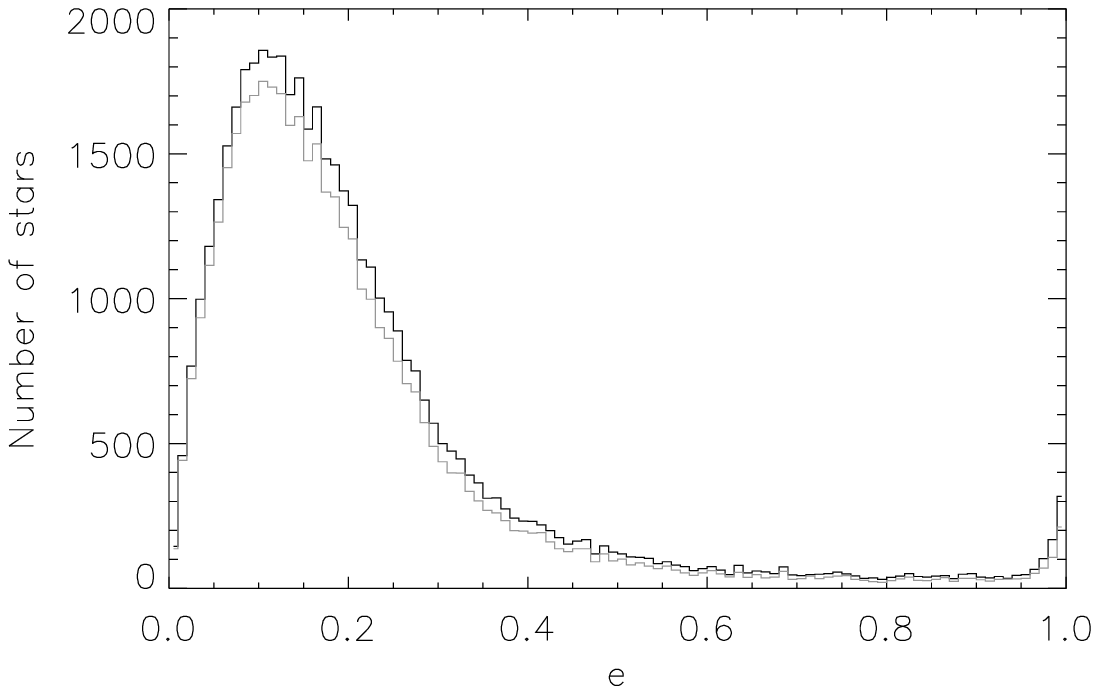}
\caption{
Same as Fig.\,\ref{eccentricity_gac} but for the M\,31-M\,33 sample.
\label{eccentricity_m31}}
\end{figure}

Fig.\,\ref{M31_ebv_dist} shows distributions of values of \ebv~and distances for the M\,31-M\,33 sample.
The values adopted are as for the LSS-GAC main sample.
Most of the stars have \ebv~values lower than 0.3 mag. The median value is 0.093 mag, consistent with the SFD extinction map.
The median distance of the M\,31-M\,33 sample is about 1.2 kpc, slightly smaller than the main sample. 
Only 23.7, 8.9 and 3.1 per cent targets have distances larger than 2.0, 4.0 and 8.0 kpc, respectively. 
Spatial distributions of the M\,31-M\,33 sample in the (X, Y), (X, Z) and (Y, Z) planes are displayed in Fig.\,\ref{GAC_xyz}.
Similar to the main sample, most stars in the M\,31-M\,33 sample are 
from the Galactic disk.
Fig.\,\ref{eccentricity_m31} shows histogram distributions of eccentricities for the M\,31-M\,33 sample.
The distributions peak at $e = 0.11$ and have median values of 0.16. 
Again, a small fraction of targets have false values close to unity.

\begin{figure*}\includegraphics[width=180mm]{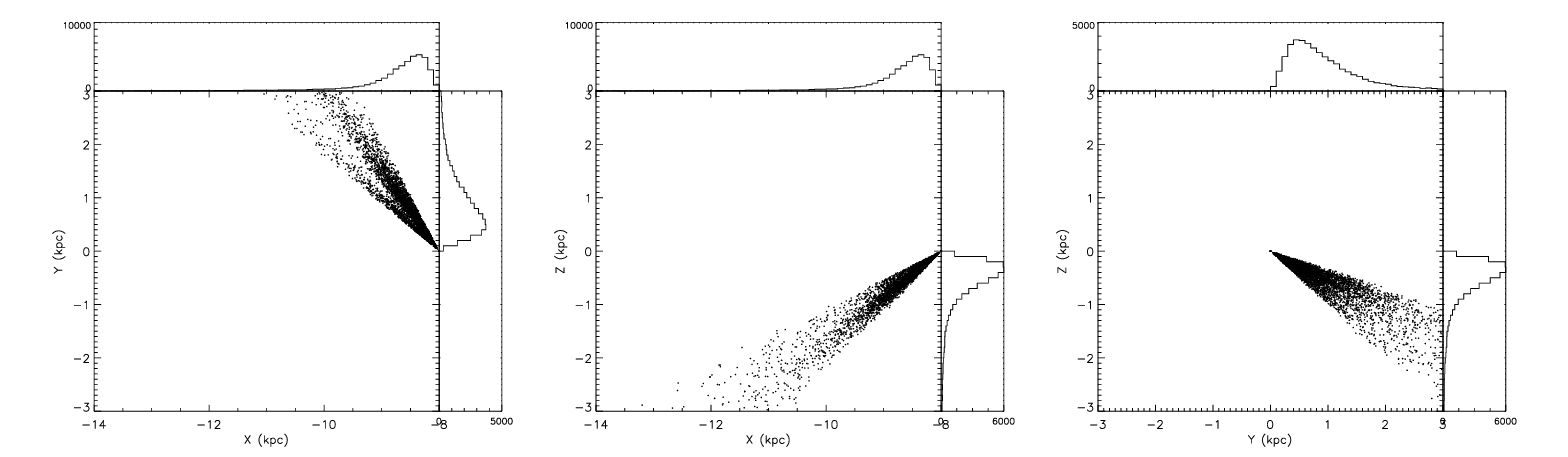}
\caption{
Same as Fig.\,\ref{GAC_xyz} but for the M\,31-M\,33 sample.
\label{M31_xyz}}
\end{figure*}

\subsection{The VB sample}
\begin{figure*}\includegraphics[width=120mm]{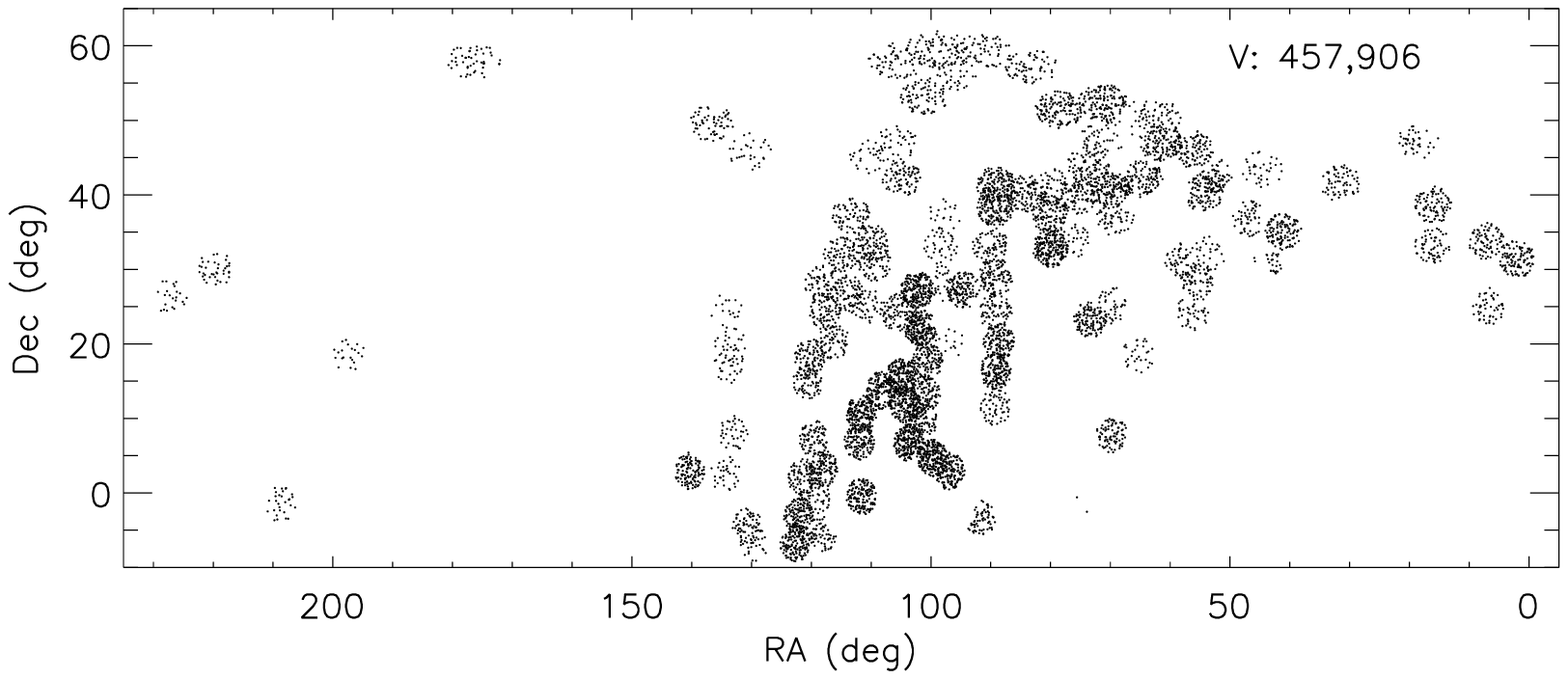}
\caption{
Same as Fig.\,\ref{spatial_vb_bluered} for the VB sample.
\label{spatial_vb_blue}}
\end{figure*}

\begin{figure}\includegraphics[width=90mm]{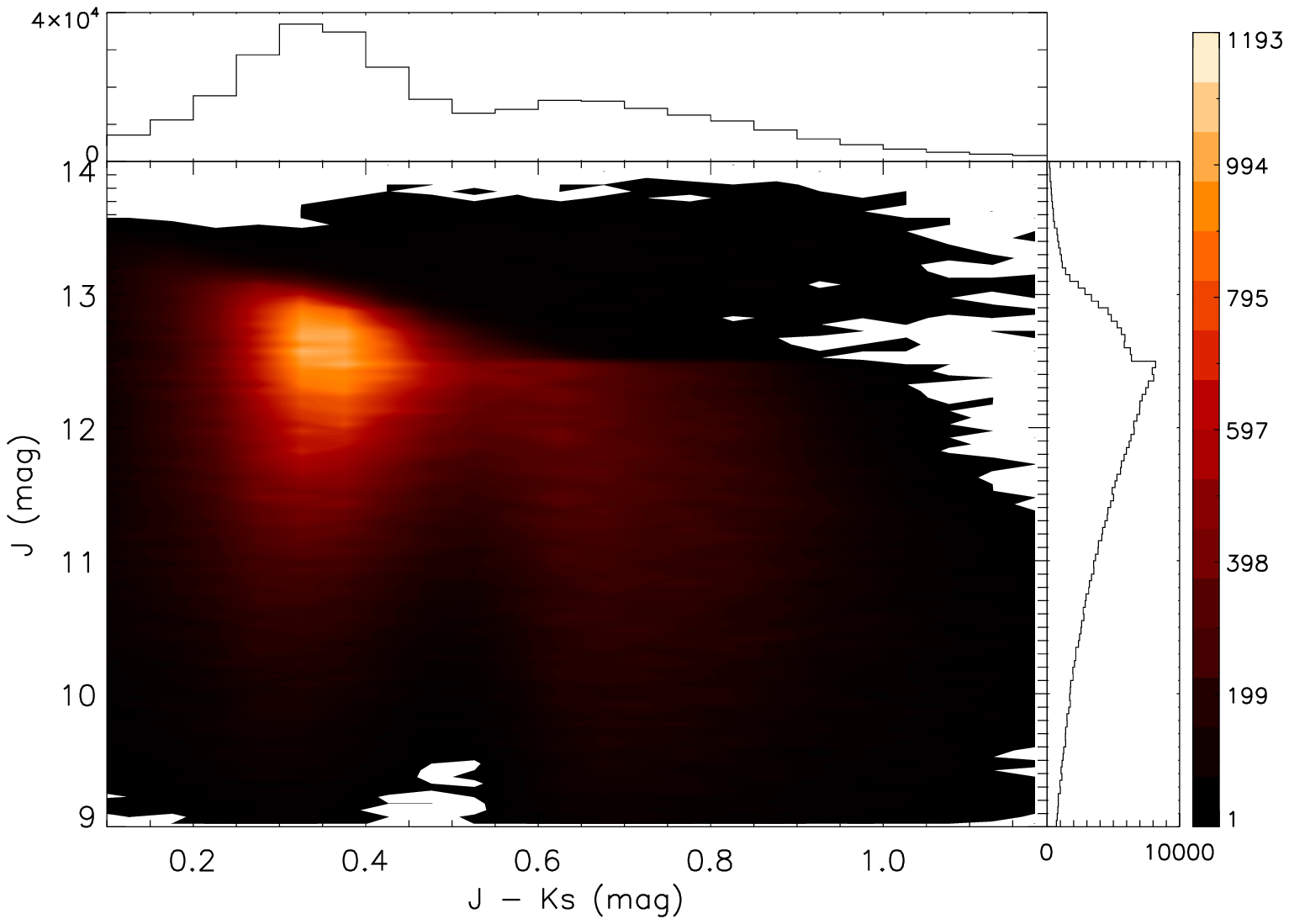}
\caption{
Same as Fig.\,\ref{VB_CMD_bluered} for the VB sample.
\label{VB_CMD_blue}}
\end{figure}

Observations of VB plates began on January 5, 2012.
By June 2013, 259 VB plates have been observed, yielding a total number of  
791,530 spectra of 638,836 unique stars. Amongst them, 
457,906 spectra of 385,672 unique targets have S/N(4650\AA)  higher than 10,
and they consist the current VB sample. 
Their spatial distribution is shown in Fig.\,\ref{spatial_vb_blue}.
Most of them are in the GAC direction.
About 84.5, 12.8, 2.1, 0.4 and 0.1 per cent targets are observed by 1 to 5 times, respectively.
Unlike those duplicate targets in the main and M\,31-M\,33 samples, most duplicate 
observations of the VB sample are taken in the same night. 
The distribution of the VB sample in ($J$, $J-K_{\rm s}$) Hess diagram is shown in Fig.\,\ref{VB_CMD_blue}.
Compared to the sample targeted by the VB survey (Fig.\,\ref{VB_CMD_observed}), the 
actual sample that meets the S/N requirement slightly biases 
to blue and bright stars. The distributions in 
$J-K_{\rm s}$ and $J$ reflect well what have been designed except that 
the numbers of stars are less by 40 per cent.
Note the branch of red giants is clearly visible.

\begin{figure*}\includegraphics[width=180mm]{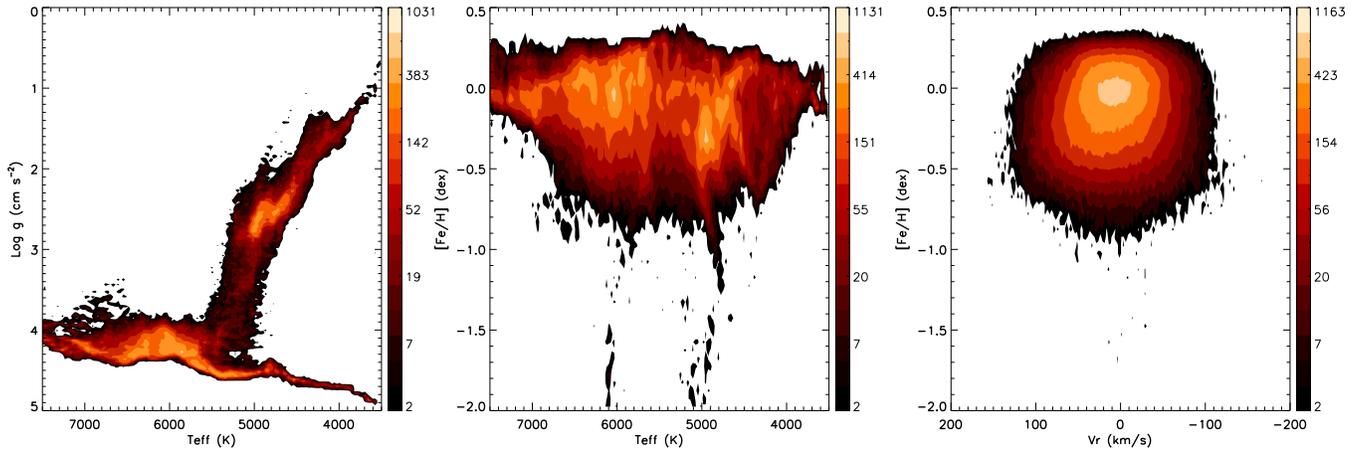}
\caption{
Same as Fig.\,\ref{GAC_HR} but for the VB sample.
\label{VB_HR}}
\end{figure*}

Fig.\,\ref{VB_HR} shows the distributions of the VB sample in the \teff~-- \logg, \teff~-- \feh~and 
$V_{\rm r}$ -- \feh~planes. 
Given the brightness of the VB sample, 
a large 27.1 per cent of the sample are giants, twice that of the main sample. 
About 70 per cent of the giants are possibly RC stars. 
Due to the same reason, the VB sample contains more hot dwarfs.
About 17.8, 39.0, 38.0, 4.5 and 0.7 per cent dwarfs fall in the \teff~range of
$>$ 7,000\,K, 6,000 -- 7,000\,K, 5,000 -- 6,000\,K, 4,000 -- 5,000\,K and $<$ 4,000\,K,  respectively.
The sample is also metal-rich.  About 1.3 per cent of the stars have \feh~values lower 
than $-$1.0 dex and 6.7 per cent lower than $-$0.5 dex.

\begin{figure}\includegraphics[width=90mm]{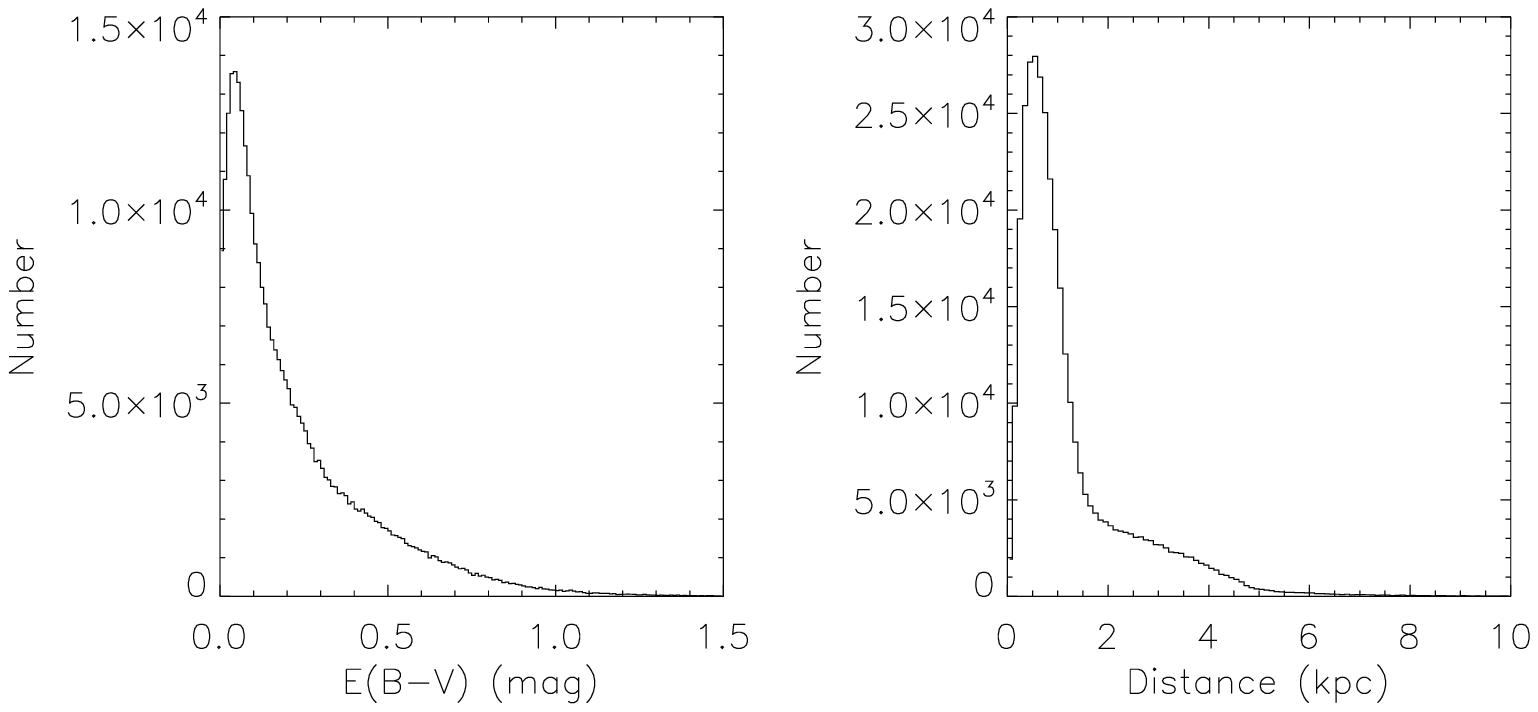}
\caption{
Same as Fig.\,\ref{GAC_ebv_dist} but for the VB sample.
\label{VB_ebv_dist}}
\end{figure}

\begin{figure*}\includegraphics[width=180mm]{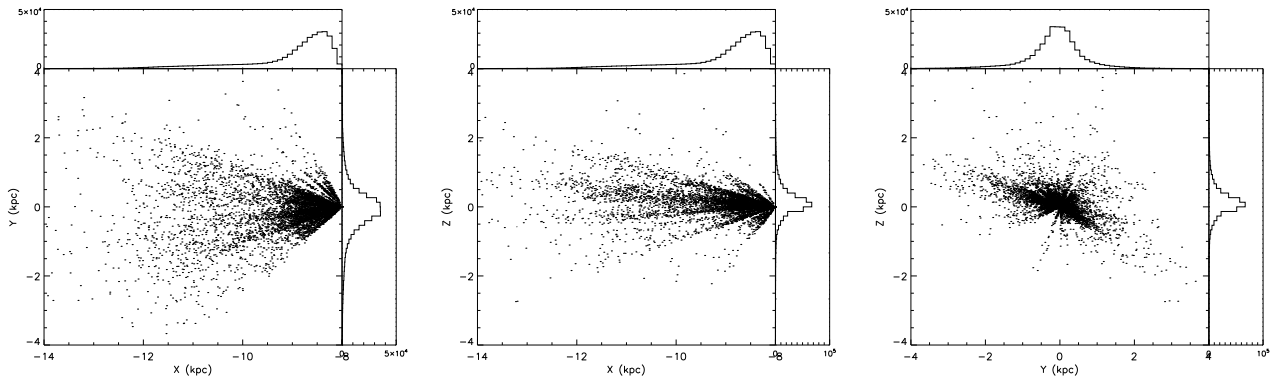}
\caption{
Same as Fig.\,\ref{GAC_xyz} but for the VB sample.
To avoid overcrowding, only one-in-fifty stars are shown.
\label{VB_xyz}}
\end{figure*}

\begin{figure}\includegraphics[width=90mm]{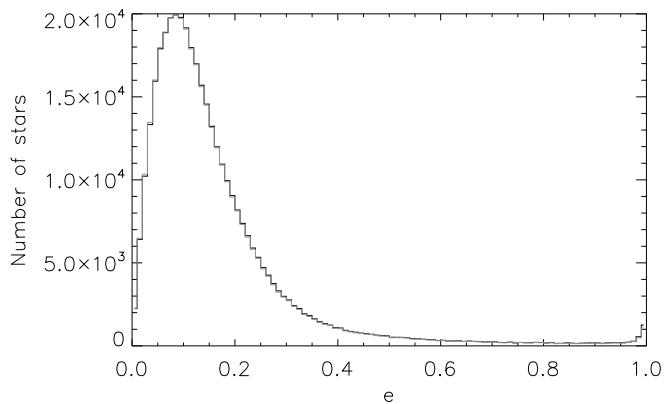}
\caption{
Same as Fig.\,\ref{eccentricity_gac} but for the VB sample.
\label{eccentricity_vb}}
\end{figure}

Fig.\,\ref{VB_ebv_dist} shows histogram distributions of \ebv~and distances for the VB sample.
The median value of \ebv~ is 0.15 mag.
About 28.1, 13.1 and 1.4 per cent stars have \ebv~values higher than 0.3, 0.5 and 1.0 mag, respectively.
The median distance of the VB sample is about 900 pc. Only 20.6, 4.1 and 0.3 per cent targets
have distances larger than 2.0, 4.0 and 8.0 kpc, respectively. 
Spatial distributions of the VB sample in the (X, Y), (X, Z) and (Y, Z) planes are displayed in Fig.\,\ref{VB_xyz}.
Most stars are from the Galactic thin disc ($|Z| <$ 0.5 kpc), consistent with their
kinematics and chemistry as shown in Fig.\,\ref{VB_HR}.
Fig.\,\ref{eccentricity_vb} shows histogram distributions of eccentricities for the VB sample.
The distributions peak at $e = 0.08$ and have median values of 0.12, smaller that those 
for the main sample. Again, a small fraction of targets show flase values close to unity.

\subsection{Value-added catalogues}
The first release of LSS-GAC value-added catalogues consist of three binary tables in FITS format, 
i.e., LSS-GAC\_DR1\_main.fits, LSS-GAC\_DR1\_M31.fits and LSS-GAC\_DR1\_VB.fits, 
corresponding respectively to the main, the M\,31-M\,33 and the VB samples described above.
The catalogues include basic information of the LAMOST observations, 
stellar atmospheric parameters and corresponding errors and flags yielded by the LSP3,
multi-band photometric data from the far-UV to the  mid-IR collected from the GALEX, XSTPS-GAC, 2MASS and WISE surveys,
values of the interstellar reddening and distances determined with the methods described in Sections 5 and 6, 
proper motions collected from the PPMXL and UCAC4 catalogues after correcting for the systematic errors, 
proper motions newly derived by comparing the XSTPS-GAC and 2MASS position measurements, as well as orbital parameters of sample stars. 
The catalogues are accessible from http://162.105.156.249/site/LSS-GAC-dr1/ along 
with a descriptive readme.txt file.
Table\,\ref{value-added} gives a detailed description of the content of catalogues.
An example entry from the catalogue  of the LSS-GAC main sample is shown in Table\,\ref{value-added3}.

\begin{table*}{}
\small
\centering
\begin{minipage}{180mm}
\caption{Description of the first release of LSS-GAC value-added catalogues.
\label{value-added} } \end{minipage}
\tabcolsep 1mm
\begin{tabular}{lll}
  \hline\noalign{\smallskip}
Col.     &  Name  & Description   \\
 \hline\noalign{\smallskip}
1  & spec\_id &  LAMOST unique spectral ID, in format of date-plateid-spectrographid-fibreid   \\
2  & date     &  Date of observation \\
3  & plate    &  LAMOST plate ID, not necessarily unique \\
4  & sp\_id   &  LAMOST spectrograph ID, ranging from 1 to 16\\
5  & fibre\_id & LAMOST fibre ID for a given spectrograph, ranging from 1 to 250 \\
6  & platetype & Plate type (`B', `M', `F' and `V') \\
7  & objid    & Object ID in the input catalogues \\
8  & objtype  & Initial object type from the survey input catalogues, could be stars (`star' and `Star'), \\
   &          & emission line objects (`EMOs') and young stellar objects (`yso') \\  
9  & snr\_b   & S/N(4650\AA) per pixel \\ 
10 & snr\_r   & S/N(7450\AA) per pixel \\
11 & ra       & Right Ascension of J2000.0 (deg) \\
12 & dec      & Declination of J2000.0 (deg) \\
13 & l        & Galactic longitude (deg) \\
14 & b        & Galactic latitude (deg) \\
15 & uqflag   & Uniqueness flag. If the target has been observed $n$ times, then uqflag runs from 1 to $n$, \\ 
   &          & with 1 denoting the spectrum of the highest snr\_b \\ 
16 & vr      & Radial velocity yielded by the LSP3, after corrected for a systematic offset of $-$3.5 \kms~(\kms) \\
17 & teff1  &  \teff~yielded by the LSP3 from the weighted mean algorithm (K) \\ 
18 & logg1  &  \logg~yielded by the LSP3 from the weighted mean algorithm (cm s$^{-2}$) \\
19 & feh1  &  \feh~yielded by the LSP3 from the weighted mean algorithm (dex) \\
20 & teff2  &  \teff~yielded by the LSP3 from the $\chi^2$ minimization algorithm (K) \\ 
21 & logg2  &  \logg~yielded by the LSP3 from the $\chi^2$ minimization algorithm (cm s$^{-2}$) \\
22 & feh2   &  \feh~yielded by the LSP3 from the $\chi^2$ minimization algorithm (dex) \\
23 & teff      & Final \teff~yielded by the LSP3 from the weighted mean algorithm, calibrated to photometric temperatures (K) \\
24 & logg      & Final \logg~ yielded by the LSP3 from the weighted mean algorithm, same to logg1 in the current version (cm s$^{-2}$) \\
25 & feh     & Final \feh~yielded by the LSP3 from the weighted mean algorithm, same to feh1 in the current version (dex) \\
26 & vr\_err &  Error of radial velocity yielded by the LSP3 (\kms) \\
27 & teff\_err & Error of the final \teff~yielded by the LSP3 (K) \\
28 & logg\_err & Error of the final \logg~yielded by the LSP3 (cm s$^{-2}$) \\
29 & feh\_err  & Error of the final \feh~yielded by the LSP3 (dex) \\
30 & flags[9]  & Flags assigned to the final LSP3 parameters$^a$ \\
31 & fuv    & GALEX $FUV$ band magnitude (mag)\\ 
32 & nuv    & GALEX $NUV$ band magnitude (mag)\\
33 & g     & XSTPS-GAC $g$ band magnitude (mag) \\
34 & r     & XSTPS-GAC $r$ band magnitude (mag)\\
35 & i     & XSTPS-GAC $i$ band magnitude (mag)\\
36 & j     & 2MASS $J$ band magnitude (mag)\\
37 & h     & 2MASS $H$ band magnitude (mag)\\
38 & k     & 2MASS $K_{\rm s}$ band magnitude (mag)\\
39 & w1    & WISE $W1$ band magnitude (mag)\\
40 & w2    & WISE $W2$ band magnitude (mag)\\
41 & w3    & WISE $W3$ band magnitude (mag)\\
42 & w4    & WISE $W4$ band magnitude (mag)\\
43 & err\_fuv    & Error of the GALEX $FUV$ magnitude (mag)\\ 
44 & err\_nuv    & Error of the GALEX $NUV$ magnitude (mag)\\
45 & err\_g     & Error of the XSTPS-GAC $g$ band magnitude (mag)\\
46 & err\_r     & Error of the XSTPS-GAC $r$ band magnitude (mag)\\
47 & err\_i     & Error of the XSTPS-GAC $i$ band magnitude (mag)\\
48 & err\_j     & Error of the 2MASS $J$ band magnitude (mag)\\
49 & err\_h     & Error of the 2MASS $H$ band magnitude (mag)\\
50 & err\_k     & Error of the 2MASS $K_{\rm s}$ band magnitude (mag)\\
51 & err\_w1    & Error of the WISE $W1$ band magnitude (mag)\\
52 & err\_w2    & Error of the WISE $W2$ band magnitude (mag)\\
53 & err\_w3    & Error of the WISE $W3$ band magnitude (mag)\\
54 & err\_w4    & Error of the WISE $W4$ band magnitude (mag)\\
55 & ph\_qual\_2mass & 2MASS photometric quality flag$^b$ \\ 
56 & ph\_qual\_wise & WISE photometric quality flag$^c$ \\ 
57 & var\_flg\_wise & WISE variability flag$^c$ \\ 
58 & ext\_flg\_wise & WISE extended source flag$^c$ \\ 
59 & cc\_flg\_wise & WISE contamination and confusion flag$^c$\\ 
\noalign{\smallskip}\hline
\end{tabular}
\flushleft
$^{\rm a}$ See Table\,2 of Xiang et al. (2014b). \\
$^{\rm b}$ See http://www.ipac.caltech.edu/2mass/releases/allsky/doc/sec1\_6b.html\#origphot. \\ 
$^{\rm c}$ See http://wise2.ipac.caltech.edu/docs/release/allsky/expsup/sec2\_2a.html\#ext\_flg.  
\end{table*}

\setcounter{table}{5}
\begin{table*}{}
\small
\centering
\begin{minipage}{180mm}
\caption{\it -- continued.
\label{value-added2} } \end{minipage}
\tabcolsep 2mm
\begin{tabular}{lll}
  \hline\noalign{\smallskip}
Col.     &  Name  & Description   \\
 \hline\noalign{\smallskip}
60  & ebv\_sfd &  \ebv~from the SFD extinction map (mag) \\
61  & ebv\_sp & \ebv~ derived from the star pair method (mag) \\
62  & ebv\_mod &  \ebv~derived by comparing the observed and synthetic model atmosphere colours (mag) \\
63  & ebv\_phot & \ebv~derived by fitting multi-band photometry to the empirical stellar loci (Chen et al. 2014) (mag) \\
64  & dist\_em & Distance derived using the empirical relations of absolute magnitudes as a function of stellar atmospheric \\ 
     &         & parameters as yielded by the MILES stars except for RCs. Distances of RCs are derived assuming they are \\
     &         & standard candles (kpc) \\ 
65  & dist\_mod & Distance derived from the theoretical isochrones of stellar models (kpc) \\
66  & dist\_phot & Distance derived from the photometric parallax method (Chen et al. 2014) (kpc) \\
67 & pmra\_ppmxl & Proper motion $\mu_\alpha\cos\delta$ in RA from the PPMXL catalogues (mas yr$^{-1}$) \\
68 & pmdec\_ppmxl  & Proper motion $\mu_\delta$ in Dec from the PPMXL catalogues (mas yr$^{-1}$) \\
69 & epmra\_ppmxl  & Proper motion error in RA from the PPMXL catalogues (mas yr$^{-1}$) \\
70 & epmdec\_ppmxl  & Proper motion error in Dec from the PPMXL catalogues (mas yr$^{-1}$) \\
71  & pmra\_ppmxl\_corr  & Proper motion $\mu_\alpha\cos\delta$ in RA from the PPMXL catalogues after corrected for systematics \\ 
    &                    & using the formula of Carlin et al. (2013) (mas yr$^{-1}$) \\
72 & pmdec\_ppmxl\_corr  & Proper motion $\mu_\delta$ in Dec from the PPMXL catalogues after corrected for systematics \\
    &                    &using the formula of Carlin et al. (2013) (mas yr$^{-1}$) \\
73 & pmra\_ucac4  & Proper motion $\mu_\alpha\cos\delta$ in RA from the UCAC4 catalogues (mas yr$^{-1}$) \\
74 & pmdec\_ucac4  & Proper motion $\mu_\delta$ in Dec from the UCAC4 catalogues (mas yr$^{-1}$) \\
75 & epmra\_ucac4 & Proper motion error in RA from the UCAC4 catalogues (mas yr$^{-1}$) \\
76 & epmdec\_ucac4 & Proper motion error in Dec from the UCAC4 catalogues (mas yr$^{-1}$) \\
77 & pmra\_ucac4\_corr & Proper motion $\mu_\alpha\cos\delta$ in RA from the UCAC4 catalogues after corrected for systematics \\
    &                    &using the formula of Huang et al. (2014) (mas yr$^{-1}$) \\
78 & pmdec\_ucac4\_corr  & Proper motion in $\mu_\delta$ in Dec from the UCAC4 catalogues after corrected for systematics \\   
 &                    &using the formula of Huang et al. (2014) (mas yr$^{-1}$) \\
79 & pmra\_xuyi2mass  & Proper motion $\mu_\alpha\cos\delta$ in RA derived by comparing the XSTPS-GAC and 2MASS positions (mas yr$^{-1}$) \\
80 & pmdec\_xuyi2mass & Proper motion $\mu_\delta$ in Dec derived by comparing the XSTPS-GAC and 2MASS positions (mas yr$^{-1}$) \\
81 & x                & x coordinate in a  Galactocentric Cartesian reference system, positive towards the Galactic centre (kpc) \\ 
82 & y                & y coordinate in a  Galactocentric Cartesian reference system, positive in the direction of disc rotation (kpc) \\
83 & z                & z coordinate in a  Galactocentric Cartesian reference system, positive towards the North Galactic Pole (kpc) \\ 
84 & u[2]             &  Galactic space velocities in the x direction computed from dist\_em and the corrected PPMXL and UCAC4 \\ 
   &                       &proper motions, respectively, positive towards the Galactic centre (\kms) \\
85 & v[2]            &  Galactic space velocities in the y direction computed from dist\_em and the corrected PPMXL and UCAC4 \\
   &                      &  proper motions, respectively, positive in the direction of Galactic rotation (\kms) \\
86 & w[2] &  Galactic space velocities in the z direction computed from dist\_em and the corrected PPMXL and UCAC4\\
   &           & proper motions, respectively, positive towards the North Galactic Pole (\kms) \\
87 & v\_r[2]             &  Galactic space velocities in the $r$ direction in a Galactocentric cylindrical polar coordinate system, \\ 
   &                       & computed from dist\_em and the corrected PPMXL and UCAC4 proper motions, respectively (\kms) \\
88 & v\_phi[2]             &  Galactic space velocities in the $\phi$ direction in a Galactocentric cylindrical polar coordinate system, \\
   &                       &  computed from dist\_em and the corrected PPMXL and UCAC4 proper motions, respectively, \\
   &                      &  positive in the direction of counter disc rotation (\kms) \\ 
89 & v\_z[2]             &  Galactic space velocities in the $z$ direction in a Galactocentric cylindrical polar coordinate system, \\ 
   &                       & computed from dist\_em and the corrected PPMXL and UCAC4 proper motions, respectively,\\
   &                       & positive towards the North Galactic Pole (\kms) \\
90 & e[2] &  Orbital eccentricities computed from dist\_em and the corrected PPMXL and UCAC4 proper motions, respectively  \\
91 & Rapo[2] & Maximum Galactic radii reached by the orbits from dist\_em and the corrected  PPMXL and UCAC4 \\ 
   &         & proper motions, respectively (kpc) \\ 
92 & Rperi[2] & Minimum Galactic radii reached by the orbits from dist\_em and the corrected  PPMXL and UCAC4 \\ 
   &         & proper motions, respectively (kpc) \\
93 & Rg[2]  & Galactic guiding radii from dist\_em and the corrected  PPMXL and UCAC4 proper motions, respectively (kpc) \\
\noalign{\smallskip}\hline
\end{tabular}
\flushleft
\end{table*}

\begin{table*}{}
\small
\centering
\begin{minipage}{180mm}
\caption{An example entry in the value-added catalogue of the LSS-GAC main sample.
\label{value-added3} } \end{minipage}
\tabcolsep 2mm
\begin{tabular}{lllll}
  \hline\noalign{\smallskip}
spec\_id    &  date  & plate  & sp\_id  & fibre\_id   \\
`20111003-PA09B\_keda1-1-5'  & `20111003' & 'PA09B\_keda1' & 1 & 5 \\ \hline 
platetype    & objid   & objtype & snr\_b & snr\_r   \\
`B'  & `S1174-010845' & `star' & 31.7  &91.3 \\ \hline
ra  & dec  & l  & b & uqflag  \\
64.17234 & 27.47495 & 169.16076 & $-$16.45076 & 1 \\ \hline
vr  & teff1  & logg1 &  feh1 &   teff2  \\
-17.8 & 6102.0 & 4.22 & $-$0.14 &  6189.0 \\\hline
logg2 & feh2& teff  & logg &  feh \\
4.48 & $-$0.06 & 6036.0 & 4.22 & $-$0.14 \\\hline
vr\_err & teff\_err & logg\_err & feh\_err & flags \\
5.2     &107.0    & 0.19 &     0.10 &  [1, 1, 1, 1, 1, 2, 1, 2, 1] \\\hline
 &  & fuv & nuv &  g  \\
 &  & $-$99.00  &$-$99.00 & 14.64 \\\hline
r & i & j &  h &  k \\
13.78 & 13.37 & 12.15 &11.78 & 11.64 \\\hline
w1 & w2 & w3 &  w4 &  err\_fuv \\
11.52 & 11.54 & 11.84  & 9.01 & 0.00 \\\hline
err\_nuv & err\_g &err\_r & err\_i &err\_j \\
0.00 & 0.007 & 0.007  & 0.006 & 0.022 \\\hline
err\_h & err\_k &err\_w1 & err\_w2 &err\_w3 \\
0.022 & 0.023 & 0.023 & 0.023 & 0.253 \\\hline
err\_w4 & ph\_qual\_2mass & ph\_qual\_wise & var\_flg\_wise & ext\_flg\_wise \\
0.00  & `AAA' & `AABU' & `00nn' & `0' \\\hline
cc\_flg\_wise &  ebv\_sfd & ebv\_sp & ebv\_mod & ebv\_phot \\
`0000' & 0.98 & 0.50 & 0.50 & 0.37 \\\hline
dist\_em & dist\_mod & dist\_phot & pmra\_ppmxl & pmdec\_ppmxl \\
0.64   & 0.57    & 0.42 & -0.8 &  3.2 \\\hline
epmra\_ppmxl & epmdec\_ppmxl & pmra\_ppmxl\_corr & pmdec\_ppmxl\_corr & pmra\_ucac4 \\
4.3 & 4.3 & 0.4 & 6.2 & $-$3.6 \\\hline
pmdec\_ucac4  & epmra\_ucac4 & epmdec\_ucac4 & pmra\_ucac4\_corr & pmdec\_ucac4\_corr \\
$-$1.0 & 1.7 & 1.9 & $-$3.8 & 0.6 \\\hline
pmra\_xuyi2mass  &  pmdec\_xuyi2mass & x  & y  & z \\
$-$4.0 & 6.9 & $-$8.62 & 0.12 & $-$0.19 \\\hline
u  & v & w &  v\_r & v\_phi \\
14.9, 20.2 & 10.4, 5.9 & 18.6, $-$1.9 & $-$11.8, $-$17.1 & 230.6, 226.1 \\\hline
v\_z & e & Rapo  & Rperi & Rg \\
18.6, $-$1.9 &0.078, 0.078 & 9.86, 9.65 & 8.43, 8.25  & 9.15, 8.95\\\hline
\end{tabular}
\flushleft
\end{table*}

\section{Summary}
Taking advantage of the 4,000 fibres provided by the LAMOST and tailored to the geographical location of the telescope and the 
seasonal variations of the site weather, 
the LSS-GAC main survey aims to survey a significant volume
of the Galactic thin and thick discs and the halo in a large, contiguous sky area centred on the GAC
($|b| \leq 30^{\circ}$, $150 \leq l \leq 210^{\circ}$), and obtain
$\lambda\lambda$3700 -- 9000 low resolution ($R \sim 1,800$) spectra, 
spectral classifications, radial velocities and atmospheric parameters  
for a statistically complete sample of $\sim$ 3 million stars of all colours down to a limiting magnitude of $r \sim 17.8$\,mag (18.5\,mag for selected fields) 
that are uniformly and randomly selected from the ($r$, $g - r$) and ($r$, $r - i$) Hess diagrams. 
Together with distances, proper motions and radial velocities provided by Gaia, the LSS-GAC will yield a unique and promising dataset to
advance our understanding of the structure and assemblage of the Galaxy, particularly the Galactic disk(s).
In addition to the main survey, the LSS-GAC also targets hundreds of thousands of objects
in the vicinity fields of M\,31 and M\,33, the other two spirals in the Local Group.
The targets include foreground Galactic stars, background quasars and various objects of special interests in M\,31 and M\,33,
such as PNe, H\,{\sc ii} regions, supergiants and globular clusters.
Supplementary to the main survey, the LSS-GAC VB survey covers a significant fraction of 
randomly selected very bright stars ($r \le 14$ mag) in the whole northern celestial hemisphere,
yielding an excellent sample of local stars of a wide sky coverage, selected with a 
simpler algorithm, to probe the solar neighbourhood.

In this paper, we present a detailed description of the survey design, target selection, observation and data reduction of the LSS-GAC. 
Properties of data collected so far are also presented and discussed. By June 2013, end of the first year Regular Surveys, 
a total number of 1,040,649 [672,721] spectra of  S/N(7450\AA) $\ge$ 10 [S/N(4650\AA) $\ge$ 10] have been collected,
including 441,495 [219,045], 123,916 [60,869] and 475,238 [392,807] from the main, the M\,31-M\,33 and the VB surveys, respectively.
About 15 -- 25 per cent spectra are from duplicate targets, opening up the possibility of time-domain spectroscopy.

For stars of spectral S/N(4650\AA) $\ge$ 10, 
the pipelines for spectral flux-calibration and for stellar parameter determinations that have been developed at Peking University 
(cf. the two companion papers by Xiang et al. 2014a,b)
have been able to deliver accurately flux-calibrated spectra and robust stellar parameters, including 
radial velocities $V_{\rm r}$, and atmospheric parameters \teff, \logg~and \feh, 
with an accuracy of 5\,km\,s$^{-1}$, 150\,K, 0.25\,dex and 0.15\,dex, respectively.
Three samples consisting of stars of spectral S/N(4650\,{\AA}) $\geq 10$ are defined, corresponding to, respectively, 
the footprints of the LSS-GAC main, M31-M33 and VB surveys. Spectral classifications and 
stellar parameters yielded by the LSP3 are presented in publically accessible online value-added catalogues, supplementary to the official LAMOST DR1. 
The  value-added catalogues also contain  multi-band photometric data from the far-UV to the  mid-IR 
collected from the GALEX, XSTPS-GAC, 2MASS and WISE surveys.
The values of the interstellar extinction $E(B-V)$, as well as distances of the sample stars have been determined.
An accuracy of 0.04 mag in \ebv~ is achieved.
Proper motions collected from the PPMXL and UCAC4 catalogues, 
after corrected for systematics using quasars, are also provided. The values are 
typically accurate to 5 mas yr$^{-1}$.  
New determinations of proper motions are deduced by combining measurements of the XSTPS-GAC and 2MASS surveys 
and are accurate to about  6 mas yr$^{-1}$. 
Finally, from the distances, proper motions and radial velocites, the orbital parameters (eccentricity $e$, guiding radius $R_{\rm g}$) are 
calculated and included in the value-added catalogues for all sample stars.

Samples released in this first release of value-added catalogues are biased to brighter and bluer stars than those 
observed with the LAMOST hitherto, in particular for stars targeted with M plates. This is partly caused by 
the fact that some of the observations fail to reach the designed depth of the plates. With better observing 
strategy and more strict quality control, the situation is much improved in the second year Regular Surveys 
from September 2013 to May 2014. Restricted by the lack of suitable spectral templates in the red wavelength range 
(beyond 7400\,{\AA}), the current implementation of the LSP is also unable to deliver robust stellar atmospheric 
parameters for many red stars targeted by the surveys. As such, a significant chunk of (almost half) stars observed 
hitherto that have S/N(4650\,{\AA}) $\le 10$ but S/N(7450\,{\AA}) $\ge 10$ (many of them are cool dwarfs or giants) 
are missing in the current release of value-added catalogues. We plan to improve the situation in the next data release.

\vspace{7mm} \noindent {\bf Acknowledgments}{
We would like to thank the referee for his/her helpful comments.
This work is supported by National Key Basic Research Program of China 2014CB845700. 
Guoshoujing Telescope (the Large Sky Area Multi-Object Fibre Spectroscopic Telescope, LAMOST) is 
a National Major Scientific Project built by the Chinese Academy of Sciences. Funding for the project 
has been provided by the National Development and Reform Commission. LAMOST is operated and managed by the National 
Astronomical Observatories, Chinese Academy of Sciences.
This work has made use of data products from the Sloan Digital Sky Survey (SDSS),
Galaxy Evolution Explorer (GALEX), Two Micron All Sky Survey (2MASS), Wide-field Infrared Survey
Explorer (WISE), NASA/IPAC Infrared Science Archive and the cross-match service provided by CDS, Stra{\ss}bourg.
}

\label{lastpage}

\end{document}